\documentclass[aps,prx,preprint,bibliography,superscriptaddress]{revtex4-1}
\usepackage[utf8]{inputenc}
\usepackage{amsmath}

\usepackage{physics}
\usepackage{mathtools}
\usepackage{amsfonts}
\usepackage{amssymb}
\usepackage{graphicx}
\usepackage{wrapfig}
\usepackage{textcomp}
\usepackage{color}
\usepackage[dvipsnames]{xcolor}
\usepackage{comment}
\usepackage{multirow}

\begin{document}

\title{Altermagnetism from the viewpoint of chemistry}

\author{Nayana Devaraj}
\affiliation{Solid State and Structural Chemistry Unit, Indian Institute of Science, Bangalore 560012, India}
\author{Anumita Bose} 
\affiliation{Scuola Internazionale Superiore di Studi Avanzati (SISSA), I-34136 Trieste, Italy.}
\author{Md Afsar Reja}
\affiliation{Solid State and Structural Chemistry Unit, Indian Institute of Science, Bangalore 560012, India}
\author{Arka Bandyopadhyay}
\affiliation{Institut f\"{u}r Theoretische Physik und Astrophysik and W\"{u}rzburg-Dresden Cluster of Excellence ctd.qmat, Universit\"{a}t W\"{u}rzburg, Am Hubland Campus S\"{u}d, 97074 W\"{u}rzburg, Germany.}
\author{Awadhesh Narayan}
\email{awadhesh@iisc.ac.in}
\affiliation{Solid State and Structural Chemistry Unit, Indian Institute of Science, Bangalore 560012, India}

\date{\today}

\begin{abstract}
Magnetism has been a central theme of research in chemistry, physics, and materials science, with chemical composition and bonding playing key roles in determining magnetic behavior. Altermagnets are a newly identified class of magnetic materials that combine features of conventional ferromagnets and antiferromagnets, arising from specific symmetry and electronic-structure motifs. In this review, we present a chemistry-driven viewpoint on altermagnetism, highlighting how crystal chemistry, bonding, and electronic structure enable this unconventional magnetic order. We begin by introducing the fundamental concepts required to understand altermagnets, with an emphasis on symmetry considerations, orbital character, and electronic-structure signatures. We then survey the diverse material families in which altermagnetism has been identified, including pnictides, oxides, and metal-organic and covalent-organic frameworks, drawing attention to coordination environments and structure–property relationships that favor altermagnetic order. We subsequently present experimental approaches which are useful for the characterization of altermagnetic materials. We examine \textit{ab initio} materials discovery as a promising strategy for identifying new altermagnets, emphasizing how chemical constraints, such as symmetry and bonding, can guide computational searches. Other than their intrinsic importance, altermagnets provide interesting possibilities for technology. For this reason, we highlight possible applications that may be enabled through altermagnetic materials, along with their coupling with existing orders such as ferroelectricity and superconductivity. In conclusion, we point out some challenges and prospects, where chemically-based design guidelines can play an important role towards advancing altermagnetism research. In summary, this review offers an account of recent developments in altermagnetism, from basic concepts to the current state-of-the-art.
\end{abstract}

\maketitle

\tableofcontents

\section{Introduction}

Magnetism has been a central theme of research in chemistry, physics, and materials science for well over a century~\cite{spaldin2010magnetic}. Over the years, the understanding and exploitation of magnetic order have proceeded hand in hand with advances in chemical synthesis, crystallographic characterization, and electronic structure theory. Throughout this history, two fundamental classes of collinear magnetic order have been at the foundation of investigations, namely, ferromagnetism, in which atomic moments align parallel to each other to produce a net magnetization, and antiferromagnetism, in which moments on neighboring atoms align in an antiparallel manner, resulting in zero net magnetization.

This two-fold classification has been very useful. Ferromagnetism, understood since the early work of Weiss, underpins technologies ranging from permanent magnets and magnetic recording to spintronic memories~\cite{coey2010magnetism}. Antiferromagnetism, identified by N\'eel in the 1930s~\cite{neel1948proprietes}, was long regarded as magnetically `invisible', as a consequence of lacking the net moment and spin-polarized transport signatures that made ferromagnets technologically useful. The recognition that antiferromagnets possess their own distinctive advantages, including robustness to stray fields and ultrafast dynamics, has more recently led to a revival in interest in the field of antiferromagnetic spintronics~\cite{jungwirth2016antiferromagnetic,baltz2018antiferromagnetic}.

The chemical foundations of this classification are well established. The Goodenough-Kanamori rules~\cite{goodenough1963magnetism,kanamori1959superexchange}, formulated in the 1950s, provide a remarkably successful framework for predicting whether the exchange interaction between magnetic ions bridged by ligands will be ferromagnetic or antiferromagnetic. These rules are based on the orbital occupancy, bond angles, and metal-ligand covalency. Crystal-field theory, ligand-field theory, and the language of coordination chemistry have been the natural tools for rationalizing magnetic behavior in solids, and the interplay between crystal structure, electronic structure, and magnetic order is a central theme of solid-state chemistry.

Against this backdrop, the recent identification of a different fundamental class of collinear magnetic order, termed altermagnetism, represents a conceptual advance with remarkable implications for both fundamental science and technology~\cite{tamang2025altermagnetism}. The possibility of such a phase was identified independently by several theoretical groups~\cite{hayami2019momentum,vsmejkal2020crystal,yuan2020giant}. Subsequently, \v{S}mejkal, Sinova, and Jungwirth coined the term `altermagnetism' and provided the systematic classification based on spin-symmetry groups that has since become the standard framework~\cite{vsmejkal2022beyond,vsmejkal2022emerging}.

The defining feature of an altermagnet is subtle but consequential. Similar to a conventional antiferromagnet, an altermagnet has two sublattices with antiparallel spins and zero net magnetization. Notably, while in the case of a usual antiferromagnet, the sublattices with opposite spins have the symmetry operation of lattice translation or inversion, for altermagnets, it is a crystal rotation that links these sublattices together. The presence of just one such symmetry feature yields important electronic effects. Now, the bands of the spin up and spin down states have a gap between them with positive and negative signs in momentum space. Consequently, the material demonstrates both antiferromagnetic character and ferromagnetism features at once. Earlier, it was believed that such a combination was impossible due to symmetry considerations.

Remarkably, altermagnetism does not need unusual chemistry or spin-orbit interactions. This phenomenon is non-relativistic and arises from the interplay between conventional antiferromagnetic exchange and the geometry of the ligand field environment. In chemical language, if the coordination polyhedra surrounding opposite-spin metal centres are rotated relative to each other (rather than being translationally equivalent or centrosymmetrically related), the material is an altermagnet. Long studied antiferromagnets, such as the rutile structure of MnF$_2$ and the NiAs structure of MnTe and CrSb, are now recognized as altermagnets. While this reclassification does not change their chemistry, but it reveals that they possess electronic properties that had been overlooked for decades. The experimental confirmation of altermagnetic band splitting has followed rapidly. In 2024, Krempasky \textit{et al.} used spin-resolved photoemission spectroscopy to directly observe the altermagnetic lifting of Kramers spin degeneracy in MnTe~\cite{krempasky2024altermagnetic}. On the transport side, Reichlova \textit{et al.} observed a spontaneous anomalous Hall effect at zero net magnetization, which is a signature that is forbidden in conventional antiferromagnets but permitted in altermagnets~\cite{reichlova2024observation}.

\begin{figure}[t]
\centerline{\includegraphics[scale=0.60]{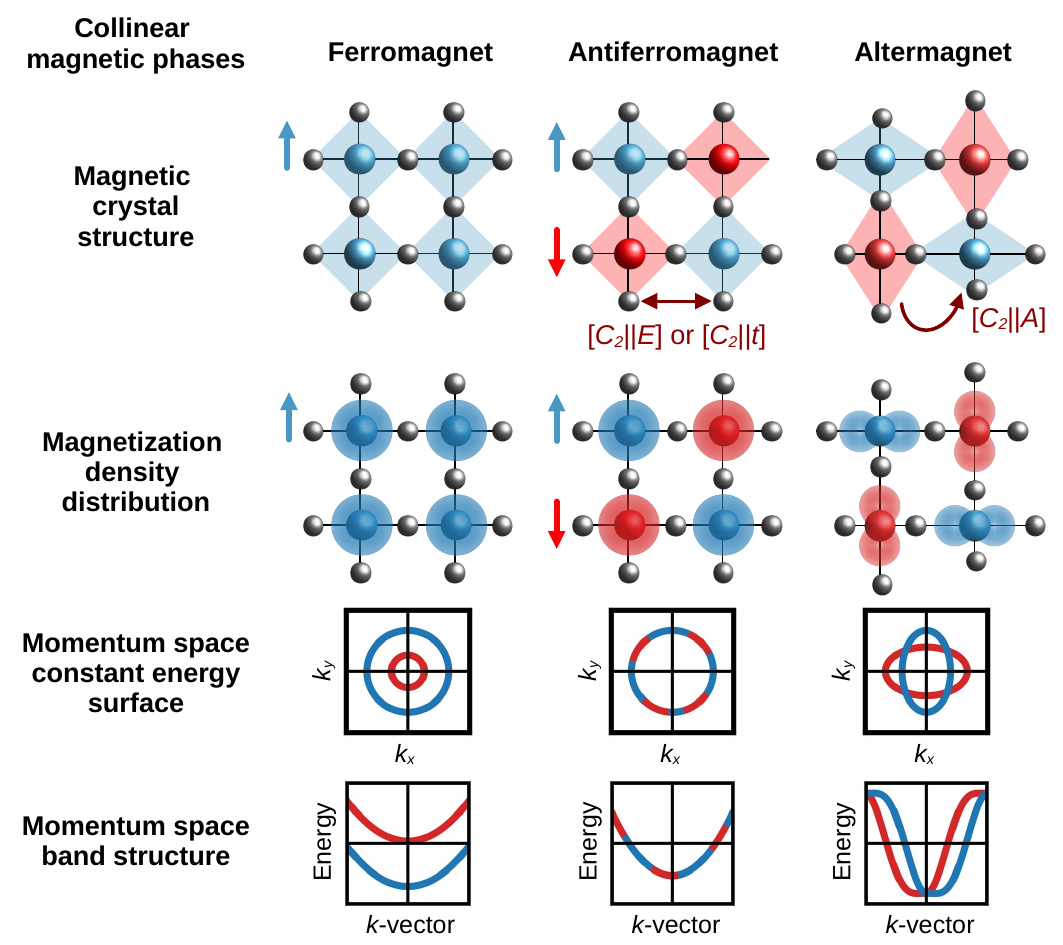}}
\caption{\textbf{A comparison of three main collinear magnetic phases.} The crystal structures of ferromagnets, antiferromagnets, and altermagnets are shown in the first row, where atoms with spin-up and spin-down magnetization are represented by blue and red colors, respectively. Non-magnetic atoms are shown in gray color. The spin polarized magnetization density distributions, Fermi surfaces, and band structures of these magnetic classes are presented in the second, third, and fourth rows, respectively.}
\label{fig_Mag_class}
\end{figure}

In this review, we present a chemistry-focused account of altermagnetism, emphasizing how crystal chemistry, bonding, and electronic structure collectively make this unconventional magnetic order possible. Our goal is to make altermagnetism accessible from a chemistry perspective and to highlight that the underlying principles for discovering and controlling altermagnets are underpinned by well-established chemical concepts. Following this introduction, the review is organized as follows. Section~\ref{sec:basic_concepts} introduces the basic concepts of altermagnetism, including real and momentum space pictures, while highlighting the salient differences and similarities with ferromagnets and antiferromagnets. Next, in Section~\ref{sec:minimal_models_altermagnetism}, we present a simple yet illustrative microscopic tight-binding model on a square lattice. We use this model to highlight the interplay of magnetic exchange and ligand field to realize the altermagnetic phase. Subsequently, we present a few important theoretical concepts, including the spin group theory, canonical form factors, and effects of spin-orbit coupling (Section~\ref{sec:spin_group_theory}). In Section~\ref{sec:chemical_design}, we turn to the key chemical design principles in engineering altermagnetism, including the role of coordination environment as well as non-magnetic atoms. Next, we summarize the various known altermagnetic material classes in Section~\ref{sec:material_classes}, with a detailed analysis of NiAs and rutile structure types. We also catalog altermagnetism in two-dimensional materials, along with metal- and covalent-organic frameworks. In Section~\ref{sec:synthesis} we discuss growth methods - flux growth, chemical vapour transport, thin-film epitaxy, and detwinning - used to synthesize altermagnetic crystals. In Section~\ref{sec:experimental_characterization}, we summarize the experimental techniques that are directly useful for characterizing altermagnets. We highlight the central ideas behind these techniques and how they enable characterization of altermagnets. We then outline a general workflow to search for altermagnets combining symmetry and first-principles computations in Section~\ref{sec:theoretical_search}. Effects of structural and external perturbations on altermagnetism are discussed in Section~\ref{sec:effect_external_perturbation}. In Section~\ref{sec:potential_applications}, we move on to potential applications that altermagnets may lead to, emphasizing their cross-coupling with other properties such as ferroic order, topological phases, and superconductivity. We then summarize the future directions and prospects in this rapidly growing field in Section~\ref{sec:future_directions}, and end with an outlook towards the role of chemical design principles in this field.

\begin{table*}[ht]
\centering
\caption{Comparison of the three collinear magnetic phases from structural and orbital perspectives. Here FM stands for ferromagnet, AFM for antiferromagnet, while AM denotes altermagnet.}
\label{tab:comparison}
\renewcommand{\arraystretch}{1.45}
\tiny
\begin{tabular}{@{} p{3.2cm} p{4.0cm} p{4.0cm} p{4.0cm} @{}}
\hline
\hline
\textbf{Property} & \textbf{Ferromagnet} & \textbf{Antiferromagnet} & \textbf{Altermagnet} \\
\hline
\hline
Spin arrangement
  & All moments parallel
  & Antiparallel on two sublattices, related by translation or inversion
  & Antiparallel on two sublattices, related by a crystal rotation/mirror \\

Net magnetization
  & Nonzero
  & Zero
  & Zero \\

Dominant exchange mechanism
  & Direct exchange or double exchange; Hund's rule coupling within shells
  & Superexchange through bridging ligands (Goodenough-Kanamori rules);
    $180^{\circ}$ metal-ligand-metal pathways
  & Same superexchange as AFM, but with an additional requirement on
    the spatial arrangement of the ligand environment \\

Crystal-field requirements
  & No special requirement on coordination-cage orientation; all sites
    may be symmetry-equivalent
  & No special requirement; opposite-spin sites related by translation
    or inversion (coordination cages are identical or centrosymmetrically
    related)
  & Coordination polyhedra on opposite-spin sublattices must be
    rotated (not translated or inverted) relative to each other,
    creating anisotropic orbital overlap \\

Key structural motif
  & Any crystal structure supporting magnetic order
  & Any structure where opposite-spin sites are related by a translation, glide, or inversion center
  & Structures with alternating orientation of ligand cages \\

Role of non-magnetic atoms
  & Ligand identity affects exchange strength but not the magnetic class
  & Same as FM
  & Ligand cage geometry determines whether the
    sublattice-connecting symmetry is a rotation (AM) or
    translation/inversion (AFM). Changing the ligand arrangement
    can switch between AFM and AM \\

Orbital character
  & Spin splitting is isotropic in $\mathbf{k}$-space. All
    $d$-orbitals on all sites see nearly the same exchange field
  & Bands are spin-degenerate; orbital character irrelevant to spin
    splitting, which is absent
  & Spin splitting has $d$/$g$/$i$-wave nodal structure inherited from
    the anisotropic crystal-field environment. Orientation of
    $d$-orbital lobes relative to lattice vectors directly controls
    splitting magnitude and direction \\

Band structure signature
  & Rigid spin-up/spin-down band offset; splitting is roughly uniform
    across the Brillouin zone
  & Spin-up and spin-down bands are degenerate at every $\mathbf{k}$-point
  & Spin-split bands with momentum-dependent sign reversal. Splitting
    vanishes on 2, 4, or 6 nodal planes ($d$-, $g$-, $i$-wave)\\

\hline
\hline
\end{tabular}
\end{table*}

\section{Basic concepts of altermagnets}
\label{sec:basic_concepts}

Magnetism in materials comes from the orbital motion and spin of electrons, which generate atomic magnetic moments. The collective ordering of these moments, through exchange interactions and crystal symmetry, gives rise to different magnetic phases. Traditionally, magnetic materials have been categorized according to the spatial arrangement of magnetic moments and the resulting net magnetization. In the case that all magnetic moments are collinear, this classification yields two types of magnets, ferromagnets and antiferromagnets (see Fig.~\ref{fig_Mag_class}). In ferromagnets, magnetic moments are aligned parallel to each other, resulting in a finite net magnetization. This magnetic order breaks time reversal symmetry, leading to a non-zero spin polarization and spin-split bands in the momentum space. In contrast, antiferromagnets exhibit antiparallel alignment of magnetic moments that fully compensate each other, resulting in a zero net magnetization. The combined action of time-reversal symmetry and crystal symmetry in these systems leads to spin degenerate bands.

For many decades, this classification was sufficient to explain the magnetic behavior of most materials. However, several materials, which were considered conventional antiferromagnets, were predicted to exhibit phenomena typically associated with time-reversal-symmetry-breaking magnetic systems, such as anomalous Hall effect, Kerr effect, spin-split band structure, spin torque, tunnel and giant magnetoresistance~\cite{vsmejkal2020crystal,mazin2021prediction,vsmejkal2022anomalous,naka2020anomalous,samanta2020crystal,zhou2021crystal,naka2021perovskite,yuan2020giant,vsmejkal2022giant,shao2021spin}. Some of these effects were also observed experimentally~\cite{feng2022anomalous,bose2022tilted,bai2022observation}. The emergence of these effects challenged the conventional understanding of magnetic order and motivated a closer examination of the role of crystal symmetry in determining magnetic properties. This led to the identification of a new magnetic class, termed `altermagnets', as well as a symmetry-based reclassification of collinear magnets~\cite{vsmejkal2022beyond,vsmejkal2022emerging}. The key distinguishing characteristics of ferromagnets, antiferromagnets, and altermagnets are compared schematically in Fig.~\ref{fig_Mag_class}, and a summary of their characteristics from structural and orbital perspectives is presented in Table~\ref{tab:comparison}.

Understanding this new classification requires moving beyond the conventional description of magnetism, which considers only the arrangement of magnetic atoms and their spin orientations. Instead, one must consider spin sublattices, which consist of magnetic atoms together with their surrounding chemical environment formed by non-magnetic atoms~\cite{vsmejkal2022beyond,vsmejkal2022emerging,fender2025chemical,wei2024crystal}. The symmetry and arrangement of these non-magnetic atoms play a crucial role in determining the magnetic properties of the material. When non-magnetic atoms create a symmetric environment around magnetic atom, the magnetization density remains symmetric such as that in ferromagnets and antiferromagnets as illustrated in Fig.~\ref{fig_Mag_class}. Consequently, non-magnetic atoms do not play any role in determining magnetic properties in ferromagnets and antiferromagnets. However, the surrounding ligand environment is anisotropic in the case of altermagnets. This, in turn, leads to an anisotropic magnetization density distribution and also fundamentally changes the symmetry relations connecting spin sublattices.

The unconventional symmetry relations in altermagnets cannot be fully described within the conventional magnetic space-group formalism, which treats spin and crystal-lattice symmetries as coupled. Describing altermagnetism, a fundamentally non-relativistic phenomenon, requires the spin space group formalism, in which spin and crystal symmetries are treated as independent symmetry operations~\cite{brinkman1966theory,litvin1974spin,litvin1977spin}. We will discuss this in detail in the forthcoming Section~\ref{sec:spin_group_theory}.

Although both antiferromagnets and altermagnets exhibit zero net magnetization resulting from the compensation of opposite-spin sublattices, the symmetry relations connecting these sublattices are fundamentally different. This distinction forms the basis for their contrasting electronic and magnetic properties. In antiferromagnets, opposite sublattices are connected by translation or inversion~\cite{nunez2006theory,vsmejkal2017electric}. In altermagnets, however, these symmetries do not connect the opposite-spin sublattices. Instead, they are connected by proper or improper rotational symmetries~\cite{vsmejkal2022beyond,vsmejkal2022emerging,mazin2022altermagnetism,jungwirth2026symmetry,jungwirth2025altermagnetism}. Such rotational relations cannot occur in arbitrary crystal structures and require a specific arrangement of ligand cages, where the local environments surrounding opposite-spin magnetic atoms are rotated with respect to one another. Consequently, altermagnetism can only emerge in materials possessing the required crystallographic symmetry. In contrast, ferromagnetism and antiferromagnetism can occur in a much broader range of crystal structures.

This unique rotational relation between opposite-spin sublattices is reflected in both real and momentum space. Just as the spin sublattices are connected by a rotation in real space, the Fermi surfaces corresponding to opposite spin channels are related by a rotation in momentum space as illustrated schematically in Fig.~\ref{fig_Mag_class}. In contrast, the spin-resolved Fermi surfaces of antiferromagnets are identical.

The broken time-reversal symmetry lifts Kramers degeneracy in altermagnets and as a result they exhibit band structures with spin splitting. Although ferromagnets also exhibit spin split band structures, the nature of spin splitting occurring in altermagnets is distinct from that in ferromagnets. In altermagnets, spin splitting is highly anisotropic in nature, and vanishes along certain symmetry protected regions of the Brillouin zone called nodal planes, where the up and down spin bands remain degenerate. Everywhere else, bands are spin split and, further, the magnitude and the sign of splitting varies with momentum. The spin splitting can be defined as $\Delta E(\mathbf{k}) = E_{\uparrow}(\mathbf{k}) - E_{\downarrow}(\mathbf{k})$, and its sign alternates across momentum space, changing across nodal planes. The alternating spin polarization in real space and momentum space motivates to call this particular magnetic class as altermagnets. In contrast, in ferromagnets, the spin splitting is typically isotropic and does not exhibit such sign-changing behavior across the Brillouin zone.

\begin{figure}[t]
\centerline{\includegraphics[scale=0.40]{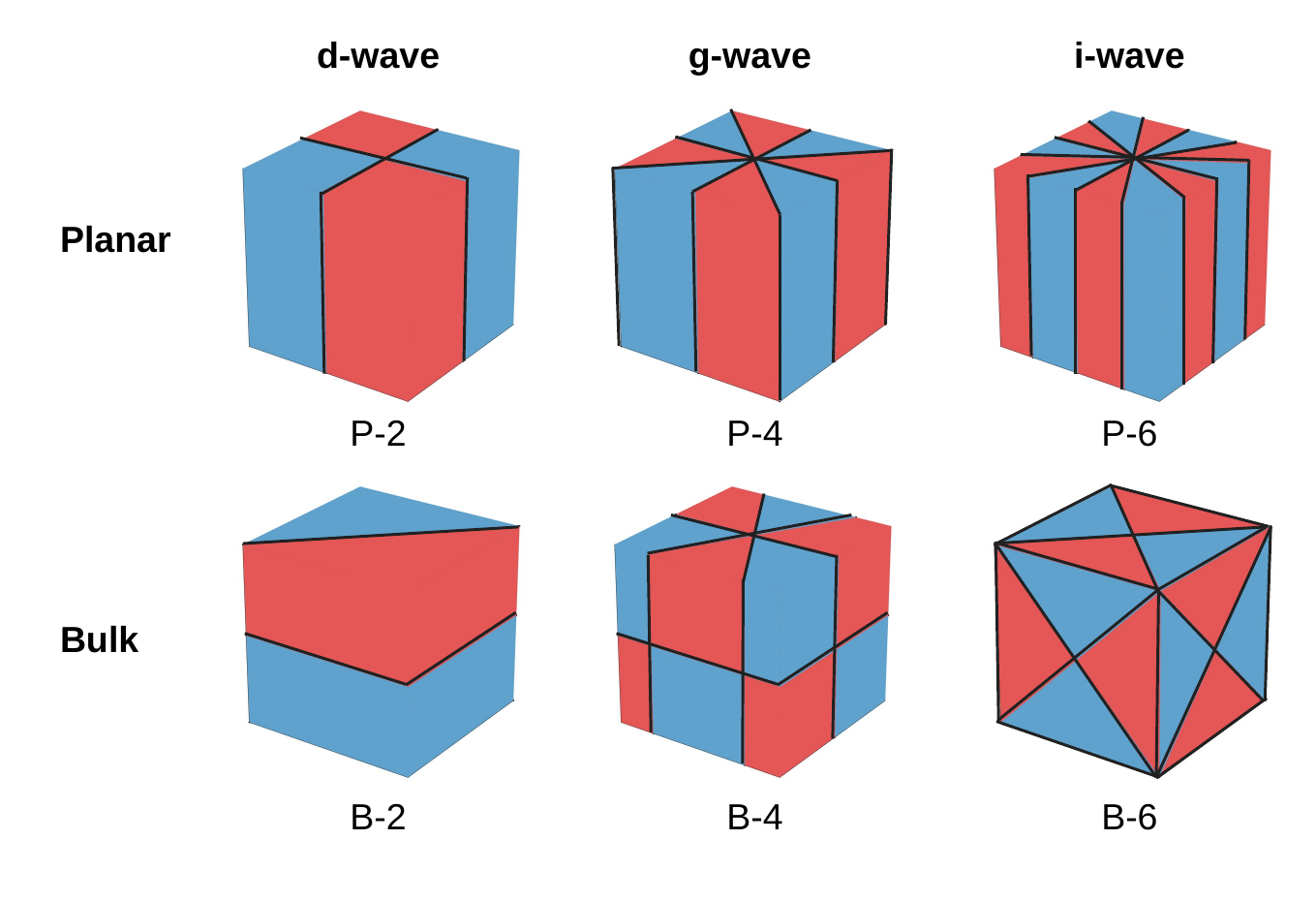}}
\caption{\textbf{Schematic of spin splitting in  $d$-,$g$-, and $i$- wave altermagnets.} Illustration of spin splitting in planar and bulk $d$-, $g$-, and $i$- wave altermagnets are shown in the first and second rows, respectively. Blue and red color regions represent the spin up and spin down regions and the nodal planes are shown in black lines.}
\label{fig_d-g-i}
\end{figure}

The momentum-dependent spin splitting in altermagnets follows characteristic $d$-, $g$-, and $i$-wave symmetries, originating from the anisotropic crystal-field environment and the underlying crystal symmetry~\cite{vsmejkal2022beyond,vsmejkal2022emerging,bhowal2025non}. Accordingly, altermagnets are classified into $d$-, $g$-, and $i$-wave types, characterized by 2, 4, and 6 nodal planes passing through the Brillouin zone center ($\Gamma$ point), respectively. Based on the orientation of these nodal planes, they are further categorized as planar or bulk altermagnets. Planar altermagnets (P-2, P-4, and P-6) possess only nodal planes parallel to the rotation axis, whereas bulk altermagnets (B-2, B-4, and B-6) contain an additional nodal plane perpendicular to the rotation axis. Schematic illustrations of these altermagnetic wave types are shown in Fig.~\ref{fig_d-g-i}, while their characteristics and representative material examples are summarized in Tables~\ref{tab:3D} and \ref{tab:2D}.
 
\section{A microscopic lattice model to understand altermagnetism}
\label{sec:minimal_models_altermagnetism}

\begin{figure}[t]
\centerline{\includegraphics[scale=0.95]{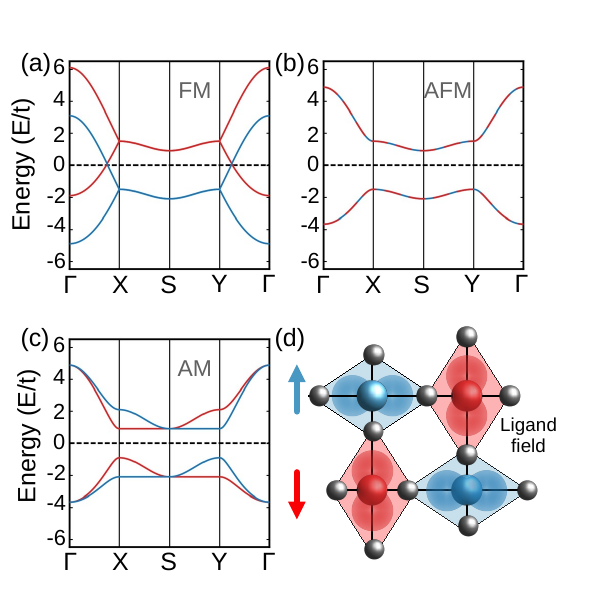}}
\caption{\textbf{Local ligand-field mechanism and spin-resolved band structures of the altermagnetic model.} (a-c) Spin-resolved band structures along the high-symmetry path
$\Gamma$-$X$-$S$-$Y$-$\Gamma$, with the energy measured in units of the nearest-neighbor hopping amplitude $t$. The red and blue curves denote the two opposite spin channels. The horizontal dashed line marks $E=0$, while the vertical lines indicate the high-symmetry momenta. (a) Ferromagnetic (FM) case with $\Delta_{\mathrm{F}}=1.50$, $\Delta_{\mathrm{AF}}=0.00$, and $\delta=0.00$, showing the conventional exchange-induced spin splitting. (b) Antiferromagnetic (AFM) case with $\Delta_{\mathrm{F}}=0.00$, $\Delta_{\mathrm{AF}}=1.50$, and $\delta=0.00$,
where the two spin channels remain degenerate due to symmetry. (c) Altermagnetic (AM) case with $\Delta_{\mathrm{F}}=0.00$, $\Delta_{\mathrm{AF}}=1.50$, and $\delta=0.30$, exhibiting momentum-dependent spin splitting in the absence of net magnetization. (d) Schematic illustration of the local ligand-field mechanism. The red and blue sites represent compensated antiparallel spin sublattices, whereas the anisotropic ligand environments indicate local crystal fields with different orientations on the two sublattices. This sublattice-dependent ligand-field anisotropy lifts the spin degeneracy without generating a net magnetic moment.}
\label{fig:altermagnetic_ligand_field}
\end{figure}

To illustrate the microscopic origin of altermagnetic spin splitting, we first introduce a simple tight-binding model that continuously connects ferromagnetic, conventional antiferromagnetic, and altermagnetic limits. We consider a square lattice with two magnetic sublattices, $A$ and $B$, whose magnetic ions may be viewed as transition-metal sites surrounded by ligand coordination cages, as sketched in Fig.~\ref{fig:altermagnetic_ligand_field}(d). The key structural ingredient is that the local ligand environment on sublattice $B$ is rotated by $90^\circ$ relative to that on sublattice $A$. This rotation interchanges the dominant orbital overlaps along the $x$ and $y$ directions and produces a sublattice-dependent hopping anisotropy. Such local site-symmetry reduction is the microscopic ingredient that allows a staggered exchange field to generate a momentum-dependent spin splitting without producing a net magnetization~\cite{hayami2019momentum,yuan2020giant,bhowal2024ferroically,mcclarty2024landau,roig2024minimal}.

For each spin projection $\sigma=\pm1$, the Bloch Hamiltonian in the sublattice basis
$(A,B)$ can be written as
\begin{equation}
H_{\sigma}(\mathbf{k})
=
\begin{pmatrix}
\varepsilon_{A\sigma}(\mathbf{k}) & \gamma(\mathbf{k})\\
\gamma(\mathbf{k}) & \varepsilon_{B\sigma}(\mathbf{k})
\end{pmatrix},
\label{eq:microscopic_matrix}
\end{equation}
with
\begin{align}
\varepsilon_{A\sigma}(\mathbf{k})
&=
-\sigma(\Delta_F+\Delta_{AF})
+
2t_x'\cos k_x
+
2t_y'\cos k_y ,
\label{eq:eps_A_sigma}
\\
\varepsilon_{B\sigma}(\mathbf{k})
&=
-\sigma(\Delta_F-\Delta_{AF})
+
2t_y'\cos k_x
+
2t_x'\cos k_y ,
\label{eq:eps_B_sigma}
\\
\gamma(\mathbf{k})
&=
4t\cos\frac{k_x}{2}\cos\frac{k_y}{2}.
\label{eq:gamma_k}
\end{align}
Here $t$ is the nearest-neighbor hopping between the two sublattices, while $t_x'$
and $t_y'$ are anisotropic next-nearest-neighbor hopping terms within each sublattice. We parametrize them as
\begin{equation}
t_x'=t'_{\rm avg}+\frac{\delta}{2},
\qquad
t_y'=t'_{\rm avg}-\frac{\delta}{2}.
\label{eq:tprime_delta}
\end{equation}
The parameter $\delta$ measures the orbital-overlap anisotropy induced by the
rotated ligand cages. The exchange parameters $\Delta_F$ and $\Delta_{AF}$ denote
the ferromagnetic and antiferromagnetic components of the local exchange field,
respectively.

It is useful to rewrite Eq.~\eqref{eq:microscopic_matrix} in terms of Pauli matrices
in the sublattice space. Defining
\begin{equation}
\epsilon_{\rm av}(\mathbf{k})
=
2t'_{\rm avg}(\cos k_x+\cos k_y),
\qquad
g_d(\mathbf{k})
=
\cos k_x-\cos k_y,
\label{eq:epsav_gd}
\end{equation}
one obtains
\begin{equation}
H_{\sigma}(\mathbf{k})
=
\left[
\epsilon_{\rm av}(\mathbf{k})-\sigma\Delta_F
\right]\tau_0
+
\gamma(\mathbf{k})\tau_x
+
\left[
\delta g_d(\mathbf{k})-\sigma\Delta_{AF}
\right]\tau_z .
\label{eq:compact_microscopic_model}
\end{equation}
This form makes the microscopic mechanism quite transparent. The term proportional to $\Delta_{AF}$ is the staggered exchange field. The term proportional to $\delta g_d(\mathbf{k})$ is the sublattice-dependent orbital anisotropy. Altermagnetic spin splitting appears when these two terms act together.

The band energies are obtained to be
\begin{equation}
E_{\sigma,\nu}(\mathbf{k})
=
\epsilon_{\rm av}(\mathbf{k})
-\sigma\Delta_F
+
\nu
\sqrt{
\gamma^2(\mathbf{k})
+
\left[
\delta g_d(\mathbf{k})-\sigma\Delta_{AF}
\right]^2
},
\label{eq:exact_bands_microscopic}
\end{equation}
where $\nu=\pm1$ labels the bonding and antibonding sublattice bands. This result
shows explicitly how ferromagnetism, conventional antiferromagnetism, and
altermagnetism arise in different parameter regimes of the same microscopic model.

First, if $\Delta_F\neq0$ and $\Delta_{AF}=0$, the system behaves like a ferromagnet [Fig.~\ref{fig:altermagnetic_ligand_field}(a)]. The two spin sector bands are shifted by the uniform exchange field $\Delta_F$. In the simplest case, the spin splitting is essentially uniform in the momentum space.

Second, if $\Delta_F=0$, $\Delta_{AF}\neq0$, and $\delta=0$, the model describes a
conventional compensated antiferromagnet [Fig.~\ref{fig:altermagnetic_ligand_field}(b)]. In this limit,
\begin{equation}
E_{\sigma,\nu}(\mathbf{k})
=
\epsilon_{\rm av}(\mathbf{k})
+
\nu
\sqrt{
\gamma^2(\mathbf{k})
+
\Delta_{AF}^2
},
\label{eq:ordinary_af_degenerate}
\end{equation}
which is independent of $\sigma$. The bands are therefore spin degenerate at every
momentum point.

Finally, if $\Delta_F=0$, $\Delta_{AF}\neq0$, and $\delta\neq0$, the system is
altermagnetic, as shown in Fig.~\ref{fig:altermagnetic_ligand_field}(c). The spin splitting within band $\nu$ is
\begin{align}
\Delta E_{\nu}(\mathbf{k})
&=
E_{+,\nu}(\mathbf{k})-E_{-,\nu}(\mathbf{k})
\nonumber\\
&=
\nu
\left\{
\sqrt{
\gamma^2(\mathbf{k})
+
\left[
\delta g_d(\mathbf{k})-\Delta_{AF}
\right]^2
}
-
\sqrt{
\gamma^2(\mathbf{k})
+
\left[
\delta g_d(\mathbf{k})+\Delta_{AF}
\right]^2
}
\right\}.
\label{eq:exact_alt_splitting}
\end{align}

It is interesting to note that this splitting vanishes when $g_d(\mathbf{k})=0$ and changes sign when $g_d(\mathbf{k})$ changes sign. Thus the spin splitting has a $d$-wave nodal structure. For weak orbital anisotropy, $|\delta g_d(\mathbf{k})|\ll
\sqrt{\gamma^2(\mathbf{k})+\Delta_{AF}^2}$, Eq.~\eqref{eq:exact_alt_splitting}
reduces to
\begin{equation}
\Delta E_{\nu}(\mathbf{k})
\simeq
-2\nu
\frac{\Delta_{AF}\delta}
{\sqrt{\gamma^2(\mathbf{k})+\Delta_{AF}^2}}
g_d(\mathbf{k}).
\label{eq:weak_delta_alt_splitting}
\end{equation}
Therefore, after projection onto band $\nu$, the low-energy Hamiltonian has the
altermagnetic form
\begin{equation}
H_{\rm eff}^{(\nu)}(\mathbf{k})
=
\epsilon_{\nu}(\mathbf{k})\sigma_0
+
J_{\nu}(\mathbf{k})g_d(\mathbf{k})\sigma_z ,
\label{eq:projected_altermagnetic_model}
\end{equation}
where
\begin{equation}
J_{\nu}(\mathbf{k})
=
-\nu
\frac{\Delta_{AF}\delta}
{\sqrt{\gamma^2(\mathbf{k})+\Delta_{AF}^2}}.
\label{eq:Jnu_effective}
\end{equation}
This derivation illustrates an important general principle: the altermagnetic
splitting is not produced by local staggered exchange alone. It appears when
staggered exchange is projected through a momentum-dependent orbital and bonding
structure that distinguishes the two opposite-spin sublattices. This minimal model therefore motivates the spin-group formulation developed below, which provides the symmetry language for classifying such momentum-dependent exchange textures and for constructing the corresponding model Hamiltonians.

\section{Spin group theory of altermagnets and other concepts}
\label{sec:spin_group_theory}

The symmetry-based classification of altermagnetism requires moving beyond the 
conventional relativistic magnetic groups, as these groups treat spin and spatial degrees of freedom as coupled and cannot distinguish between different non-relativistic magnetic phases. The spin-group formalism, on the other hand, treats rotations in spin space and real space as entirely independent. This is an exact symmetry in the non-relativistic limit where spin-orbit coupling is negligible. A spin group is written as a direct product of a spin-only group, which acts purely in spin space, and a nontrivial spin group, whose elements are pairs of transformations acting independently on spin space and real space. For collinear magnets, the spin-only group has two important consequences. First, spin becomes a good quantum number with a common quantization axis across the entire Brillouin zone, so the band structure separates into non-mixing spin-up and spin-down channels. Second, there is always an effective inversion symmetry in reciprocal space, meaning the bands satisfy $\varepsilon(s, \mathbf{k}) = \varepsilon(s,-\mathbf{k})$, regardless of whether the crystal has a real-space inversion center.

Building on this theory, \v{S}mejkal, Sinova, and Jungwirth showed that all non-relativistic collinear magnets fall into exactly three symmetry types~\cite{vsmejkal2022beyond}. The first type describes ferromagnets, where the two spin channels are split throughout the entire Brillouin zone with no degeneracy enforced anywhere. This yields 11 distinct spin Laue groups. The second type 
describes conventional antiferromagnets. In this case a symmetry operation combining a spin-space $180^\circ$ rotation with a real-space transformation enforces global Kramers-like spin degeneracy across the Brillouin zone, again yielding 11 groups. The opposite-spin sublattices in these materials are connected by either a translation or by an inversion. The third type is the altermagnetic class, characterized by the coset decomposition of the form,

\begin{equation}
    R_s^{\mathrm{III}} = [E \| H] + [C_2 \| \mathbf{G} - H].
\end{equation}

Here, $H$ is a halving (index-two) subgroup of the real-space group $\mathbf{G}$, such that $\mathbf{G}$ is partitioned into the two cosets $H$ and $\mathbf{G}-H$, which are paired with the spin-space operations $E$ and $C_2$, respectively. The double vertical bar separates the spin-space transformation (left) from the real-space transformation (right), reflecting the independence of the two spaces in the non-relativistic spin-group formalism. In other words, the halving subgroup $H$ collects all real-space transformations that map atoms within the same spin sublattice onto each other, while the coset $\mathbf{G} - H$ collects the transformations that connect atoms between opposite spin sublattices only via proper or improper rotations (not by translation or inversion). This distinction leads to the alternating-sign spin splitting across momentum space while keeping the net magnetization strictly zero. There are precisely 10 nontrivial altermagnetic spin Laue groups, which can be constructed from only 8 of the  11 crystallographic Laue groups. The remaining three Laue groups, based on $\bar{1}$, $\bar{3}$, and $m\bar{3}$, cannot host altermagnetism. The wave symmetry class of each group is determined by the number of spin-degenerate nodal surfaces that cross the $\Gamma$-point in the Brillouin zone, where the spin splitting is forced to vanish by symmetry (Fig.~\ref{fig_d-g-i}). When the opposite-spin sublattices are connected by a symmetry operation that generates 2 such nodal surfaces, the spin splitting changes sign twice around $\Gamma$, giving $d$-wave character. When the connecting operation generates 4 nodal surfaces the result is $g$-wave, and with 6 nodal surfaces it is $i$-wave. The three planar $d$-wave groups, namely $^2m^2m^1m$, $^24/^1m$, and $^24/^1m^2m^1m$, each have 2 spin-degenerate nodal surfaces crossing $\Gamma$. These arise because the halving subgroup $H$ contains a mirror perpendicular to the principal rotation axis, which together with the opposite-spin-sublattice rotation generates exactly one pair of nodal planes in reciprocal space. The single planar $g$-wave group $^14/^1m^2m^1m$ has 4 nodal surfaces, generated by a four-fold rotation axis in ($\mathbf{G} - H$). The single planar $i$-wave group $^16/^1m^2m^2m$ has 6 nodal surfaces, generated by a six-fold rotation. On the bulk side, where the nodal structure extends throughout the full three-dimensional Brillouin zone rather than being confined to a plane. The bulk $d$-wave group $^22/^2m$ has 2 nodal surfaces. The three bulk $g$-wave groups $^1\bar{3}^2m$, $^26/^2m$, and $^26/^2m^2m^1m$ have 4 nodal surfaces each. Finally, the single bulk $i$-wave group $^1m^1\bar{3}^2m$ has 6 nodal surfaces. Together, these 10 spin Laue groups encompass 37 nontrivial altermagnetic spin point groups, and the full enumeration of collinear spin space groups yields 1421 distinct groups, providing an exhaustive symmetry framework for identifying altermagnetic materials in three dimensions. The complete classification of three-dimensional altermagnets based on spin Laue symmetry is summarized in Table~\ref{tab:3D}. The table lists the corresponding wave type, number of symmetry-enforced nodal planes, crystallographic Laue group, and representative material realizations, providing a direct link between the underlying symmetry and the resulting altermagnetic spin-splitting pattern.

For two-dimensional systems, the three-dimensional spin space group formalism requires certain changes. The symmetry setting for quasi-two-dimensional materials 
is the spin layer group, introduced by Zeng and Zhao~\cite{zeng2024description}. This spin-group construction is based on the 80 layer groups rather than the 230 space 
groups. The construction proceeds analogously to the three-dimensional case, with one change. In two dimensions, four of the crystal symmetries, namely, translation $\tau$, real-space inversion $\bar{E}$, the out-of-plane mirror $M_z$, and the out-of-plane two-fold rotation $C_{2z}$, enforce Kramers spin degeneracy throughout the entire Brillouin zone if they connect opposite-spin sublattices. They must therefore be excluded from the altermagnetic coset $\mathbf{G} - H$. The first two are already known from the three-dimensional case. The latter two are specific to the two-dimensional situation, since $M_z$ and $C_{2z}$ map every in-plane $\mathbf{k}$-vector onto itself. Consequently, in any two-dimensional altermagnet, the opposite-spin sublattices must be 
connected exclusively by in-plane rotations or by mirrors through planes perpendicular to the system. Taking the direct product of the space-inversion group with the 80 layer 
groups yields 10 distinct reciprocal-space symmetry groups, of which three are excluded 
from hosting altermagnetism for the reasons above. This leaves seven distinct altermagnetic spin layer groups, which are two more than the five plane-type spin Laue groups expected from the three-dimensional classification. The two additional groups are $^22/^2m_x$ (layer groups 8-18), which exhibits spin-momentum locking identical to the previously known $^2m^2m^1m$, giving rise to $d$-wave altermagnetism with 2 nodal surfaces, and $^1\bar{3}^2m$ (layer groups 67-72), which hosts $i$-wave altermagnetism with 6 nodal surfaces. Among the five previously known groups, $^2m^2m^1m$ (layer groups 19-48), $^24/^1m$ (layer groups 49-52), and $^24/^1m^2m^1m$ (layer groups 53-64) all host $d$-wave altermagnetism with 2 nodal surfaces, $^14/^1m^2m^2m$ (layer groups 53-64) hosts $g$-wave altermagnetism with 4 nodal surfaces, and $^16/^1m^2m^2m$ (layer groups 76-80) hosts $i$-wave altermagnetism with 6 nodal surfaces. These seven spin layer groups provide the complete symmetry classification of altermagnetic order in two-dimensional materials, in analogy with the 10 spin Laue groups of the three-dimensional case.
The classification of two-dimensional altermagnets is summarized in Table~\ref{tab:2D}, which lists the seven altermagnetic spin layer groups together with their associated wave type, nodal-line structures, and representative material candidates reported in the literature.

The spin-group classification is a very useful symmetry framework to understand altermagnetism. Once $H$ and $\mathbf{G} - H$ are specified, the transformation properties of the spin-resolved Bloch Hamiltonian are constrained. These constraints determine the allowed symmetry character of the altermagnetic exchange field, including its sign-changing form factor and symmetry-enforced nodal structure. The microscopic Hamiltonian then determines the material-dependent magnitude and detailed momentum dependence of the splitting through orbital character, hopping amplitudes, crystal fields, and exchange interactions.

\subsection{Hamiltonian constraints from spin-group symmetry}
\label{sec:symmetry_based_models}

In a non-relativistic collinear magnet, spin is a conserved quantum number and the Bloch Hamiltonian can be written in spin blocks, $H_{\uparrow}(\mathbf{k})$ and $H_{\downarrow}(\mathbf{k})$. This block structure is the Hamiltonian representation of the spin-group separation between spin and real-space operations~\cite{brinkman1966theory,litvin1974spin,litvin1977spin,liu2022spin,vsmejkal2022beyond}. Operations in the halving subgroup $H$ preserve a spin sublattice, whereas operations in $G - H$ exchange opposite-spin sublattices. The corresponding constraints are
\begin{equation}
\begin{aligned}
U_h(\mathbf{k})H_{\uparrow}(\mathbf{k})U_h^\dagger(\mathbf{k})
&=
H_{\uparrow}(h\mathbf{k}),
&&h\in H,\\
U_a(\mathbf{k})H_{\uparrow}(\mathbf{k})U_a^\dagger(\mathbf{k})
&=
H_{\downarrow}(a\mathbf{k}),
&&a\in G - H .
\end{aligned}
\label{eq:block_constraints_compact}
\end{equation}
Here $U_h$ and $U_a$ are representation matrices acting on orbital, sublattice, or band degrees of freedom. In a Bloch basis they may also contain momentum-dependent phase factors. The second relation is the essential altermagnetic condition of opposite spin states being symmetry-related at transformed momenta, not necessarily at the same $\mathbf{k}$. Consequently, spin splitting is symmetry-allowed at generic momenta even though the net magnetization vanishes. At momenta satisfying $a\mathbf{k}=\mathbf{k}+\mathbf{G}_{\rm rec}$, where $\mathbf{G}_{\rm rec}$ is a reciprocal-lattice vector, this relation can enforce spin degeneracy. These symmetry-invariant momenta form the nodal lines or nodal surfaces of the altermagnetic splitting.

These constraints also provide a constructive route to minimal Hamiltonians. In the symmetry-based construction of Roig \textit{et al.}, the relevant microscopic input is the relation between the full crystallographic point group $P$ and the magnetic-atom site-symmetry group $S$~\cite{roig2024minimal}. For the multiplicity-two Wyckoff setting considered in their construction, the two magnetic atoms are related by an operation $h\in P$ that is not contained in $S$, so that one may write schematically $P=S\cup hS$. The N\'eel order then transforms according to a one-dimensional irreducible representation $\Gamma_N$ whose character is even on $S$ and odd on the coset $hS$,
\begin{equation}
\chi_{\Gamma_N}(g)=+1\quad (g\in S),
\qquad
\chi_{\Gamma_N}(g)=-1\quad (g\in hS).
\label{eq:gamma_n_character}
\end{equation}
In this formulation, the momentum-space form factor is not introduced as a phenomenological ansatz. It is fixed at the symmetry level by the mismatch between the local site symmetry of the magnetic atom and the full crystallographic point group.

For a two-sublattice model, with Pauli matrices $\tau_i$ acting in sublattice space and $\boldsymbol{\sigma}$ in spin space, a representative minimal Hamiltonian for the centrosymmetric multiplicity-two Wyckoff setting can be written as
\begin{equation}
H(\mathbf{k})
=
\epsilon_0(\mathbf{k})\tau_0
+
t_x(\mathbf{k})\tau_x
+
t_z(\mathbf{k})\tau_z
+
\tau_z\mathbf{J}\cdot\boldsymbol{\sigma}
+
\tau_y\boldsymbol{\lambda}(\mathbf{k})\cdot\boldsymbol{\sigma}.
\label{eq:roig_minimal_model}
\end{equation}
Here $\mathbf{J}$ is the staggered exchange field and $\boldsymbol{\lambda}(\mathbf{k})$ is the symmetry-allowed spin-orbit coupling. The sublattice-even terms $\epsilon_0(\mathbf{k})\tau_0$ and $t_x(\mathbf{k})\tau_x$ are invariant under the full point group, while the sublattice-odd bilinears $\tau_y$ and $\tau_z$ transform as $\Gamma_N$. The nonmagnetic term $t_z(\mathbf{k})\tau_z$ must therefore carry the same symmetry character as the altermagnetic order parameter. Physically, $t_z(\mathbf{k})$ encodes local site-symmetry reduction, such as opposite ligand-field, orbital, or multipolar environments on the two magnetic sublattices~\cite{bhowal2024ferroically,mcclarty2024landau,roig2024minimal}.

The microscopic origin of the spin splitting becomes transparent when spin-orbit coupling is neglected and the N\'eel vector is chosen as the spin quantization axis. For spin projection $s=\pm1$, Eq.~\eqref{eq:roig_minimal_model} gives
\begin{equation}
E_{\alpha s}(\mathbf{k})
=
\epsilon_0(\mathbf{k})
+
\alpha
\sqrt{
t_x^2(\mathbf{k})+
\left[t_z(\mathbf{k})+sJ\right]^2
},
\qquad
\alpha=\pm1 .
\label{eq:minimal_model_no_soc_spectrum}
\end{equation}
Thus, as we have also seen from the microscopic model in Section~\ref{sec:minimal_models_altermagnetism}, a staggered exchange field alone does not generate the characteristic momentum-dependent splitting. The altermagnetic splitting appears through the cooperation of the primary N\'eel order $J$ and the sublattice-odd hopping or crystal-field term $t_z(\mathbf{k})$. For exchange small compared with the sublattice hybridization scale, $J\ll \sqrt{t_x^2(\mathbf{k})+t_z^2(\mathbf{k})}$, the leading splitting in band $\alpha$ is
\begin{equation}
\Delta E_{\alpha}(\mathbf{k})
=
E_{\alpha,+}(\mathbf{k})-E_{\alpha,-}(\mathbf{k})
\simeq
2\alpha J
\frac{t_z(\mathbf{k})}
{\sqrt{t_x^2(\mathbf{k})+t_z^2(\mathbf{k})}} .
\label{eq:leading_splitting_roig}
\end{equation}
This expression shows explicitly that the nodal structure of the altermagnetic splitting is controlled by the zeros of the symmetry-allowed $t_z(\mathbf{k})$ form factor, or more generally by the vanishing of the corresponding projected sublattice-odd matrix element. In nonsymmorphic systems, symmetry-enforced band degeneracies can further enhance the altermagnetic susceptibility and amplify Berry-curvature responses once spin-orbit coupling is included~\cite{roig2024minimal}.

After projection to an isolated low-energy band, or to a band subspace that does not mix with other states under the relevant symmetries, this microscopic structure reduces to the scalar form-factor Hamiltonian
\begin{equation}
H_{\rm alt}(\mathbf{k})
=
\epsilon_0(\mathbf{k})\sigma_0
+
Jg(\mathbf{k})\hat{\mathbf n}\cdot\boldsymbol{\sigma},
\qquad
\Delta_{\rm alt}(\mathbf{k})=2Jg(\mathbf{k}) .
\label{eq:alt_effective_compact}
\end{equation}
The projected form factor obeys
\begin{equation}
g(h\mathbf{k})=g(\mathbf{k}),
\qquad
g(a\mathbf{k})=-g(\mathbf{k}),
\qquad
h\in H,\; a\in G - H .
\label{eq:g_symmetry_compact}
\end{equation}
For the usual even-parity non-relativistic altermagnets described by spin-Laue symmetry, and in the absence of external magnetic fields, spin-orbit coupling, or complex orbital fluxes, the spin-resolved spectrum is even under $\mathbf{k}\rightarrow-\mathbf{k}$. The altermagnetic splitting is therefore even in momentum but changes sign between momentum sectors related by coset operations. This distinguishes it from Rashba or Dresselhaus spin splitting, which is relativistic and typically odd in momentum, and also from ferromagnetic exchange splitting, which does not alternate in sign across the Brillouin zone.

\subsection{Projection viewpoint and microscopic origin of the form factor}
\label{sec:projection_viewpoint}

The projected form-factor language is the band-space version of the symmetry constraints discussed above. It is useful because the experimentally visible spin splitting is not determined only by the local exchange field, but by how this exchange field is distributed over the Bloch eigenstates. This viewpoint connects earlier microscopic descriptions of momentum-dependent spin splitting in collinear compensated magnets with recent symmetry-dictated minimal models of altermagnetism~\cite{hayami2019momentum,yuan2020giant,vsmejkal2022emerging,roig2024minimal}.

For a multi-orbital tight-binding Hamiltonian, the staggered exchange term may be written as
\begin{equation}
H_{\rm ex}
=
\Delta
\sum_{i,\mu}
\eta_i\,
c_{i\mu}^{\dagger}
\hat{\mathbf n}\cdot\boldsymbol{\sigma}
c_{i\mu},
\qquad
\eta_i=\pm1,
\label{eq:staggered_exchange_compact}
\end{equation}
where $\eta_i$ distinguishes the two opposite-spin sublattices. For an isolated nondegenerate band $n$, first-order projection gives an effective altermagnetic exchange field
\begin{equation}
J_n^{\rm eff}(\mathbf{k})
=
\Delta
\langle u_{n\mathbf{k}}|\hat{\eta}|u_{n\mathbf{k}}\rangle .
\label{eq:projected_exchange_compact}
\end{equation}
The two spin eigenvalues are shifted by $\pm J_n^{\rm eff}(\mathbf{k})$. Therefore, the corresponding spin splitting is
\begin{equation}
\Delta E_n(\mathbf{k})
=
E_{n,+}(\mathbf{k})-E_{n,-}(\mathbf{k})
=
2\Delta
\langle u_{n\mathbf{k}}|\hat{\eta}|u_{n\mathbf{k}}\rangle .
\label{eq:projected_exchange_splitting_compact}
\end{equation}
Thus, the form factor is determined by the sublattice-odd weight of the Bloch wave function. If the Bloch state has equal weight on the two opposite-spin sublattices, the projected exchange field vanishes. If the sublattice and orbital weights of the Bloch state vary with momentum due to the underlying ligand-field and hopping structure, the same local exchange produces a momentum-dependent spin splitting.

For a set of nearly degenerate bands, the projection must be performed as a matrix within the low-energy subspace,
\begin{equation}
\mathcal{M}_{mn}(\mathbf{k})
=
\langle u_{m\mathbf{k}}|\hat{\eta}|u_{n\mathbf{k}}\rangle,
\qquad
m,n\in \mathcal{L},
\label{eq:projected_eta_matrix}
\end{equation}
where $\mathcal{L}$ denotes the chosen low-energy manifold. The effective exchange Hamiltonian is then
\begin{equation}
H_{\rm ex}^{\rm eff}(\mathbf{k})
=
\Delta\,
\mathcal{M}(\mathbf{k})
\otimes
\hat{\mathbf n}\cdot\boldsymbol{\sigma},
\label{eq:projected_exchange_matrix}
\end{equation}
and the altermagnetic exchange fields are obtained from the eigenvalues of $\mathcal{M}(\mathbf{k})$, or more generally by diagonalizing the full projected Hamiltonian within the low-energy subspace. This matrix formulation is essential near symmetry-enforced band degeneracies, where a single-band projection is not well-defined. It also explains why nonsymmorphic degeneracies can enhance altermagnetic susceptibilities and Berry-curvature responses in realistic minimal models~\cite{roig2024minimal}.

The connection to the two-sublattice model in Eq.~\eqref{eq:roig_minimal_model} is the following. In the absence of spin-orbit coupling, the projected sublattice-odd matrix element is controlled by $t_z(\mathbf{k})$ and the inter-sublattice mixing $t_x(\mathbf{k})$. For a nondegenerate band away from crossings, Eq.~\eqref{eq:leading_splitting_roig} shows that the effective form factor scales as
\begin{equation}
g_{\alpha}(\mathbf{k})
\propto
\alpha
\frac{t_z(\mathbf{k})}
{\sqrt{t_x^2(\mathbf{k})+t_z^2(\mathbf{k})}} .
\label{eq:projected_form_factor_roig}
\end{equation}
Therefore the nodes and sign changes of $g_{\alpha}(\mathbf{k})$ inherit the symmetry of the sublattice-odd hopping or crystal-field term. This makes explicit how the abstract spin-group form factor emerges from orbital hybridization, Wyckoff-site symmetry, and local ligand-field anisotropy.

This projection picture clarifies how the magnitude of altermagnetic spin splitting is material-dependent. Large splitting requires sizable local exchange, Bloch states with substantial magnetic-sublattice character, and a symmetry-allowed anisotropic hopping or crystal-field scale. In a schematic form,
\begin{equation}
\Delta_{\rm alt}(\mathbf{k})
\sim
\Delta\,
\frac{\delta t(\mathbf{k})}{W},
\label{eq:chemical_design_scaling_compact}
\end{equation}
where $\delta t(\mathbf{k})$ is a sublattice-odd anisotropic hopping or ligand-field scale and $W$ is a characteristic bandwidth. This expression is only a design guide, but it captures why metal-ligand covalency, orbital order, local multipolar environments, and electronic correlations can strongly influence the observable altermagnetic band splitting~\cite{bhowal2024ferroically,roig2024minimal}. We next collect the form factors associated with different types of altermagnets.

\subsection{Canonical form factors for $d$-, $g$-, and $i$-wave altermagnets}
\label{sec:canonical_form_factors}

The altermagnetic form factor $g(\mathbf{k})$ is the momentum-space representation of the symmetry channel that exchanges the two opposite-spin sublattices. Its nodal structure is constrained by the spin group and the relevant little-group symmetries, while its detailed lattice harmonics depend on the space group, Wyckoff position, orbital content, and hopping range. The expressions below are representative lowest-order harmonics, not complete bases of all allowed terms. This is why realistic minimal models generally contain material-dependent functions transforming according to the same irreducible representation as the N\'eel order~\cite{vsmejkal2022beyond,vsmejkal2022emerging,roig2024minimal}.

For a square or tetragonal lattice, the two most common $d$-wave representatives are
\begin{equation}
g_{x^2-y^2}(\mathbf{k})
=
\cos k_x-\cos k_y,
\qquad
g_{xy}(\mathbf{k})
=
2\sin k_x\sin k_y .
\label{eq:canonical_d_compact}
\end{equation}
Near the Brillouin-zone center, these reduce to $g_{x^2-y^2}\simeq -(k_x^2-k_y^2)/2$ and
$g_{xy}\simeq 2k_xk_y$, respectively. Thus, the leading $d$-wave altermagnetic splitting is quadratic in momentum and changes sign upon the corresponding crystallographic
rotation. The nodal lines of these form factors mark momenta where the non-relativistic altermagnetic splitting vanishes by symmetry.

Higher-wave altermagnets are obtained when the spin splitting changes sign more frequently around $\Gamma$. A representative two-dimensional $g$-wave form is
\begin{equation}
g_g(\mathbf{k})
=
\sin k_x\sin k_y(\cos k_x-\cos k_y)
\simeq
-k_xk_y(k_x^2-k_y^2)/2 ,
\label{eq:g_wave_compact}
\end{equation}
which has four nodal lines passing through $\Gamma$ in two dimensions. In hexagonal or three-dimensional systems, equivalent $g$-wave representations may take different polynomial
forms, such as $k_xk_z(k_x^2-3k_y^2)$ or $k_yk_z(3k_x^2-k_y^2)$, depending on the irreducible representation selected by the magnetic Wyckoff position.

Similarly, an idealized in-plane $i$-wave representative can be written as
\begin{equation}
g_i(\mathbf{k})
\propto
k^6\sin(6\phi)
\quad
\mathrm{or}
\quad
k^6\cos(6\phi),
\label{eq:i_wave_compact}
\end{equation}
while a cubic representative may take the form
\begin{equation}
g_i^{\rm cubic}(\mathbf{k})
\propto
k_x^4(k_y^2-k_z^2)
+
k_y^4(k_z^2-k_x^2)
+
k_z^4(k_x^2-k_y^2).
\label{eq:i_wave_cubic}
\end{equation}
These examples illustrate the central point that the labels $d$-, $g$-, and $i$-wave refer to the symmetry-enforced nodal structure and sign-changing character of the
non-relativistic spin splitting. In real materials, the leading harmonic can be
supplemented by higher-order terms and orbital matrix elements, but all allowed contributions must transform in the same symmetry channel. Consequently, symmetry fixes
the nodal topology of the altermagnetic exchange texture, whereas chemical bonding, ligand fields, correlations, and band hybridization determine the magnitude and detailed
momentum dependence of the observable splitting.

\subsection{Role of spin-orbit coupling}
\label{sec:soc_altermagnetism}

As we have already discussed, altermagnetism is defined by a momentum-dependent exchange splitting that exists already in the non-relativistic limit, where spin and real-space rotations can be treated independently~\cite{vsmejkal2022beyond,vsmejkal2022emerging}. Spin-orbit coupling (SOC) is therefore not the microscopic origin of the primary altermagnetic exchange texture. It is, however, unavoidable in real materials and determines how this texture is realized experimentally. SOC locks spin to the lattice, selects the magnetic easy axis, and can generate symmetry-allowed canting. Furthermore, it mixes spin components in momentum space, reconstructs non-relativistic nodal manifolds, and enables relativistic Berry-curvature responses \cite{jungwirth2026symmetry}.

The relation between SOC and weak ferromagnetism has its historical origin in the symmetry analysis of Dzyaloshinsky and in Moriya's microscopic theory of anisotropic exchange~\cite{dzialoshinskii1957thermodynamic,dzyaloshinsky1958thermodynamic,moriya1960anisotropic}. For a bilinear two-spin interaction, the exchange tensor can be decomposed as
\begin{equation}
H_{ij}
=
J_{ij}\,\mathbf{S}_i\!\cdot\!\mathbf{S}_j
+
\mathbf{D}_{ij}\!\cdot\!(\mathbf{S}_i\times\mathbf{S}_j)
+
\mathbf{S}_i\!\cdot\!\boldsymbol{\Gamma}_{ij}\!\cdot\!\mathbf{S}_j,
\label{eq:soc_exchange_tensor}
\end{equation}
where the Dzyaloshinskii-Moriya interaction (DMI) is the antisymmetric bilinear contribution and is linear in SOC to leading order. Importantly, global centrosymmetry does not require every bond-resolved \(\mathbf{D}_{ij}\) to vanish. Rather, \(\mathbf{D}_{ij}=0\) is enforced when the corresponding bond possesses an inversion center at its midpoint. In centrosymmetric altermagnets, symmetry-related local DMI vectors may therefore form a staggered pattern whose crystallographic sum vanishes while still producing symmetry-allowed canting~\cite{autieri2025staggered}. SOC can also generate weak-ferromagnetic couplings beyond the simplest DMI picture. Roig \textit{et al.} showed that an SOC-enabled uniaxial spin-space quasisymmetry can strongly constrain the allowed uniform spin magnetization~\cite{roig2025quasisymmetry}, while Sandratskii \textit{et al.} found that retaining only the SOC component parallel to the N\'eel vector can eliminate spin weak ferromagnetism even though an orbital weak-ferromagnetic contribution survives~\cite{sandratskii2026spin}.

These examples highlight an important distinction between the dependence of a phenomenological magnetic coupling on the order parameters and the perturbative order of its microscopic coefficient in SOC. MnTe provides a particularly clear example. For an in-plane N\'eel vector, a symmetry-allowed canting contribution may be written as
\begin{equation}
F_{\mathrm{cant}}
=
\gamma_3 M_z N_y\left(3N_x^2-N_y^2\right),
\label{eq:mnte_canting_invariant}
\end{equation}
which is cubic in the components of the N\'eel order parameter and linear in the induced magnetization \(M_z\). By contrast, describing the apparent ferromagnetism of MnTe as a ``third-order SOC effect'' refers to the microscopic scaling of the effective coupling, which emerges only at third order in the small SOC strength~\cite{mazin2024origin,chen2026dominant}. These are therefore two conceptually distinct notions of order.

A second central consequence of SOC is that the relativistic spin texture is generally multicomponent. In a locally projected two-level sector one may write
\begin{equation}
H_{\mathrm{eff}}(\mathbf{k})
=
\epsilon_0(\mathbf{k})\sigma_0
+
\mathbf{d}(\mathbf{k})\!\cdot\!\boldsymbol{\sigma},
\qquad
E_{\pm}
=
\epsilon_0(\mathbf{k})\pm |\mathbf{d}(\mathbf{k})|,
\label{eq:soc_projected_field}
\end{equation}
where \(\boldsymbol{\sigma}\) denotes the Pauli matrices in the active two-level subspace. In the non-relativistic collinear limit, \(\mathbf{d}(\mathbf{k})\) is effectively parallel to the N\'eel-vector direction, so the zero of a single scalar form factor can define an extended nodal manifold, including the nodal planes encountered in three-dimensional altermagnets. With SOC, the components \(d_x\), \(d_y\), and \(d_z\) can acquire different momentum dependences and wave symmetries~\cite{fernandes2024topological,autieri2025relativistic}, and a true two-level degeneracy requires all symmetry-allowed splitting terms to vanish simultaneously.

It is therefore useful to distinguish a \emph{non-relativistic nodal plane}, where the scalar exchange splitting vanishes in the spin-rotation-symmetric limit, from a \emph{component-resolved zero}, where only one spin projection or one relativistic-field component vanishes, and from a true \emph{relativistic band-degeneracy manifold}, where all symmetry-allowed splitting terms vanish. Non-relativistic nodal planes are consequently generically lifted by SOC or reduced to lower-dimensional nodal lines or points~\cite{fernandes2024topological}. They are not forbidden in every relativistic model. Special nodal planes can remain symmetry enforced for multidimensional irreducible representations, including the \(E_g\) model of \(O_h\) discussed by Fernandes \textit{et al.}~\cite{fernandes2024topological}. Such cases, however, must be established from the full relativistic magnetic symmetry rather than from the non-relativistic form factor.

Equation~\eqref{eq:soc_projected_field} should be viewed as a projected illustration rather than a generic description of a multiorbital altermagnet. In multisublattice or multiorbital systems, spin is entangled with other internal degrees of freedom and its expectation value must be evaluated from the full eigenstate. Symmetry-allowed relativistic terms can, for example, take the form
\(\tau_y\boldsymbol{\lambda}(\mathbf{k})\!\cdot\!\boldsymbol{\sigma}\),
with \(\boldsymbol{\tau}\) acting in sublattice or orbital space; such terms can generate Berry curvature already at linear order in SOC~\cite{roig2024minimal}. Yang, Fernandes, and Birol provide a complementary classification of SOC-induced spin splitting into Rashba-, Dresselhaus-, Weyl-, and Ising-type structures~\cite{yang2026symmetries}. Although developed for non-magnetic crystals, it provides a useful reference because magnetic and crystallographic symmetries similarly constrain the allowed momentum-dependent subdominant spin components in altermagnets. This also explains why a small integrated canted moment does not imply a small momentum-resolved spin projection. Brillouin-zone and sublattice cancellations can suppress the net moment even when individual Bloch states carry sizable subdominant spin components~\cite{autieri2025relativistic}.

At the opposite extreme, SOC does not necessarily destroy spin collinearity. Campos \textit{et al.} introduced persistent altermagnetic spin polarization (PASP), in which an out-of-plane mirror or glide-mirror symmetry of the magnetic structure enforces a globally collinear spin polarization throughout the two-dimensional Brillouin zone even in the presence of SOC~\cite{campos2026persistent}. In the strong-PASP example V$_2$Te$_2$O, the large altermagnetic splitting (\(\sim1.5\)~eV) originates from non-relativistic exchange, while SOC preserves the mirror-protected collinearity. For comparison, the same work discusses a conventional SOC-driven persistent spin texture in NbSe$_2$, where the relativistic splitting is of order \(0.01\)-\(0.1\)~eV. In the weak-PASP example La$_2$CuO$_4$, the relevant non-relativistic bands are spin degenerate and SOC itself lifts this degeneracy while the collinear texture remains symmetry protected. Thus, depending on symmetry and band character, SOC can either reconstruct a pre-existing exchange splitting or generate an additional relativistic one. The classification contains 158 altermagnetic spin layer groups, divided into 92 strong and 66 weak cases, and the same work demonstrates ferroelectrically switchable PASP in VSI$_2$~\cite{campos2026persistent}.

Returning to MnTe from the complementary perspective of momentum-space spin structure, hexagonal \(\alpha\)-MnTe provides a transparent example of SOC-driven relativistic reconstruction. In the non-relativistic limit it is commonly described by a bulk \(g\)-wave altermagnetic texture, and complete spin degeneracy occurs on the \(k_z=0\) and \(k_y=0\) planes within the corresponding non-relativistic symmetry analysis~\cite{rooj2025altermagnetic}. Once SOC and the experimentally relevant in-plane N\'eel-vector orientation are included, these relations are reduced. For \(\mathbf{N}\parallel\hat{\mathbf{y}}\), fully spin-resolved calculations find \(d_{xz}\)-, \(d_{yz}\)-, and \(s\)-wave spin-momentum locking for \(S_x\), \(S_y\), and \(S_z\), respectively~\cite{autieri2025relativistic}. This component-resolved structure resolves the apparent nodal-plane issue: on \(k_z=0\), both \(d\)-wave components vanish but the \(s\)-wave \(S_z\) component need not; on \(k_y=0\), the \(d_{yz}\) component vanishes while \(d_{xz}\) and/or \(S_z\) may remain finite. Hence, the non-relativistic \(k_z=0\) and \(k_y=0\) nodal planes are not generic total relativistic degeneracy planes in MnTe with SOC.

SOC further fixes the magnetic anisotropy and enables transverse responses in MnTe. Antiferromagnetic resonance measurements extract an out-of-plane single-ion anisotropy of approximately \((40\pm10)\,\mu\mathrm{eV}\), quantifying the easy-plane character that constrains the N\'eel-vector orientation~\cite{dzian2025antiferromagnetic}. A spontaneous anomalous Hall effect has been observed in epitaxial MnTe, while bulk measurements established the coexistence of a hysteretic anomalous Hall response with only a very small weak net magnetization~\cite{gonzalez2023spontaneous,kluczyk2024coexistence}. These results illustrate that SOC-induced Berry-curvature responses need not scale directly with the magnitude of the uniform spin moment.

A further striking relativistic prediction concerns orbital magnetization. For \(\mathbf{N}\parallel\hat{\mathbf{y}}\), Ye \textit{et al.} evaluated the intrinsic orbital magnetization using the modern Berry-phase formulation and obtained
\(|M_z^{\mathrm{orb}}|\simeq0.176\,\mu_{\mathrm B}\) per unit cell, compared with only \(|M_z^{\mathrm{spin}}|\simeq0.002\,\mu_{\mathrm B}\) per unit cell~\cite{chen2026dominant}. The orbital contribution is remarkably large compared with the extremely small remanent moment reported experimentally in bulk MnTe,
\(M_{\mathrm{REM}}\simeq3\times10^{-5}\,\mu_{\mathrm B}\) per Mn~\cite{kluczyk2024coexistence}. A direct comparison requires caution because the calculation describes an intrinsic single-domain orbital magnetization, whereas bulk magnetometry probes a domain-averaged macroscopic response. Domain compensation and itinerant Berry-phase contributions may therefore substantially suppress the measured uniform moment, but a complete quantitative understanding is presently lacking. Establishing how the predicted orbital magnetization manifests experimentally thus remains an important open problem. Complementary relativistic calculations likewise emphasize the coexistence of spin and orbital weak-ferromagnetic contributions and the reconstruction of Bloch states near SOC-induced avoided crossings~\cite{sandratskii2026spin}.

Thus, the defining altermagnetic exchange texture is non-relativistic in origin, whereas SOC controls its experimental realization through anisotropic exchange, symmetry-allowed canting, multicomponent spin textures, nodal reconstruction, magnetic anisotropy, Berry-curvature responses, and spin/orbital magnetization. This distinction provides the bridge between spin-group symmetry, realistic electronic structure, and experiment. 

\begin{figure}[t]
\centerline{\includegraphics[scale=0.7]{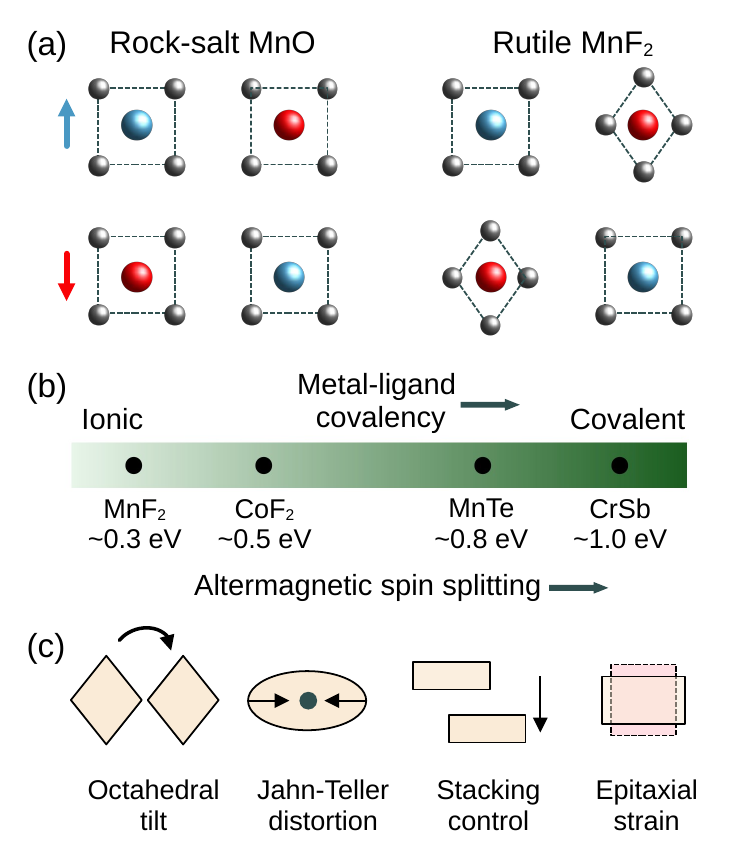}}
\caption{(a) A comparison of the coordination environment for rock-salt MnO and rutile MnF$_2$. In the rock-salt structure the octahedra surrounding spin-up and spin-down Mn$^{2+}$ ions are related by a lattice translation. In the rutile case, the octahedra on the two sublattices are rotated by 90$^{\circ}$ about the tetragonal $c$-axis. (b) Altermagnetic spin splitting is often enhanced by increasing metal-ligand covalency, aince covalency strengthens ligand-mediated hopping anisotropy. In highly ionic compounds, the altermagnetic splitting is typically weaker. (c) Design strategies for altermagnets using octahedral tilts, Jahn-Teller distortion, stacking of layered systems, and epitaxial strain.}
\label{fig_chem_design}
\end{figure}

\section{Chemical design principles}
\label{sec:chemical_design}

Now that we have covered the main concepts of theory required for understanding altermagnets, we can now delve into the chemistry of designing these materials.

\subsection{Coordination environment and ligand-cage rotation}

As we have seen, the central chemical insight behind altermagnetism is that the coordination polyhedra surrounding opposite-spin magnetic ions must be related by a rotation, not by a translation or inversion. This single requirement distinguishes altermagnets from conventional antiferromagnets at the level of local coordination chemistry. As an illustration, let us consider a transition-metal oxide or fluoride in which each metal center sits inside an octahedron of anions as depicted in Fig.~\ref{fig_chem_design}(a). In a conventional antiferromagnet such as rock-salt MnO, the octahedra surrounding spin-up and spin-down Mn$^{2+}$ ions are related by a lattice translation. The octahedra are identical in orientation, size, and distortion. The combined symmetry of sublattice translation and time reversal (the operation that flips spins) protects Kramers degeneracy, and the electronic bands are spin-degenerate at every crystal momentum point in the Brillouin zone. Now consider the rutile structure adopted by MnF$_2$, RuO$_2$, and many other binary compounds. Each metal center again sits inside an anion octahedron, but the octahedra on the two sublattices are rotated by 90$^{\circ}$ about the tetragonal $c$-axis. When antiferromagnetic order sets in, the spin-up and spin-down sublattices are no longer related by translation. In fact, they are related by this 90$^{\circ}$ rotation. The combined translation and time-reversal symmetry that protected spin degeneracy is broken, and a momentum-dependent spin splitting with $d$-wave symmetry emerges. The same principle operates in other coordination environments. 

In the NiAs-type structure (which as we will see in the next section is adopted by MnTe and CrSb), the metal ions occupy edge-sharing octahedra with a 60$^{\circ}$ rotation between sublattices, producing $g$-wave altermagnetism. In layered perovskites such as La$_2$CuO$_4$, the CuO$_6$ octahedra undergo cooperative tilts that rotate the coordination cages on opposite-spin sublattices with respect to each other. In each case, the essential ingredient is the same. There is a rotational relationship between the ligand environments of opposite-spin ions. For the synthetic chemist, this provides a concrete design target. Rather than searching for exotic electronic states, one should look to engineer crystal structures in which the metal coordination polyhedra alternate in orientation between sublattices. Corner-sharing, edge-sharing, and face-sharing connectivities each offer different geometric handles for achieving this rotation.

\subsection{Bonding and exchange pathways}

The existence of altermagnetism requires two ingredients, namely, antiferromagnetic exchange between sublattices and anisotropic orbital overlap within each sublattice. Both are controlled by chemical bonding. The antiferromagnetic exchange itself follows the familiar Goodenough–Kanamori rules. Superexchange through 180$^{\circ}$ M-X-M pathways (where X is an anion bridge) favors antiparallel spin alignment when the magnetic orbitals overlap with the same ligand $p$-orbital on both sides. This is identical to the exchange mechanism in conventional antiferromagnets and does not, by itself, produce altermagnetism. As we saw through the simple model in Section~\ref{sec:minimal_models_altermagnetism}, the altermagnetic ingredient is the anisotropic next-nearest-neighbor hopping. Consider two same-spin metal ions on the same sublattice, connected through a network of ligands. The orbital overlap integral for this next-nearest-neighbor pathway depends on the orientation of the $d$-orbitals relative to the bond direction. If the coordination cage is elongated along $x$ (for example, due to a Jahn–Teller distortion or an intrinsic structural anisotropy), the next-nearest-neighbor hopping along $x$ will differ from that along $y$. Crucially, because the coordination cages on the opposite sublattice are rotated, the anisotropy is swapped. What was the "strong" direction on sublattice A becomes the "weak" direction on sublattice B. This swapped anisotropy is the microscopic origin of the $d$-wave spin splitting. 

The magnitude of the altermagnetic splitting scales with both the exchange splitting and the next-nearest-neighbor hopping anisotropy. This means that one can tune the splitting by modifying either ingredient independently, i.e., strengthening the exchange (for example, by choosing more covalent ligands or shorter M–X bonds) or increasing the orbital overlap anisotropy. This may be achieved, for instance, by applying epitaxial strain, chemical pressure, or substituting ligands of different size along different crystallographic directions. An important subtlety concerns the role of metal–ligand covalency [Fig.~\ref{fig_chem_design}(b)]. In highly ionic compounds, the $d$-orbitals are well localized and the next-nearest-neighbor hopping is small, leading to weak altermagnetic splitting even if the structural anisotropy is large. In more covalent compounds, the $d$-orbitals hybridize strongly with ligand $p$-orbitals, amplifying the next-nearest-neighbor overlap and producing larger splittings. This explains why CrSb (with substantial Cr–Sb covalency) exhibits one of the largest predicted altermagnetic splittings ($\sim$1 eV), while more ionic compounds such as MnF$_2$ show substantially smaller values~\cite{bhowal2024ferroically}.

\subsection{Role of non-magnetic atoms}

A distinctive feature of altermagnetism, compared with conventional ferromagnetism and antiferromagnetism, is the key role played by non-magnetic atoms. In a ferromagnet, replacing one ligand with another of similar electronegativity typically changes the exchange strength but not the fundamental magnetic class. In an altermagnet, the ligand sublattice geometry can change the altermagnetic state entirely. This is because the symmetry operation connecting opposite-spin sublattices is a property of the full crystal structure, not just the magnetic sublattice. Two compounds can have identical magnetic-ion sublattices and yet belong to different magnetic classes if the ligand arrangement differs. For instance, MnF$_2$ in the rutile structure is an altermagnet because the fluorine octahedra are rotated between sublattices. A hypothetical MnF$_2$ polymorph in the rock-salt structure would be a conventional antiferromagnet, even though the Mn$^{2+}$ ions carry the same moment and interact through similar superexchange. This observation has practical implications for materials design. It means that the altermagnetic state can, in principle, be switched on or off by structural modifications that affect only the non-magnetic sublattice, for instance, by anion ordering in mixed-anion compounds, by intercalation, or by epitaxial strain that selectively rotates the ligand polyhedra.

\subsection{Dimensionality and structural distortions}

Altermagnetism is not restricted to three-dimensional bulk crystals. Two-dimensional materials and quasi-two-dimensional layered structures offer additional design flexibility, because the reduced dimensionality introduces new symmetry operations (such as in-plane mirrors and horizontal reflections) that can connect opposite-spin sublattices in ways that have no direct analogs in three-dimensions. In two-dimensions, the classification uses spin layer groups rather than spin Laue groups, and the number of allowed altermagnetic classes increases from the naively expected 5 to 7, as we saw in Section~\ref{sec:spin_group_theory}. The two additional classes arise from layer-specific symmetry operations, for example, an in-plane C$_2$ rotation combined with a horizontal mirror can connect sublattices in a way that produces a spin layer group not derivable from any three-dimensional planar spin Laue group. This has been predicted for materials such as monolayer RuF$_4$~\cite{milivojevic2024interplay} and Janus Mn$_2$P$_2$S$_3$Se$_3$~\cite{mazin2023induced}. 

Structural distortions provide a powerful handle for tuning between magnetic phases. Several mechanisms can be particularly important as schematically illustrated in Fig.~\ref{fig_chem_design}(c).

\begin{itemize}
\item Octahedral tilts and rotations in perovskites can convert a conventional antiferromagnet into an altermagnet by introducing a rotational relationship between opposite-spin coordination cages~\cite{bernardini2025ruddlesden}.

\item Jahn-Teller distortions elongate or compress the coordination polyhedra, increasing the orbital overlap anisotropy and thereby strengthening the altermagnetic splitting. In Cu$^{2+}$ compounds ($d^9$ configuration), the strong Jahn–Teller effect produces large tetragonal elongations that naturally generate the required anisotropy~\cite{vsmejkal2022beyond}.

\item Stacking control in van der Waals layered materials allows the relative orientation of magnetic layers to be tuned. For example, in MnPSe$_3$, different stacking sequences exhibit altermagnetic order, or suppress altermagnetism entirely in favor of conventional antiferromagnetism~\cite{gonzalez2025engineering}.

\item Chemical pressure and epitaxial strain can continuously tune the rotation angle of coordination polyhedra, offering a route to controlling the magnitude of the altermagnetic splitting without changing the chemical composition~\cite{meier2026net}.
\end{itemize}

\subsection{Design rules for altermagnetic materials}
\label{sec:design_rules}

The preceding discussion establishes the physical ingredients of altermagnetism. Next we present a set of design principles that can guide the selection and modification of candidate materials. We emphasise that these principles are not independent. In fact, several are in partial tension with one another, and the design problem is best understood as a constrained optimization in an complex design space.

The first consideration is crystallographic and acts as a hard filter. Altermagnetism requires that the opposite-spin sublattices occupy Wyckoff positions related by a rotation. This in turn requires at least two symmetry-inequivalent magnetic sites within the primitive cell. If all magnetic ions in the primitive cell are related by translation or inversion, no amount of ligand modification will produce altermagnetism, because the crystallographic prerequisite is absent. This test should therefore be applied before any other, and it can be performed without recourse to electronic-structure calculation. The magnetic structure databases MAGNDATA and the associated symmetry tools on the Bilbao crystallographic server allow the sublattice-connecting operation to be identified directly from any experimentally determined magnetic structure~\cite{gallego2016magndata,perez2015symmetry}.

Once this filter is passed, the question becomes one of local coordination geometry, and here a reformulation due to Yuan and Zunger provides a directly usable criterion available to a synthetic chemist. Yuan and Zunger showed that the formal spin-symmetry conditions can be recast in terms of easily visualized spin-structure motif pairs~\cite{yuan2023degeneracy}. These are the coordination polyhedra, namely, octahedra, tetrahedra, square planes, surrounding the two opposite-spin sublattices. If the two motifs interconvert by neither translation nor inversion, spin splitting is allowed. The significance of this reformulation is that it replaces an abstract group-theoretic test with a structural inspection that can be carried out on a crystal-structure drawing. In practical terms, one looks for coordination environments that are mutually rotated rather than mutually translated. The rutile, NiAs, and tilted-perovskite structure types all satisfy this condition.

The same framework also governs the magnitude of the splitting. The larger the angle between the local anisotropy axes of the two motifs, the larger the resulting spin splitting~\cite{yuan2023degeneracy}. In rutile-type structures this angle is 90$^{\circ}$ within the $ab$ plane, which accounts for the pronounced d-wave splitting found in MnF$_2$, NiF$_2$, and RuO$_2$-type compounds. In perovskites the corresponding angle is set by the Glazer tilt system, and is therefore accessible through the tolerance factor by A-site substitution. Notably, this is a synthetic handle that is already thoroughly established in the oxide chemistry literature~\cite{glazer1972classification}. The practical prescription is to choose or engineer structures in which the two motifs appear maximally different when viewed down a low-index direction.

A useful corollary follows from this, and it is a valuable practical tool for assessing the type of momentum space spin splitting of candidate structure. Since the real-space motif inequivalence maps onto the reciprocal-space splitting pattern, the momentum dependence of the splitting can be read off the crystal structure by inspection alone~\cite{yuan2023degeneracy}. Viewing the structure along a direction in which the two coordination motifs appear identical identifies a nodal direction, along which the bands remain spin-degenerate. On the other hand, viewing along a direction in which they appear maximally different identifies the direction of largest splitting. In rutile RuO$_2$, for example, the motifs are indistinguishable when viewed along [100] and maximally distinct along [110]. These correspond respectively to spin degeneracy along $\Gamma-X$ and maximum splitting along $\Gamma-M$~\cite{wei2024crystal}. No calculation is required, and the heuristic can be applied to any structure for which coordinates are available.

Beyond geometry, the electronic character of the metal–ligand bond controls the amplitude of the spin splitting. The altermagnetic splitting scales with the next-nearest-neighbour hopping integral, which depends in turn on the degree of $d–p$ hybridization. Ionic compounds with well-localised $d$ orbitals exhibit small splittings even when the structural anisotropy is substantial, whereas covalent compounds exhibit large ones. This trend is consistent with the identification of the ligand-to-metal valence electron ratio as one of the descriptors emerging from recent machine-learning analyses of the design space~\cite{chun2026continuous}. The corresponding synthetic strategy is to move down the chalcogenide and pnictide groups and to prefer softer, more polarizable ligands. Note, however, the trade-off discussed below.

Chemistry also offers a route to generating the required crystal-field anisotropy without external intervention. Ions with an orbitally degenerate ground state, such as Cu$^{2+}$ ($d^9$), high-spin Mn$^{3+}$ ($d^4$), and low-spin Ni$^{3+}$ ($d^7$), spontaneously distort their coordination environments through the Jahn-Teller effect. Cooperative Jahn-Teller ordering in perovskites and layered perovskites therefore provides a means of engineering motif inequivalence through composition~\cite{glazer1972classification,kanamori1960crystal}. The caveat is that the distortion pattern must be the correct one for generating altermagnetism.

Finally, and most importantly for materials selection, these considerations do not all drive in the same direction. The N\'eel temperature is set by the strength of the superexchange, which favours short metal-ligand bonds, M-X-M angles near 180$^{\circ}$, and relatively high-symmetry coordination. The altermagnetic splitting, on the other hand, is set by the crystal-field anisotropy, which requires distorted and mutually rotated coordination environments. These requirements are in competition, and structural distortions that enhance the splitting frequently weaken the exchange. CrSb is exceptional in achieving both a high N\'eel temperature (nearly 705 K) and a large splitting (more than 1 eV). Establishing why this combination occurs and how it can be optimized is among the most important open questions in altermagnetic materials chemistry. \\

\section{Altermagnetic material classes}
\label{sec:material_classes}

Altermagnetism has been proposed in a wide range of materials, including bulk~\cite{guo2023spin,bhattarai2025high,wan2025high, sufyan2026high} and two-dimensional~\cite{sodequist2024two, zeng2024description, wang2025pentagonal,haddadi2026exploring,bose2026symmetry} systems, spanning metals, semiconductors, and insulators~\cite{vsmejkal2022emerging,guo2023spin,sufyan2026high}. Several altermagnetic candidates identified through symmetry analysis have been experimentally explored, whereas even a greater number have been investigated using density functional theory (DFT) computations. NiAs-type materials and rutile oxides are two prominent classes in which altermagnetism has been predicted, and many compounds within these families have been extensively studied both experimentally and theoretically. Even before being identified as altermagnets, these materials had been well studied under the label of antiferromagnets, but their altermagnetic characteristics remained unnoticed in experimental, computational, and theoretical studies until recently. Several two-dimensional materials have been recently identified as altermagnets, which we also discuss in this section. We close this section with a detailed account of metal-organic frameworks (MOFs) and covalent-organic frameworks (COFs), both of which have emerged as promising platforms to realize altermagnetism.

\begin{table*}[ht]
\centering
\caption{Three-dimensional altermagnetism occurs in 10 spin Laue groups constructed from 8 of the 11 crystallographic Laue groups. The three excluded Laue groups are $\bar{1}$ (triclinic), $\bar{3}$ (trigonal), and $m\bar{3}$ (cubic). Prototypical material examples are noted for each case.}
\label{tab:3D}
\begin{tabular}{@{}lclllll@{}}
\hline
\hline
Wave type & Nodal planes & Sub-type & Laue group $G$ & Crystal system & Symmetry & Examples \\
\hline
\hline
\multirow{4}{*}{$d$-wave} & \multirow{4}{*}{2}
  & P-2 & $mmm$  &  Orthorhombic& $C_2$ & La$_2$CuO$_4$ parent \\
   & & P-2 & $4/m$  &  Tetragonal& $C_4$  & KRu$_4$O$_8$  \\
  & & P-2 & $4/mmm$  &  Tetragonal& $C_4$  & RuO$_2$ (debated), MnF$_2$  \\
& & B-2 &  $2/m$ &  Monoclinic  & $C_2$ & CuF$_2$ \\

\hline
\multirow{4}{*}{$g$-wave} & \multirow{4}{*}{4}
  & P-4 & $4/mmm$ & Tetragonal& $C_2$  &KMnF$_3$ \\
& & B-4 & $\bar{3}m$  &   Trigonal& $C_2$& CoF$_3$, Fe$_2$O$_3$ \\

& & B-4 &  $6/m$ & Hexagonal &  $C_6$ &  ---\\
& & B-4 & $6/mmm$ & Hexagonal & $C_6$  & CrSb, MnTe \\
\hline
\multirow{2}{*}{$i$-wave} & \multirow{2}{*}{6}
   & P-6 & $6/mmm$  &Hexagonal & $C_2$  & ---\\
& & B-6 &  $m\bar{3}m$ &  Cubic  & $C_4$ &  ---\\

\hline
\hline
\end{tabular}
\end{table*}

\begin{figure}[t]
\centerline{\includegraphics[scale=0.55]{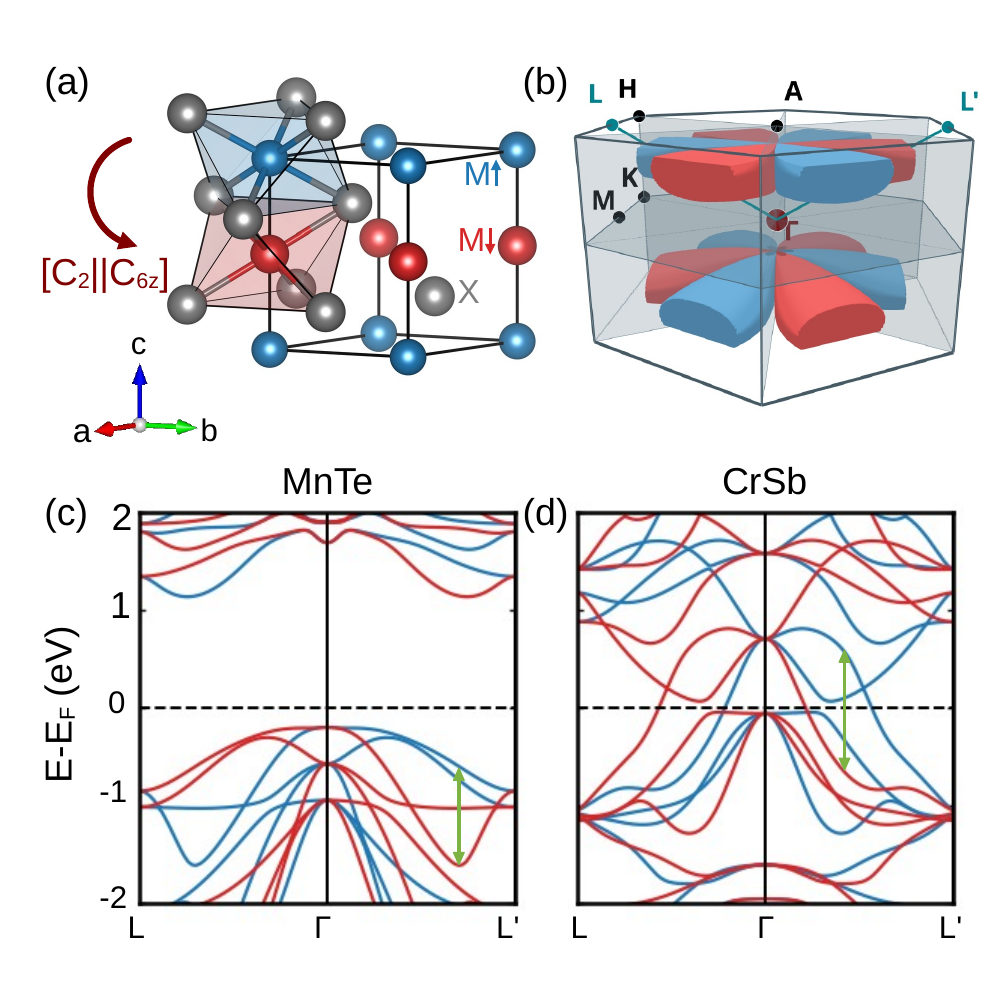}}
\caption{\textbf{Altermagnetism in NiAs-type materials.} (a) The magnetic crystal structure of NiAs-type altermagnets. Magnetic atoms with opposite spin orientations (M$\uparrow$ and M$\downarrow$) are represented by blue and red spheres, respectively, while non-magnetic atoms (X) are shown as gray spheres. Each spin sublattice consists of magnetic atoms octahedrally coordinated by surrounding non-magnetic atoms, highlighted by blue and red shaded octahedra. The two spin sublattices are related by a sixfold screw-axis rotation symmetry. The corresponding Brillouin zone is shown in panel (b), where four nodal planes passing through $\Gamma$ are depicted as gray planes. A schematic illustration of the $g$-wave Fermi surface distortion characteristic of NiAs-type altermagnets is also shown. Panels (c) and (d) present the electronic band structures of MnTe and CrSb, respectively, calculated using DFT with $U$=3 and $U$=0, along the altermagnetic path $L$–$\Gamma$–$L'$, highlighted in teal in panel (b). Spin splitting is highlighted by the green colored line.}
\label{fig-NiAs}
\end{figure}

\begin{figure}[t]
\centerline{\includegraphics[scale=0.55]{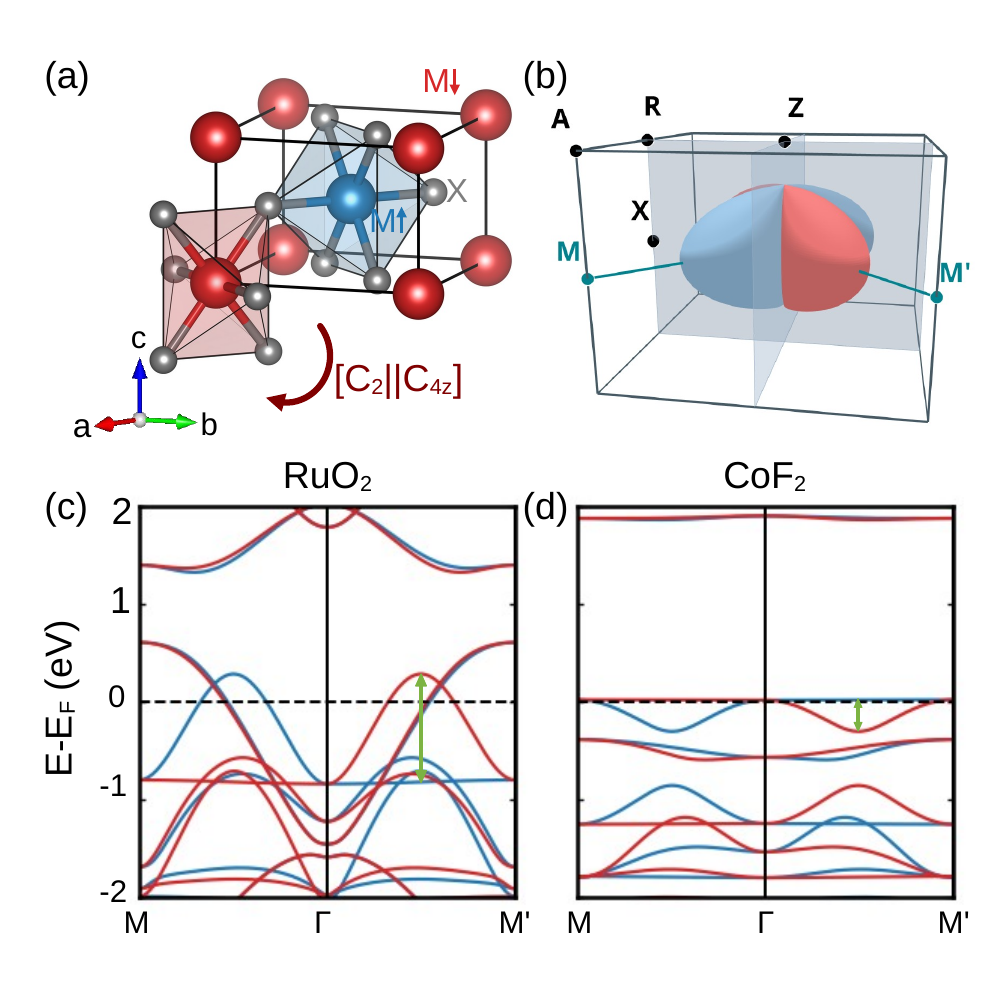}}
\caption{\textbf{Altermagnetism in rutile-type materials.} Panel (a) illustrates the magnetic crystal structure of a rutile-type altermagnet. Magnetic atoms with opposite spin orientations (M$\uparrow$ and M$\downarrow$) are depicted as blue and red spheres, respectively, whereas non-magnetic atoms (X) are shown in gray. The spin-up and spin-down magnetic sublattices, highlighted by blue and red shaded octahedra, are related through a fourfold screw-axis rotational symmetry. Panel (b) presents the corresponding Brillouin zone, where four nodal planes intersecting at $\Gamma$ are shown as gray planes. A schematic representation of the characteristic $d$-wave Fermi surface distortion in rutile-type altermagnets is also shown. Panels (c) and (d) show the electronic band structures of RuO$_2$ and CoF$_2$, respectively, calculated using DFT with $U$=2 and $U$=1.36, along the altermagnetic path $M$-$\Gamma$-$M'$, indicated by the teal line in panel (b). Spin splitting is highlighted by the green colored line.}
\label{Fig_Rutile}
\end{figure}

\subsection{NiAs structure type}

The hexagonal NiAs-type crystal structures of space group P6$_3$/mmc (No. 194) with A type antiferromagnetic order represents a prototypical platform for altermagnetism. In this structure, the transition metal magnetic atom occupies the Wyckoff position 2a, while non-magnetic chalcogen atom resides at Wyckoff position 2c. The partially filled $d$ orbitals of the transition metal play a major role in determining electrical, magnetic, and thermal properties for the materials in this family~\cite{kjekshus1964phases,d2005low,allen1977optical}. 

These structures consist of alternately stacked magnetic and non-magnetic planes, forming a simple hexagonal magnetic lattice and a hexagonal close-packed non-magnetic lattice with AB stacking. Magnetic atoms are arranged in ferromagnetic order within the $ab$-plane, and antiferromagnetic order along $c$-axis~\cite{kunitomi1964neutron,abe1984magnetic}. Each magnetic atom resides within an octahedron formed by six non-magnetic atoms. The non-magnetic atoms adopt a non-centrosymmetric arrangement, and consequently the system has no inversion center. The non-centrosymmetric arrangement of non-magnetic atoms also becomes a reason for the absence of inversion or translational symmetry operational connection between opposite sublattices that are marked in blue and red shaded region in Fig.~\ref{fig-NiAs}(a), despite equal atomic composition and identical bond lengths. Instead they are connected by non-symmorphic operations, such as a sixfold screw rotation combined with a translation, $[{C}_2 \,||\, {C}_{6z} t_{1/2}]$, or a mirror symmetry operation, $[{C}_2 \,||\, {M}_z]$. The resulting nontrivial spin Laue group is $^26/^2m^2m^1m$. Importantly, this symmetry operation preserves magnetic compensation while breaking ${PT}$ symmetry, and gives rise to the altermagnetism in these materials. These symmetry operations further enforce spin degeneracies at 3 vertical planes separated by 60\textdegree, as well as in the $k_z=n\pi/c$ planes. The resulting 4 nodal planes crossing $\Gamma$ point classifies the system as a $g$-wave altermagnet~\cite{vsmejkal2022emerging,vsmejkal2022beyond}. Moreover, the variation of spin polarization along $z$-axis arises from $[{C}_2 \,||\, {M}_z]$ symmetry, places these materials in a category of bulk altermagnet, as per our discussion of the spin space groups in Section~\ref{sec:spin_group_theory}. 

As is seen from Fig.~\ref{fig-NiAs}(b), the conventional high symmetric path $\Gamma$-M-K-$\Gamma$-A-L-H-A of this structure lies in the nodal planes and which is one of the reasons why altermagnetic spin splitting was not identified earlier even though these materials have been extensively studied.

MnTe and CrSb are among the most extensively studied altermagnetic materials in this family. As simple binary compounds that can be readily synthesized in stable, high-quality bulk and thin-film forms, they provide an ideal platform for experimental investigations. Both are $g$-wave altermagnets and exhibit the characteristic features that we discussed earlier which are common to NiAs type altermagnets. Although they share the same structural type, they exhibit  significant differences in electronic and magnetic properties. MnTe is a $p$-type semiconductor holding a band gap of $\approx$1.3 eV~\cite{allen1977optical}, whereas CrSb is a metal. In MnTe magnetic moments lie within the plane~\cite{osumi2024observation,gonzalez2023spontaneous}, while in CrSb they are oriented out-of-plane~\cite{reimers2024direct}. Both materials exhibit antiferromagnetic order at room temperature, which makes them particularly relevant for applications. The N\'eel temperature for CrSb is $\approx$705K~\cite{park2020effects} and that of MnTe is $\approx$307K~\cite{walther1967ultrasonic}. The relatively high N\'eel temperatures have facilitated experimental observations of altermagnetic spin splitting in these materials~\cite{ding2024crsb,yang2025three,krempasky2024altermagnetic,osumi2024observation,lee2024broken}. Moreover, the non-relativistic spin splitting is exceptionally large, reaching values of the order of an eV, which further emphasizes their significance.
The altermagnetic spin splitting calculated using DFT along $L$-$\Gamma$-$L'$ for MnTe and CrSb are shown in Fig.~\ref{fig-NiAs}(c) and (d). Altermagnetic spin splitting has been predicted by various studies using DFT, and were later observed in experiments using angle-resolved photoemission spectroscopy (ARPES), soft X-ray ARPES and spin-ARPES~\cite{krempasky2024altermagnetic, osumi2024observation,lee2024broken,reimers2024direct,zeng2024observation,yang2025three}. Spectroscopic studies on MnTe have been further supported by real-space vector imaging, enabling direct visualization and manipulation of altermagnetic order across multiple length scales, from micron-sized single domains to nanoscale domain walls and topological vortex structures~\cite{amin2024nanoscale}.

Although direct experimental observation of altermagnetic spin splitting emerged later, signatures of altermagnetism in MnTe had already been observed through the spontaneous anomalous Hall effect (AHE)~\cite{gonzalez2023spontaneous,wasscher1965evidence}. The presence of AHE in a compensated antiferromagnetic structure indicates that the material is an altermagnet. Following the identification of altermagnetism in MnTe, its AHE has attracted significant experimental and theoretical attention. AHE depends on N\'eel vector orientation and in MnTe in-plane N\'eel vector orientation allows AHE. In contrast, in CrSb, N\'eel vector orients out-of plane~\cite{reimers2024direct}, where symmetry forbids AHE, making spontaneous AHE absent in pristine CrSb~\cite{das2026effect}.

Similar to the spin splitting of electronic bands, a chiral splitting is predicted in magnon bands of altermagnets~\cite{vsmejkal2022emerging}. These have also been observed in MnTe and CrSb~\cite{liu2024chiral,singh2025chiral}.

Beyond identifying and observing altermagnetic characteristics, numerous experimental and theoretical studies have proposed ways to tune altermagnetic properties in these materials. It has been demonstrated that external parameters such as pressure, strain, temperature, and doping are effective routes to control altermagnetic spin splitting~\cite{devaraj2024interplay,devaraj2026unlocking,das2026effect,kimura2026band}. Furthermore, quantities such as the magnitude of the anomalous Hall current, the individual anomalous Hall tensor components, and even the direction in which the anomalous Hall effect is observed can be modulated through these approaches\cite{devaraj2024interplay,devaraj2026unlocking,liu2025strain,das2026effect}. Additionally, manipulation of altermagnetic order via crystal symmetry reconstruction provides a way for controlling AHE in altermagnets, as demonstrated in CrSb thin films through epitaxial growth along different crystallographic orientations~\cite{zhou2025manipulation}.

Identifying MnTe and CrSb as altermagnets has accelerated research on these materials in recent years. Apart from the electronic and magnetic properties, the magnetotransport~\cite{zhou2026surface,bey2024unexpected,shao2026epitaxial,zhao2026emergent,liu2025strain,gonzalez2024anisotropic,chen2026strain,ghorbani2026transport,bey2026conductivity,peng2025scaling,yu2025neel,thadathil2026electrical,das2026linear,rai2025direction,aota2025epitaxial}, thermoelectric~\cite{negi2025mnte,li2025large}, and optical properties~\cite{gray2024time,uykur2026revisiting,wu2025optical,haag2025optical,sun2025symmetry,zhou2026ultrafast} of altermagnets have been widely investigated in the context of altermagnetism. In addition, symmetry analysis~\cite{lovesey2023templates,rooj2025altermagnetic,hirakida2026multipole} and model Hamiltonian~\cite{roig2024minimal,takahashi2025symmetry,goswami2025investigation} studies aimed at understanding the altermagnetic behavior of these materials have also been carried out.  Other related materials in this class, including NiS~\cite{mandal2025deterministic}, CrTe~\cite{singh2026symmetry,gauswami2025exploration} with hexagonal NiAs-type crystal structures, are also predicted to be altermagnets.

\begin{table*}[t]
\centering
\caption{Two-dimensional altermagnetism is seen in 7 spin layer groups. The three-dimensional classification predicted 5 planar spin Laue groups in two-dimensions; the spin layer group analysis reveals 2 additional groups from layer-specific symmetries. Material examples are provided for each class.}
\label{tab:2D}
\smallskip
\begin{tabular}{@{}lccl@{}}
\hline
\hline
Wave type
  & Nodal lines
  & Spin layer groups
  & Example materials \\
\hline
\hline
$d$-wave & 2 & $^22/^2m_x$, $^2m^2m^1m$, $^24/^1m$, $^24/^1m^2m^1m$
  & RuF$_4$, VF$_4$, AgF$_2$, V$_2$Se$_2$O, MnTeMoO$_6$  \\[4pt]
$g$-wave & 4 & $^14/^1m^2m^2m$
  & VP$_2$H$_8$(NO$_4$)$_2$   \\[4pt]
$i$-wave & 6 & $^1\bar{3}^2m$, $^16/^1m^2m^2m$
  & Mn$_2$P$_2$S$_3$Se$_3$ \\
\hline
\hline
\end{tabular}
\end{table*}

\subsection{Rutile structure type}

Another major structural family in which altermagnetism has been extensively explored is the rutile-type crystal structure with collinear antiferromagnetic ordering. These compounds crystallize in the tetragonal rutile structure with space group $P4_2/mnm$ (No. 136) and generally follow the chemical formula $MX_
2$, where M is a transition-metal cation and X is an anion such as oxygen or fluorine. The magnetic atom M occupies the 2a Wyckoff position, where the magnetic moments of the transition-metal atoms located at the corner and body-centered sites of the unit cell align antiparallel to each other, forming a collinear antiferromagnetic configuration. Although the rutile crystal structure is globally centrosymmetric, the nonmagnetic X atoms occupy the noncentrosymmetric 4f Wyckoff positions. Consequently, the X atoms form inequivalent bond lengths with the two oppositely spin-polarized M atoms in the distinct spin sublattices. This non-magnetic environment prohibits translation or inversion relation between opposite spin sublattices.

As shown in Fig.~\ref{Fig_Rutile}(a), the two opposite spin sublattices are connected through the nonsymmorphic fourfold screw rotation symmetry operation, $[{C}_2 \,||\, {C}_{4z} t]$, where $t$ is a (1/2,1/2,1/2) translation. The spin point group of these structures is $^24/'m'm'm$. This symmetry configuration breaks time-reversal symmetry while preserving zero net magnetization, giving rise to momentum-dependent spin splitting in the electronic band structure, which is a characteristic signature of altermagnetism. These systems possess two spin degenerate nodal planes in momentum space, and the spin polarization remains invariant along the $k_z$ direction. Consequently, rutile altermagnets belong to the P-2 class of $d$-wave altermagnets.

Among rutile structure altermagnets, RuO$_2$ is considered as the most prominent and extensively studied material. In the very early stages of research on altermagnetism, RuO$_2$ received considerable attention due to its remarkably large predicted spin splitting of about 1.4 eV, which is significantly higher than that reported for other altermagnets, while its N\'eel temperature was also reported to be well above the room temperature. As a result of these attractive properties many of the pioneering experimental and theoretical altermagnetic studies were focused on RuO$_2$ and it was considered as one of the most suitable altermagnetic material for spintronic application~\cite{hussain2025exploring}.

However, the magnetic nature of RuO$_2$ has remained controversial for many years. Initially, RuO$_2$ was widely regarded as a paramagnetic metal. Later, a 2017 neutron diffraction study on single-crystal RuO$_2$ reported the presence of antiferromagnetic ordering persisting up to at least 300 K, with a relatively small magnetic moment of approximately 0.05 $\mu_B$ per Ru atom at room temperature~\cite{berlijn2017itinerant}. This observation subsequently led to its classification as a potential altermagnetic material.

Despite the theoretical prediction of altermagnetism in RuO$_2$, several experimental studies later questioned the existence and robustness of its magnetic order and its actual magnetic ground state still remains as an active topic of debate within the research community~\cite{choi2026exploring}. The ongoing debate and experimental inconsistencies related to the magnetic behavior of RuO$_2$ will be discussed separately in Section~\ref{sec:future_directions}.

ReO$_2$ is another rutile oxide which may host $d$-wave altermagnetism, and first-principles calculations have been used to predict that this altermagnetic phase can achieve stability under pressure~\cite{chakraborty2024strain}.

Transition metal fluorides such as CoF$_2$, MnF$_2$, FeF$_2$, and NiF$_2$, which crystallize in the rutile structure of space group $P4_2/mnm$ (No. 136) and exhibit antiferromagnetic ordering, are considered $d$-wave altermagnetic candidates~\cite{sadhukhan2026first,guo2023spin}. These materials share the same crystal symmetry as the rutile oxides, with fluorine atoms occupying the same non-centrosymmetric crystallographic positions as oxygen atoms in rutile oxides. Among these, altermagnetism in the wide-gap insulator MnF$_2$ was experimentally established in 2025 by Faure \textit{et al.} through a polarized neutron scattering experiment, where a 0.2 meV magnon splitting was observed originating from long-range dipolar coupling~\cite{faure2025altermagnetism}. First-principles calculations~\cite{adamantopoulos2024spin,cao2024designing,bonbien2022topological,samanta2025spin} and high-throughput studies~\cite{sufyan2026high} have predicted spin splitting in other altermagnetic candidates, typically ranging from tens to hundreds of meV in bulk. No experimental studies have measured spin splitting in the band structure of these materials yet. The very small predicted spin splitting magnitude hinders a direct measurement.

\begin{figure*}[t]
\centerline{\includegraphics[scale=0.6]{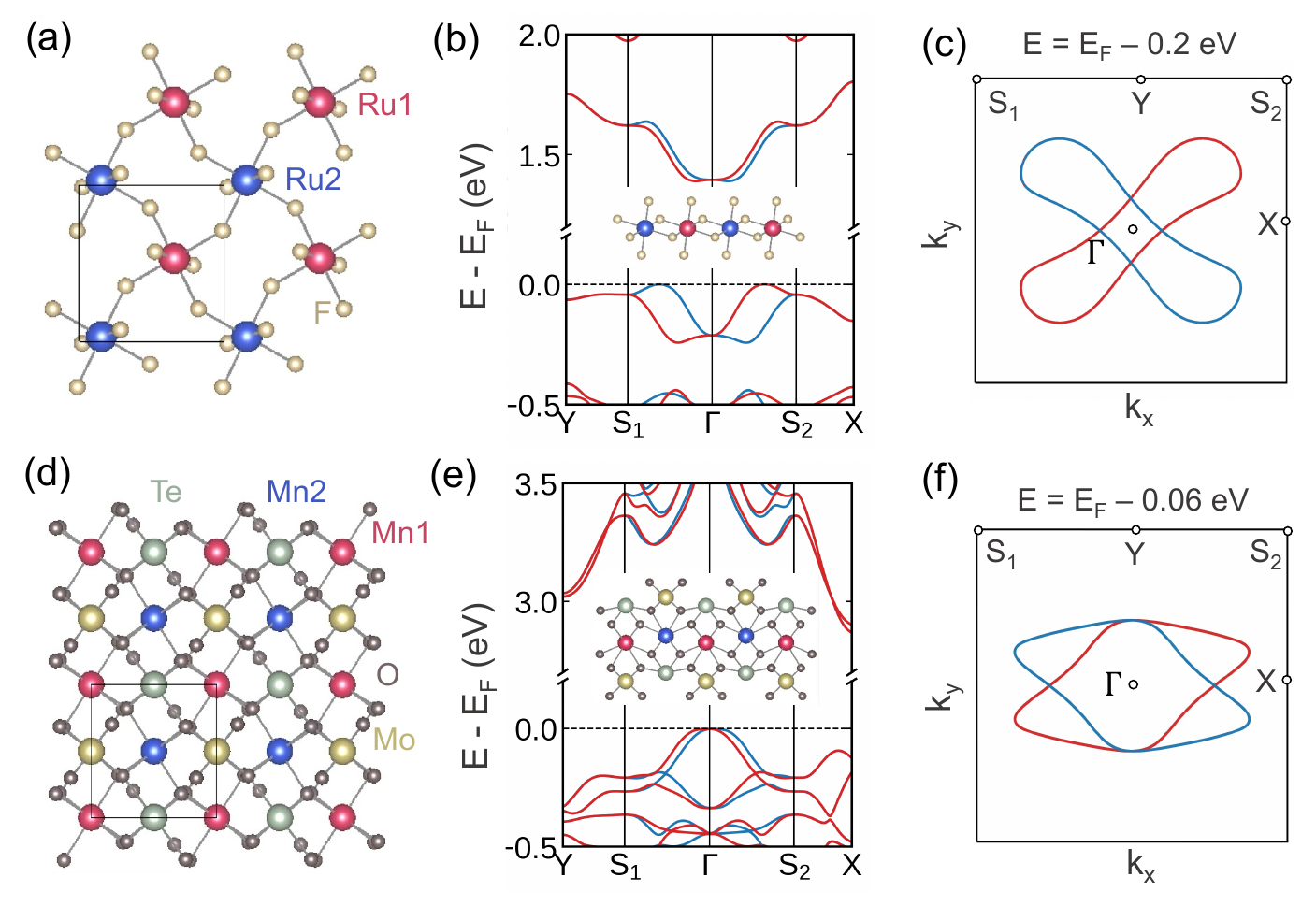}}
\caption{\textbf{Altermagnetism in two-dimensional materials.} (a) Crystal structure of RuF$_4$, shown in top view. (b) Electronic band structure of RuF$_4$ with the Fermi energy referenced to the highest occupied state. Inset shows the side view of the crystal structure. (c) Constant-energy contour at 0.2 eV below the Fermi level. The characteristic presence of two nodal planes in the spin-splitting pattern identifies RuF$_4$ as a $d$-wave altermagnet. Red and blue denote contributions from the spin-up and spin-down channels, respectively, in both the band structure and constant-energy contour. (d)-(f) Corresponding crystal structure, electronic band structure, and constant-energy contour at 0.06 eV below the Fermi level for the two-dimensional altermagnet MnMoTeO$_6$.}
\label{2D_bands_fs}
\end{figure*}

\subsection{Two-dimensional materials}
\label{sec:2D_altermagnets}

Since the experimental realization of long-range magnetic order in two-dimensional van der Waals materials, two-dimensional magnets have emerged as a major focus of spintronics research~\cite{gong2019two,li2019intrinsic,khan2020recent}. This interest is driven by the unique advantages of two-dimensional materials, including their enhanced sensitivity to external stimuli, the tunability of their physical properties, and the possibility of assembling them into designer heterostructures through van der Waals stacking~\cite{novoselov20162d,liu2016van,chen2019electrically}. These features provide a versatile platform for tailoring electronic, magnetic, and spin-dependent functionalities. Within this context, the exploration of two-dimensional altermagnets has attracted increasing attention owing to the prospect of combining the advantages of low-dimensional materials with the distinctive spin-splitting characteristics of altermagnetic order.

Despite the rapid progress in the identification of three-dimensional altermagnets, the realization of two-dimensional altermagnets remains comparatively limited~\cite{zeng2026classification}. The theoretical foundation for this field was laid by Ma \textit{et al.}, who first predicted two-dimensional altermagnetism in monolayer V$_2$Se$_2$O via crystal-symmetry-paired spin-valley locking, giving rise to giant piezomagnetism and noncollinear spin currents~\cite{ma2021multifunctional}, later realized experimentally in its bulk intercalated forms~\cite{jiang2025metallic,zhang2025crystal}. In three-dimensional magnetic systems, spin degeneracy is commonly protected by combined spin-symmetry operations such as $[{C}_2 \,||\, {P}]$ $[{C}_2 \,||\, {T}] \equiv {P}{T}$, and $[{C}_2 \,||\, \tau]$, where ${P}$, ${T}$, and $\tau$ denote spatial inversion, time-reversal symmetry, and lattice translation, respectively. These symmetries transform an energy eigenstate $E(\mathbf{k}_\parallel, s)$ into $E(\mathbf{k}_\parallel, -s)$, thereby enforcing spin degeneracy throughout the three-dimensional Brillouin zone.

In contrast, two-dimensional systems are restricted to in-plane momentum components $(\mathbf{k}_\parallel)$, with no out-of-plane component $(\mathbf{k}_\perp)$. Consequently, the breaking of the aforementioned symmetries alone is generally insufficient to lift spin degeneracy in two-dimensional materials. Additional symmetry constraints arise due to reduced dimensionality. For example, combined operations such as $[{C}_2 \,||\, {C}_{2z}][{C}_2 \,||\, {T}]$, or equivalently $[{C}_2 \,||\, {M}_z]$, ${C}_{2z}$ denotes a two-fold rotation about the out-of-plane ($z$) axis and ${M}_z$ represents mirror symmetry with respect to the $xy$-plane, can enforce band degeneracy across the entire two-dimensional Brillouin zone. These additional symmetry-imposed constraints render the realization of spin splitting and, therefore, intrinsic altermagnetism, significantly more restrictive in two-dimensional systems compared to their three-dimensional counterparts.

Nevertheless, considerable recent progress has been made in the realization, discovery, and classification of two-dimensional altermagnets. Several extrinsic routes to engineering two-dimensional altermagnetism have been proposed. These include electric field induced spin splitting~\cite{wang2024electric}, gate-voltage control mediated by spin-layer coupling~\cite{zhang2024predictable}, twisted van der Waals bilayers~\cite{liu2024twisted}, and bilayer stacking strategies~\cite{pan2024general,zeng2024bilayer}. A comprehensive overview of recent advances in low-dimensional altermagnetism was provided by Bai \textit{et al.}, who highlighted practical approaches for realizing altermagnetism through symmetry engineering, external electric fields, and Janus structural modifications~\cite{bai2024altermagnetism}.

Parallel to these developments, substantial progress has been made in the search for intrinsic two-dimensional altermagnetic materials. Using high-throughput computational screening of the C2DB repository, Sødequist and Olsen identified a number of thermodynamically stable two-dimensional materials exhibiting altermagnetic order~\cite{sodequist2024two}. In a separate study, Wang \textit{et al.} investigated a large family of pentagonal two-dimensional compounds and discovered several altermagnetic semiconductors~\cite{wang2025pentagonal}. From a symmetry perspective, Zeng and Zhao formulated a spin-group-based theoretical framework for two-dimensional magnetism, identifying seven distinct spin layer groups that provide a rigorous classification scheme for two-dimensional altermagnetic phases~\cite{zeng2024description} (detailed in Section~\ref{sec:spin_group_theory}). More recently, Sufyan \textit{et al.} conducted a large-scale symmetry-guided search of magnetic compounds contained in the MAGNDATA database and reported 180 candidate altermagnets displaying substantial momentum-dependent spin splitting. Their analysis significantly expanded the known altermagnetic materials library by revealing numerous previously unrecognized bulk and two-dimensional candidates~\cite{sufyan2026high}. In another recent study, Bose \textit{et al.} performed a high-throughput search of two-dimentional altermagnets across 2710 monolayer materials from the Materials Cloud 2D Crystals (MC2D) database. Combining chemical and symmetry-based filtering, exfoliation criteria, and DFT+$U$ calculations with self-consistently determined Hubbard-$U$ values they identified 24 robust two-dimensional altermagnetic materials, including 20 previously unreported candidates, thereby significantly expanding the library of intrinsic two-dimensional altermagnets~\cite{bose2026symmetry}.

Representative examples of the theoretically identified two-dimensional altermagnets are shown in Fig.~\ref{2D_bands_fs}. Figs.~\ref{2D_bands_fs}(a)–(c) display the crystal structure and electronic properties of RuF$_4$, which crystallizes in the monoclinic space group $P2_1/c$ (No. 14). RuF$_4$ has recently been identified as a promising two-dimensional altermagnetic material and has attracted considerable attention owing to its large momentum-dependent spin splitting and symmetry-protected nodal structure~\cite{sodequist2024two,zeng2024description,milivojevic2024interplay}. The layered crystal structure hosts a compensated magnetic configuration [Fig.~\ref{2D_bands_fs}(a)], where the red and blue spheres represent Ru atoms belonging to opposite spin sublattices. The spin-polarized band structure exhibits large spin splitting, a hallmark of altermagnetic order. The corresponding constant-energy contour at 0.2 eV below the Fermi level, as shown in Fig.~\ref{2D_bands_fs}(c), reveals two nodal planes along which the spin splitting vanishes and changes sign across symmetry-related momentum directions. This characteristic $d$-wave spin-splitting pattern reflects the underlying crystal and magnetic symmetries and identifies RuF$_4$ as a prototypical $d$-wave altermagnet.

A second representative example is MnMoTeO$_6$, shown in Figs.~\ref{2D_bands_fs}(d)–(f), which crystallizes in orthorhombic structure with space group P2$_1$2$_1$2 (No. 18). As shown in Fig.~\ref{2D_bands_fs}(d), the layered crystal structure hosts a compensated magnetic configuration, with the red and blue spheres denoting Mn atoms belonging to opposite spin sublattices. The spin-resolved band structure reveals spin splitting along $S_1$-$\Gamma$-$S_2$ path. The constant-energy contour at 0.06 eV below the Fermi level [Fig.~\ref{2D_bands_fs}(f)] exhibits symmetry-enforced nodal directions and a characteristic two-node spin texture, which further classifies MnMoTeO$_6$ as a $d$-wave altermagnet, as also predicted in previous studies~\cite{zeng2024description,bose2026symmetry}.

\begin{figure*}[htbp]
\centerline{\includegraphics[scale=0.45]{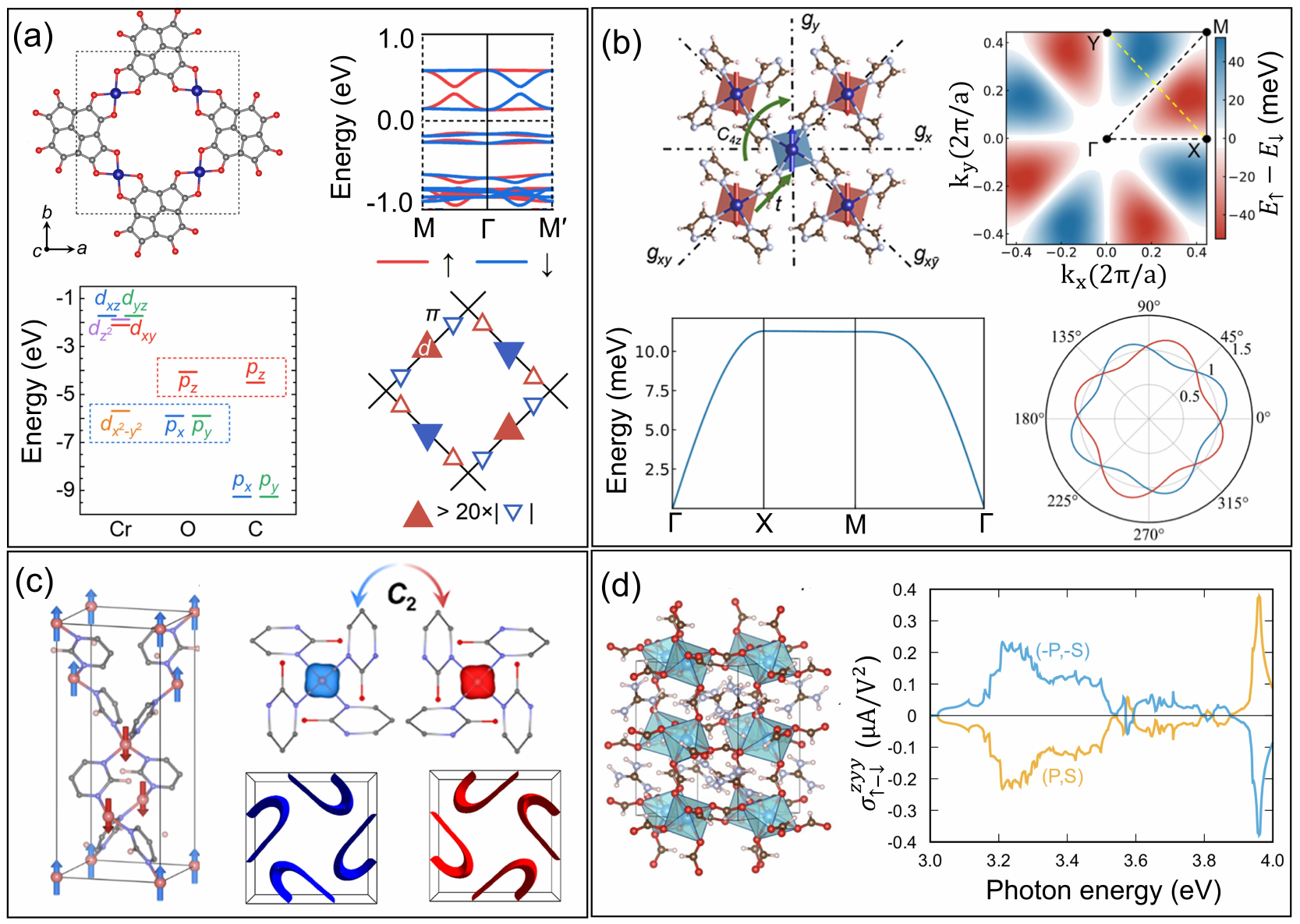}}
\caption{\textbf{Altermagnetism in representative two- and three-dimensional metal-organic frameworks.} (a) Two-dimensional \(t\)-\(\mathrm{Cr}_{2}(\mathrm{Pyc}\!-\!\mathrm{O})_{8}\) framework constructed from Cr centers and pyc-based organic linkers. The electronic structure, orbital-level alignment, and exchange-pathway schematic show that Cr \(d\) orbitals couple through ligand \(\pi\) states via a \(d\)-\(\pi\)-\(\pi\)-\(d\) channel, giving rise to momentum-dependent spin splitting in a compensated magnetic state. Blue, gray, and red spheres denote Cr, C, and O atoms, respectively; H atoms are omitted for clarity. (b) Imidazolate-based two-dimensional \(\mathrm{Cr(imz)}_{2}\) MOF. The ligand environment lowers the symmetry such that the combined spin-lattice operations protecting spin degeneracy are removed, allowing altermagnetic spin splitting near the valence-band maximum. The corresponding spin-wave spectrum and angular dependence of the nonlinear current response illustrate the \(g\)-wave character and in-plane anisotropy of the altermagnetic state. (c) Schematic spin arrangements in three-dimensional Co($X$-pymo)$_2$ showing a magnetically compensated configuration. The spin-density distribution of Co(H-pymo)$_2$ confirms an altermagnetic state, in which opposite-spin sublattices in real space are related by a twofold rotational symmetry, $C_2$.
Despite the compensated magnetic order, the conduction band exhibits spin-dependent constant-energy surfaces and corresponding contours in the first Brillouin zone, shown here at $E = 2.495$~eV. Blue and red denote the spin-up and spin-down components, respectively. (d) Polar three-dimensional Cr-based MOF exhibiting electrically switchable altermagnetism. Reversal of the polar distortion switches the coupled \((P,S)\) and \((-P,-S)\) states, leading to a sign reversal of the spin-dependent linear photogalvanic conductivity component \(\sigma^{zyy}_{\uparrow-\downarrow}\). Panels are adapted with permission from Refs.~\cite{che2025prl,lopezalcala2025chemical,ni2026jacs,gu2025prl}.
}
\label{figmof}
\end{figure*}

\subsection{Metal-organic frameworks}
\label{subsec:am_mofs}

Metal-organic frameworks (MOFs) have recently emerged as a chemically versatile platform for altermagnetism, extending the field beyond dense inorganic crystals into reticular and molecular materials. The symmetry conditions underlying altermagnetism make MOFs particularly attractive. Their metal nodes, organic linkers that mediate ligand-assisted hopping, which can be downfolded into an effective direct hopping pathway between metal centers~\cite{mazumdar2026nonlinear}, coordination geometry, stacking sequence, and network topology can be selected almost independently. This allows the magnetic exchange pathway and the symmetry relation between opposite-spin sublattices to be engineered directly by chemical design. Representative two- and three-dimensional examples are summarized in Fig.~\ref{figmof}, illustrating how reticular topology, ligand symmetry, chirality, and polarity can each provide distinct routes to altermagnetic order.

A particularly transparent design principle is provided by two-dimensional square tessellations. Che \textit{et al.} developed a symmetry-guided framework showing that altermagnetism is favored when antiparallel magnetic sites occupy Wyckoff positions of multiplicity larger than one, thereby avoiding high-symmetry positions that would otherwise protect conventional antiferromagnetic spin degeneracy~\cite{che2025prl}. By screening square tessellations, they identified the Lieb, \textit{fes}, and \textit{tts} nets as natural altermagnetic motifs. The pyracylene-based monolayer \(t\)-\(\mathrm{Cr}_{2}(\mathrm{Pyc}\! -\! \mathrm{O})_{8}\) realizes this concept in a \textit{tts} net with \(3.3.4.3.4\) tessellation [Fig.~\ref{figmof}(a)]. In this framework, Cr nodes are connected through pyc-based organic linkers, and the relevant Cr \(d\) orbitals couple through ligand \(\pi\) states via a ligand-mediated \(d\)-\(\pi\)-\(\pi\)-\(d\) exchange pathway. The combination of compensated magnetic order, linker-mediated exchange, and the absence of a spin-degeneracy-protecting translation or inversion operation gives rise to momentum-dependent spin splitting. This example innovatively connects reticular topology, Wyckoff-position engineering, and microscopic superexchange into a single chemical design rule for MOF altermagnetism.

A second route is ligand-symmetry engineering, where the organic linker actively controls the local symmetry, orbital hybridization, metal-metal distance, and ligand-mediated magnetic exchange. This mechanism is illustrated by planar tetracoordinated Cr-based MOFs~\cite{lopezalcala2025chemical}. In high-symmetry pyrazine-based lattices, spin-lattice symmetries such as glide or inversion-translation operations can enforce conventional antiferromagnetic spin degeneracy. Replacing pyrazine by lower-symmetry imidazolate-type ligands removes the degeneracy-protecting operations while preserving compensated magnetic order, thereby enabling altermagnetic spin splitting. In \(\mathrm{Cr(imz)}_{2}\), this produces a \(g\)-wave altermagnetic texture near the valence-band maximum, as shown in Fig.~\ref{figmof}(b). The corresponding spin-wave spectrum and angular dependence of the nonlinear current response further highlight the in-plane anisotropy of the altermagnetic state. Furthermore, frontier-orbital engineering in polycyclic ligands can redistribute spin density onto the ligand backbone and drive changes in the angular symmetry of the spin splitting, as exemplified by Cr(DAind)$_2$. This chemical route is powerful because it turns altermagnetism into a coordination-design problem. In particular, ligand orientation, conjugation length, and metal-ligand covalency become direct control knobs for the magnitude and wave character of altermagnetic splitting.

Three-dimensional nonpolar MOFs are also important because bulk crystals can be more accessible to synthesis, structural characterization, and magnetometry than free-standing monolayers. Co(pymo)$_2$ [as shown in Fig.~\ref{figmof}(c)] and its halogenated derivatives Co($X$-pymo)$_2$ (\(X=\mathrm{H}\), Cl, Br, I) were predicted to host \(g\)-wave altermagnetic states in which the compensated altermagnetic configuration is energetically favored among the magnetic states considered~\cite{ni2026jacs}. In Co(H-pymo)$_2$, two magnetically compensated configurations are possible. One behaves as a conventional antiferromagnet with spin-degenerate bands, whereas the other realizes altermagnetism with spin splitting along general momentum-space paths. The altermagnetic state is consistent with experimentally observed antiferromagnetic coupling between Co ions mediated by pymo ligands. The calculated spin splitting is non-relativistic and is not substantially enhanced by heavier halogen substitution, emphasizing that symmetry and exchange geometry govern the effect.

The altermagnetic design space can be further expanded by introducing chirality in the three-dimensional systems. Xie and co-workers proposed that the experimentally characterized chiral antiferromagnetic formate framework \(\mathrm{K[Co(HCOO)_3]}\) can host chiral \(g\)-wave altermagnetic spin splitting~\cite{xie2025chiral}. In this three-dimensional framework, the Co sublattice forms a chiral network in which opposite-spin sublattices are connected by a screw-rotation-type spin-space operation rather than by inversion or translation. Consequently, the system displays \(g\)-wave altermagnetic spin splitting in three-dimensional momentum space. Since the crystal is chiral, reversing the structural handedness reverses the spin character of the altermagnetic bands and the spin-polarized Fermi-surface contribution. This chirality locking can also affect electronic topology and transport~\cite{joseph2024chirality,bandyopadhyay2024non}, and has been proposed to enable enantiomer-dependent Hall and magneto-optical responses. Chiral MOFs therefore provide a route to coupling altermagnetism with handedness, topological ideas, and optical readout.

Ferroelectricity provides a complementary nonvolatile route to control altermagnetism. Gu \textit{et al.} introduced the concept of ferroelectric switchable altermagnetism, in which reversal of the electric polarization also reverses the altermagnetic spin splitting~\cite{gu2025prl}. The hybrid improper ferroelectric Cr-based MOF \(\mathrm{[C(NH_2)_3]Cr(HCOO)_3}\) serves as a representative example. In this material, the altermagnetic order parameter \(S\), defined by the spin-channel energy splitting, is coupled to the ferroelectric polarization \(P\). A low-energy structural switching path connects the \((P,S)\) and \((-P,-S)\) states without requiring reversal of the N\'eel vector. This mechanism is distinct from conventional magnetoelectric coupling, where electric polarization controls a net magnetic moment. Here, the electrically controlled quantity is the spin splitting in reciprocal space. The switching can be detected through the spin-dependent linear photogalvanic conductivity component \(\sigma^{zyy}_{\uparrow-\downarrow}\), whose sign reverses upon polarization switching, as shown in Fig.~\ref{figmof}(d). This mechanism suggests a route to nonvolatile spin filters and tunnel junctions based on electrically switchable compensated spin-polarized bands.

Beyond the representative systems highlighted in Fig.~\ref{figmof}, several additional MOF platforms further illustrate the breadth of chemical control. Early theoretical proposals showed that two-dimensional pyrazine-based frameworks \(M(\mathrm{pyz})_{2}\), with \(M=\mathrm{Ca}\) or Sr, can realize altermagnetic semiconducting states with electric-field-controlled anisotropic spin currents~\cite{che2024chemsci}. In these systems, the relevant spin-polarized frontier states are associated with the organic-inorganic framework rather than conventional transition-metal \(d\)-shell magnetism. The metal-pyrazine network forms a tetragonal structure in which opposite-spin sublattices are connected by a fourfold real-space operation combined with a spin-space rotation, rather than by inversion or translation. As a result, the net magnetization vanishes while the valence or conduction bands acquire alternating spin polarization in momentum space. This work demonstrated that light-element organic-inorganic frameworks can host non-relativistic spin splitting and pure spin-current responses.

Layered MOFs add another degree of freedom, namely the layer index. In bilayer Cr-based MOFs, altermagnetic spin splitting can be coupled to spin, valley, and layer polarization~\cite{che2025bilayer}. Starting from bilayer Cr(dcb)$_2$, which remains a conventional antiferromagnet because inversion symmetry protects spin degeneracy, chemical modification of the linkers can break the relevant symmetry and generate \(S_{4}\)-based bilayer altermagnetic states such as Cr(tcb)$_2$ and Cr(hcb)$_2$. These systems exhibit spin-split valleys at symmetry-related momenta, with opposite spin valleys localized in different layers. An out-of-plane electric field shifts the layer-polarized valleys in opposite directions, increasing the spin splitting at one valley while reducing it at the other. This provides a route to electrically tunable spin-valley-layer devices. Among these candidates, Cr(tcb)$_2$ is particularly attractive because it was predicted to be dynamically stable, whereas Cr(hcb)$_2$ is limited by its structural instability.

Beyond square nets, Cairo-pentagonal MOFs broaden the symmetry landscape of altermagnetic molecular materials. In TM$_2$(TCNQ)$_2$ monolayers, where TM denotes a transition metal and TCNQ is 7,7,8,8-tetracyanoquinodimethane, the bilaterally symmetric Cairo-pentagonal lattice supports \(g\)-wave altermagnetism~\cite{zhang2026prb}. For Ru$_2$(TCNQ)$_2$, the compensated magnetic state arises mainly from antiparallel Ru moments, while the spin sublattices are connected by mirror-related spin-space operations rather than inversion or translation. A particularly interesting feature is the strain sensitivity as a tensile strain along the \([110]\) direction lowers the symmetry and drives a transition from \(g\)-wave to \(d\)-wave altermagnetism. The same transition produces strongly anisotropic charge and spin plasmons. This work shows that flexible MOF lattices can be used not only to realize altermagnetic order, but also to tune the wave character of altermagnetism mechanically.

Together, these studies establish several important design principles for altermagnetic MOFs. First, symmetry must be designed before composition is optimized. In other words, compensated magnetic order alone is insufficient unless opposite-spin sublattices are connected by the appropriate rotational, rotoinversion, screw, glide-assisted, or spin-space operation. Second, organic linkers should be treated as active electronic components. Their point symmetry, non-bonding orbitals, radical character, \(\pi\) conjugation, and covalency can determine whether the system remains a conventional antiferromagnet or becomes an altermagnet. Third, different reticular topology such as square, Lieb, \textit{fes}, \textit{tts}, Cairo-pentagonal, chiral, polar, and layered nets provide different symmetry routes to \(d\)- or \(g\)-wave spin splitting. Fourth, multifunctionality is a distinctive advantage of MOFs. Altermagnetic spin splitting can be combined with spin-valley-layer coupling, ferroelectric switching, chirality, and nonlinear optical response.

We note that the field remains largely theoretical, and experimental realization is the next major challenge. Many proposed MOF altermagnets have low ordering temperatures, limited conductivity, or demanding synthetic requirements. Direct spin-resolved ARPES would be the most direct probe of momentum-dependent spin splitting, but high-quality oriented crystals or films are required. More practical near-term probes may include magneto-optical Kerr and Faraday effects, anomalous Hall-like responses in doped samples, or nonlinear optical spin-current generation. For experimental exploration, the most promising candidates are not necessarily those with the largest calculated spin splitting, but those combining accessible chemistry, stable magnetic order, robust framework structure, and measurable electronic or optical response. More broadly, altermagnetic MOFs show how reticular chemistry can be used to design not only exchange interactions, but also the crystallographic operations that lead to altermagnetism.

\subsection{Covalent-organic frameworks}
\label{subsec_COF}

\begin{figure}[t]
\centering
\includegraphics[width=0.45\textwidth]{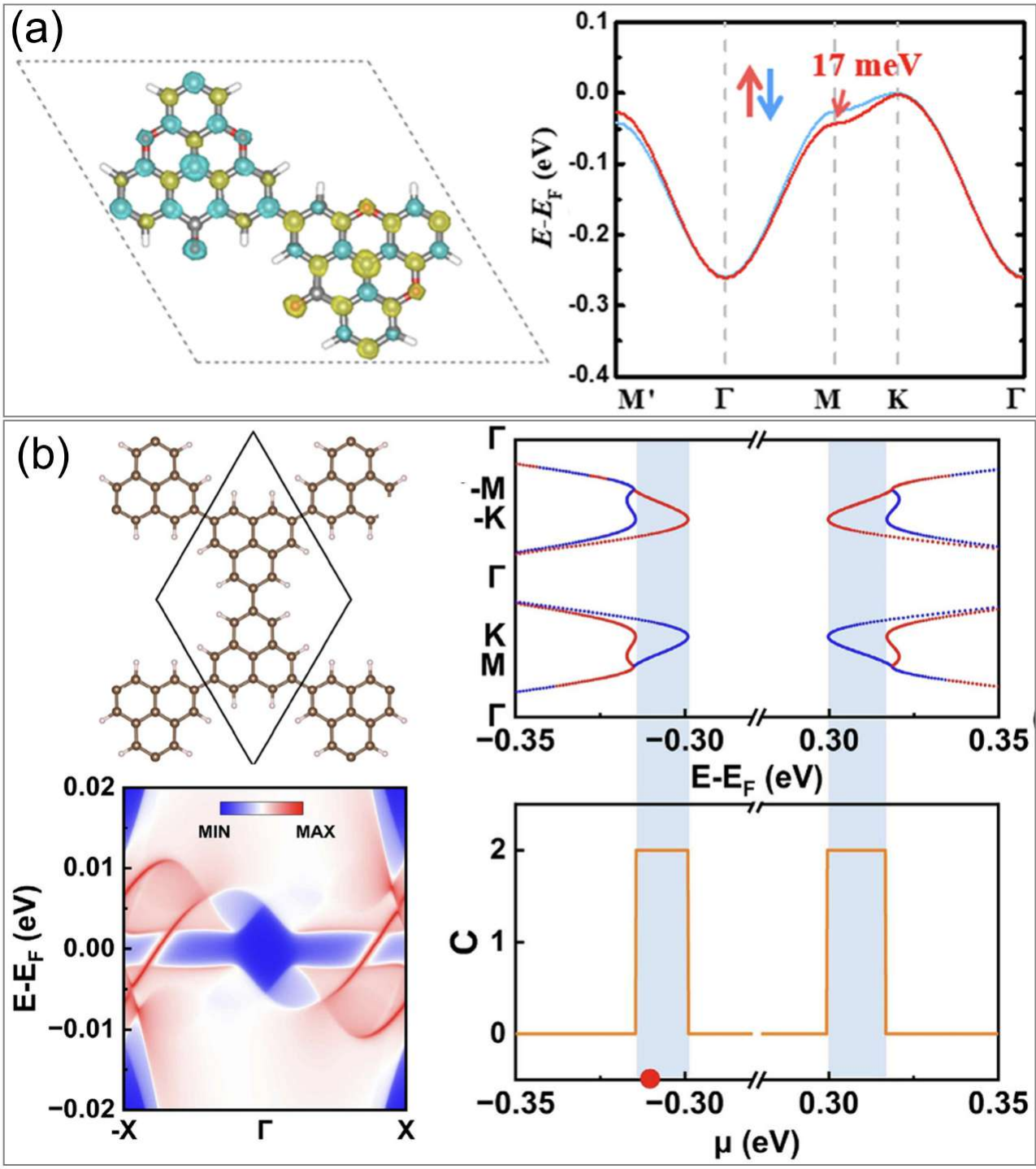}
\caption{\textbf{Organic and covalent-framework platforms for altermagnetism.} (a) Molecular design strategy for metal-free altermagnetism in honeycomb two-dimensional organic frameworks. The crystal structure and spin-density distribution of the TAM-O framework are shown together with its spin-resolved electronic band structure, where a non-relativistic spin splitting of approximately $17$ meV is obtained. (b) Light-controlled altermagnetic and topological phases in a two-dimensional [2]triangulene crystal. The atomic structure is shown together with the Floquet band structure under right-handed circularly polarized light, the spectral function, and the Chern number $C$ as a function of chemical potential $\mu$ in the presence of an $s$-wave superconducting pairing potential. The shaded regions indicate topologically nontrivial superconducting regimes supporting gapless chiral edge states. Panels are adapted with permission from Refs.~\cite{Yu2026Arxiv,Zhang2026NanoLett}.}
\label{figcof}
\end{figure}

Building on the broader framework material paradigm established by MOFs, COFs provide an all-covalent and chemically programmable route to crystalline low-dimensional materials. Since the first porous crystalline COFs were reported by C{\^o}t{\'e} \emph{et al.}, in which light elements were connected by strong covalent bonds into rigid porous architectures~\cite{Cote2005Science}, the field has developed from reticular porous chemistry into a flexible platform for electronic, optical, magnetic, and topological materials design. The extension to three-dimensional COFs~\cite{ElKaderi2007Science} and the subsequent conceptualization of COFs as an ``atom-molecule-framework'' hierarchy~\cite{Diercks2017Science} established an important principle: molecular building blocks can be selected not only for porosity and structural stability, but also for their orbital symmetry, spin state, and correlation strength.

In conventional inorganic crystals, the electronic structure is constrained by atomic orbital symmetries and by a limited set of stable crystal structures. In COFs, by contrast, the active orbitals can be encoded directly at the molecular level. The lattice connectivity, hopping anisotropy, and magnetic exchange pathways can be engineered through linker chemistry, node symmetry, conjugation length, and pore topology. As a result, two-dimensional $\pi$-conjugated COFs can realize effective tight-binding lattices that are difficult to stabilize in purely inorganic materials. This has been illustrated in theoretical and computational studies of Lieb-like COFs, where ferromagnetism, tunable topology, and doping-dependent magnetic transitions emerge from the organic network~\cite{Jiang2019NatCommun,Cui2020NatCommun}. Similarly, hexagonal $\pi$-conjugated COFs have been proposed as platforms for antiferromagnetic Mott insulating phases~\cite{Thomas2019AdvMater}, while heterotriangulene-based frameworks have been identified as organic topological insulator candidates~\cite{Ni2022JACS}. A key step towards magnetic COF materials is the development of open-shell and radical-based frameworks. Conventional closed-shell COFs tend to pair their valence electrons through covalent bonding, thereby suppressing local magnetic moments. In contrast, radical building blocks, non-Kekul'{e} nanographenes, and triangulene-derived units can retain robust $\pi$-electron spin moments when assembled into extended networks. The synthesis of $\pi$-conjugated covalent organic radical frameworks~\cite{Wu2018Angew} and the prediction of metal-free Stoner and Mott-Hubbard magnetism in triangulene-based two-dimensional polymers~\cite{Yu2024SciAdv} are important milestones. They demonstrate that magnetic functionality in COFs need not rely on transition-metal $d$ orbitals, but can instead originate from molecular topology, sublattice imbalance, and electron correlation within a conjugated carbon network.

The recent emergence of altermagnetism opens a particularly attractive direction for COFs. They are natural candidates for metal-free altermagnetism because their magnetic units can be selected from open-shell nanographenes, triangulene derivatives, or radical monomers whose spin states are controlled by molecular topology. Reticular synthesis can, in principle, fix the relative orientation of these magnetic units, allowing antiparallel spin blocks to be related by rotations or other crystalline operations rather than by inversion or translation. At the same time, the $\pi$-conjugated backbone naturally produces direction-dependent hopping amplitudes, so that the same molecular design that controls exchange coupling also controls the momentum texture of the spin splitting. In COFs, the altermagnetic form factor can therefore be encoded directly by the symmetry, connectivity, and anisotropic hopping structure of the $\pi$ network.

Recent theoretical works have begun to establish this design principle. Ortiz and coauthors proposed organic altermagnets based on nanographene frameworks, focusing on dibenzo[ef,kl]heptalene (DBH) as an $S=1$ $\pi$-conjugated building block~\cite{Ortiz2026_2DMat}. The inclusion of fused heptagonal rings avoids the symmetry restrictions that prevent ordinary alternant hydrocarbons from forming square altermagnetic networks with neighboring units related by a $\pi/2$ rotation. Their calculations show that DBH-based two-dimensional crystals can host compensated magnetic order with broken time-reversal symmetry and non-relativistic spin splitting with a $d$-wave momentum texture. Importantly, DBH units connected by $-\mathrm{C}=\mathrm{C}-$ and $-\mathrm{C}=\mathrm{N}-$ linkers form COF-like structures that retain the altermagnetic spin splitting. These DBH-based frameworks are therefore chemically realistic organic altermagnetic candidates in which the spin-degeneracy pattern can be controlled by both the magnetic molecular unit and the linker symmetry.

A complementary strategy was introduced by Yu, Brumme, and Heine, who proposed designer metal-free altermagnetism in honeycomb two-dimensional organic frameworks~\cite{Yu2026Arxiv}. Their central idea is to lower the symmetry of triangulene-derived radical monomers from $D_{3h}$ to $C_{2v}$ while preserving the bipartite character of the honeycomb lattice. In high-symmetry $D_{3h}$-based frameworks, inversion symmetry relates opposite-spin centers and protects the spin degeneracy of the antiferromagnetic state. By reducing the monomer symmetry to $C_{2v}$, this degeneracy-protecting symmetry is selectively removed while the compensated antiferromagnetic arrangement remains intact. Their DFT calculations for the proposed [TOT-O] and [TAM-O] frameworks reveal strong antiferromagnetic exchange, Mott-Hubbard insulating gaps, and altermagnetic spin splitting. In particular, the [TAM-O] framework exhibits a spin splitting of approximately $17$ meV, as shown in Fig.~\ref{figcof}(a). A minimal tight-binding description identifies anisotropic nearest-neighbor hopping, caused by direction-dependent overlap of $\pi$ orbitals, as the microscopic origin of the altermagnetic splitting. Although the large exchange scales suggest the possibility of high-temperature organic altermagnetism, the realization of true long-range magnetic order in strictly two-dimensional light-element frameworks will depend on magnetic anisotropy, interlayer coupling, substrate effects, or other mechanisms that overcome thermal spin fluctuations.

Beyond static framework design, triangulene crystals also provide a route to externally switchable altermagnetism through Floquet engineering. Zhang \emph{et al.} proposed light-induced odd-parity altermagnetism in one- and two-dimensional triangulene crystals~\cite{Zhang2026NanoLett}. In this mechanism, circularly polarized light modifies the effective hopping processes and breaks the relevant symmetries that protect spin degeneracy in the parent antiferromagnetic state. As a result, antiferromagnetic triangulene chains and sheets can be driven into odd-parity altermagnetic phases, including $p$-wave behavior in one dimension and $f$-wave behavior in two dimensions. We discuss odd-parity altermagnets later on in this review in Section~\ref{sec:future_directions}. For the two-dimensional [2]triangulene crystal, right-handed circularly polarized light produces a spin-split Floquet band structure, as illustrated in Fig.~\ref{figcof}(b). When this light-induced altermagnetic state is combined with superconductivity, topological superconducting phases can appear within selected chemical potential windows. This proposal therefore connects metal-free altermagnetism not only to organic spintronics, but also to Floquet band engineering and topological superconductivity in $\pi$-conjugated molecular frameworks.

Taken together, these studies suggest that COF-based altermagnets are promising, but they require careful symmetry engineering. Three ingredients must be optimized simultaneously: robust local or molecular spin moments, compensated magnetic order without a spin-degeneracy-enforcing symmetry, and orbital anisotropy capable of producing a momentum-dependent spin splitting. Nonalternant nanographenes, triangulene derivatives, heterotriangulene units, and symmetry-lowered radical monomers are therefore especially promising building blocks. Chemistry plays a dual role by stabilizing the open-shell magnetic state and by selecting the real-space symmetry operation that generates the altermagnetic momentum-space texture.

From a broader perspective, COF altermagnets would occupy a distinctive position among magnetic quantum materials. Their light-element composition may help isolate the non-relativistic exchange-driven origin of altermagnetic spin splitting more cleanly than in heavy-element compounds, where spin-orbit coupling can obscure the underlying mechanism. At the same time, the chemical flexibility of COFs offers routes to tune exchange coupling, band gaps, and spin splitting through linker choice, monomer symmetry, strain, substrate engineering, and external fields.

Several challenges remain before COF altermagnetism can become an experimentally established materials platform. These include the synthesis of sufficiently crystalline open-shell COFs, stabilization of long-range magnetic order, control of stacking and substrate-induced symmetry breaking, and direct detection of spin-split bands. A particularly important experimental issue will be to distinguish non-relativistic altermagnetic spin splitting from substrate-induced Rashba effects, magnetic disorder, or unintended structural symmetry breaking. Nevertheless, the interplay of reticular chemistry, molecular topology, and altermagnetic symmetry design makes COFs a promising platform for metal-free altermagnetism.

\section{Synthetic routes to altermagnetic materials}
\label{sec:synthesis}

The disparity between predicted and experimentally validated altermagnets is, in large part, a synthesis problem. High-throughput searches return candidates without regard to whether they can be prepared as phase-pure, stoichiometric, single-domain specimens of sufficient size for neutron diffraction, ARPES, or transport measurement. We next survey the growth methods that have delivered the confirmed altermagnets and identify the strategies applicable to the chemically more demanding candidates.

\subsection{Flux growth}

Metal-flux growth has been the most productive route to altermagnetic single crystals, because it operates below the melting point of the target phase, suppresses volatilization, and yields well-faceted crystals with low defect densities. For $\alpha$-MnTe, Sb-flux and Te self-flux routes have been developed. In the Sb-flux approach, Mn, Te, and a large excess of Sb are sealed under vacuum, heated to $\approx$1050$^{\circ}$C, held to homogenize, then cooled slowly to nearly 700$^{\circ}$C before the flux is removed by centrifugation, yielding crystals of several millimetres~\cite{wan2026observation}. Centimetre-scale crystals have been obtained by optimising the growth window against the binary phase diagram~\cite{wu2025bulk}. A tailored Te self-flux route has more recently produced crystals of exceptional quality. These are free of detectable diffuse scattering or mosaicity, and with stoichiometric Mn:Te composition. These samples reveal an intrinsic Anderson-insulating state in contrast to the metallic behaviour of Te-deficient specimens~\cite{huynh2026intrinsic}. Notably, the hallmark altermagnetic signatures persist in this intrinsic limit but with substantially reduced magnitude, demonstrating that defect chemistry quantitatively modifies the observables used to identify altermagnetism.

For CrSb, both Sn-flux and self-flux routes have been established, yielding hexagonal-rod and plate-like samples respectively, with residual resistivity ratios of order 10 in optimised self-flux crystals~\cite{paul2026thermodynamic,terashima2026altermagnetic}. The choice between them is dictated by the required crystal orientation. Sn-flux crystals grow with the long axis along $c$, while chemical vapour transport yields plates with the largest facet parallel to (0001). This is a practical consideration when a specific surface must be exposed for ARPES.

Flux growth is particularly well suited to the chemically demanding candidates, because the flux can be chosen to lower the growth temperature below the decomposition or volatilisation threshold of the target. The principal caveat is flux incorporation. Residual solvent atoms can act as unintended dopants, and this must be excluded by microanalysis.

\subsection{Chemical vapour transport}

Chemical vapour transport (CVT) complements flux growth for phases that are poorly soluble in available metal fluxes or that decompose peritectically. MnTe has been grown by CVT using iodine as the transport agent at $\approx$700$^{\circ}$C, with the important caveat that excessive iodine concentration suppresses growth by stabilising competing Mn–I species~\cite{de1991crystal}. CVT typically yields smaller crystals than flux growth but often with superior surface quality, making it well matched to surface-sensitive spectroscopies~\cite{liu2025strain}.

For candidate altermagnets containing volatile or easily reduced constituents, which includes many of the predicted fluorides and several chalcogenides, CVT in sealed ampoules offers the advantage of a closed system.

\subsection{Thin-film epitaxy and metastable phases}

A significant fraction of predicted altermagnets are metastable or exist only in structural polymorphs that are not thermodynamically accessible in bulk. Epitaxial growth on lattice-matched substrates provides access to these phases by using the substrate to select the polymorph and to impose the symmetry-lowering strain required for altermagnetic order. Tetragonal CuMnAs, stabilized only as a film, is a prominent example~\cite{wei2024crystal}.

Strain engineering is a useful knob which can be used to tune the altermagnetism. The RuO$_2$ case, discussed in Section 11, is instructive. Stoichiometric bulk RuO$_2$ is now understood to be non-magnetic, but the energy separation from the antiferromagnetic state is small enough that compressive strain can stabilize magnetic order. This means that substrate choice is crucial, and that films and bulk crystals of nominally the same compound may belong to different magnetic classes.

\subsection{Domain control and detwinning}

Altermagnetic crystals are generically multi-domain. Since the altermagnetic response is anisotropic and changes sign between domains related by the sublattice-connecting rotation, domain averaging suppresses the signal being sought. Single-domain regions of MnTe produce clear twofold anisotropy in constant-energy ARPES maps, whereas multi-domain regions recover apparent sixfold symmetry~\cite{krempasky2024altermagnetic}.

Approaches to domain control include growth selection (crystals nucleating from a single seed under a shallow thermal gradient tend toward fewer domains), field-cooling through $T_N$ where a magnetoelastic coupling permits, uniaxial strain applied through a piezoelectric stack or a mismatched substrate, and for the piezomagnetic altermagnets direct mechanical selection of the N\'eel vector orientation. Systematic development of detwinning protocols would, in our assessment, have a substantial impact on the rate of experimental confirmation of altermagnets. \\

\section{Experimental characterization}
\label{sec:experimental_characterization}

\begin{table*}[t]
\centering
\caption{Summary of key experimental techniques for characterizing altermagnets.
The \checkmark~indicates that the technique provides the
corresponding capability. Here ARPES stands for angle resolved photoemission spectroscopy, SARPES is spin-resolved ARPES, and AHE is the anomalous Hall effect.}
\label{tab:techniques}
\smallskip
\renewcommand{\arraystretch}{1.4}
\small
\begin{tabular}{@{} l p{4.0cm} c c c l @{}}
\hline
\hline
\textbf{Technique} & \textbf{Measures} & \textbf{Bulk?} & \textbf{$k$-resolved?} & \textbf{Spin-resolved?} & \textbf{Access} \\
\hline
\hline
Neutron diffraction
  & Magnetic structure, symmetry
  & \checkmark & $\times$ & $\times$ & Facility \\
ARPES/SARPES
  & Band splitting, spin texture
  & $\times$ & \checkmark & \checkmark & Synchrotron \\
AHE/Transport
  & Berry curvature, spin currents
  & \checkmark & $\times$ & $\times$ & Lab \\
MOKE
  & Kerr rotation screening
  & $\times$ & $\times$ & $\times$ & Lab \\
\hline
\hline
\end{tabular}
\end{table*}

Although altermagnetism was initially established through symmetry analysis and first-principles calculations, experimental validation has been crucial for confirming its existence and unique properties. In this section, we discuss the key experimental techniques that have been used to investigate altermagnets and their characteristic features (Table~\ref{tab:techniques}).

\subsection{Neutron diffraction}

Neutron diffraction is an important first step in characterizing any candidate altermagnet. Neutrons carry a magnetic moment, so they scatter from the ordered spin density in a crystal, providing direct access to the magnetic structure. This includes the arrangement of spins, the propagation vector, and the sublattice assignment~\cite{cheetham1992synchrotron}. As we are familiar by now, these properties together determine whether a material is a conventional antiferromagnet or an altermagnet.

The magnetic Bragg peaks in a neutron diffraction pattern encode the periodicity and symmetry of the ordered spin arrangement. For a commensurate antiferromagnet with propagation vector $k = 0$ (which is the most common case for altermagnets), the magnetic reflections coincide with the nuclear reflections, and the magnetic contribution must be extracted by comparing data above and below the N\'eel temperature. Rietveld refinement of the magnetic structure factor yields the ordered moment direction and magnitude on each sublattice, from which one can assign the magnetic space group and, via the symmetry checklist, determine whether the sublattice-connecting operation is a rotation (altermagnet) or a translation/inversion (conventional antiferromagnet). The MAGNDATA database of magnetic structures on the Bilbao Crystallographic Server provides a systematic resource for this analysis~\cite{gallego2016magndata}.

Single-crystal neutron diffraction is strongly preferred over powder diffraction for altermagnetic candidates, because the anisotropic magnetic form factor and the directional dependence of the spin arrangement are more precisely determined in single crystals. Polarized neutron diffraction, in which the incident beam has a well-defined spin state, provides additional sensitivity to the magnetic structure by separating nuclear and magnetic scattering contributions. For example, detailed structural studies of epitaxial $\alpha$-MnTe films, combining X-ray diffraction, transmission electron microscopy, and polarized neutron reflectometry, have revealed how thermal strain, misfit dislocations, and surface oxidation affect the structure, highlighting the sensitivity of the altermagnetic state to structural imperfections~\cite{bey2025interface}.

A practical limitation is that neutron diffraction reveals the symmetry of the magnetic order but does not directly measure the electronic spin splitting, which is the defining observable of altermagnetism. A material can have the correct magnetic structure for altermagnetism (rotation-connected sublattices) but exhibit a vanishingly small spin splitting if the crystal-field anisotropy is negligible. Conversely, a large spin splitting in band-structure calculations provides strong motivation for neutron experiments to confirm the assumed magnetic ground state. The two techniques are therefore complementary. While neutron diffraction establishes the magnetic structure, the electronic structure probes, which we discuss next, confirm the altermagnetic splitting.

\begin{figure}[t]
\centerline{\includegraphics[scale=0.55]{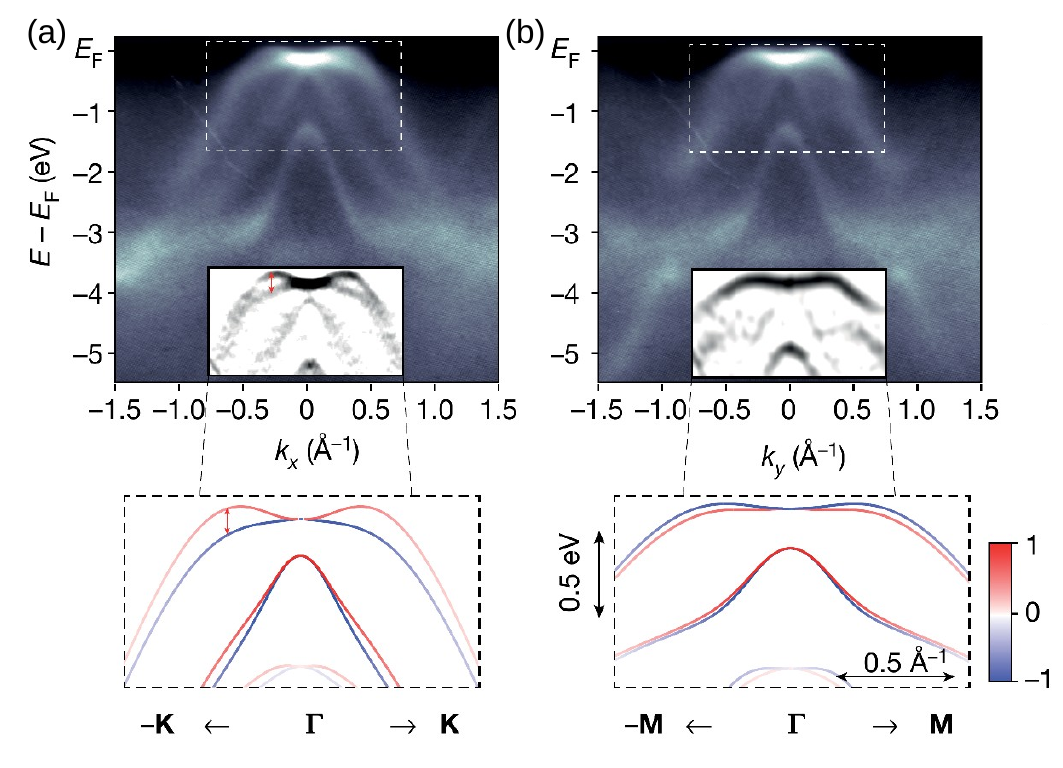}}
\caption{\textbf{Altermagnetic splitting in MnTe from ARPES experiments.} Spin-split bands along (a) $\Gamma$-K and (b) $\Gamma$-M directions. The insets are curvature plots. The bottom panels are first-principles band structures with the red and blue colors corresponding to opposite $z$-components of the spin. Reproduced with permission from Krempasky \textit{et al.}~\cite{krempasky2024altermagnetic}}
\label{fig_arpes_exp}
\end{figure}

\subsection{Angle resolved photoemission spectroscopy}

ARPES is the most direct probe of the altermagnetic band structure. By measuring the kinetic energy and emission angle of photoelectrons ejected from a crystal surface, ARPES maps the occupied electronic bands $E(\mathbf{k})$ with momentum resolution~\cite{zhang2022angle,king2020angle}. When combined with spin detection (spin-resolved ARPES, or SARPES), it can measure the spin polarization of each band at each $k$-point. This is precisely the observable that distinguishes an altermagnet from a conventional antiferromagnet.

The expected ARPES signature of a $d$-wave altermagnet is distinctive. Along certain high-symmetry directions (for example, $\Gamma\rightarrow \mathrm{X}$ in a tetragonal system), the bands should appear as two spin-split copies separated by an energy that can reach hundreds of meV in favourable cases. Along the nodal directions, such as $\Gamma\rightarrow \mathrm{M}$, where $k_x = k_y$, the splitting should vanish and the bands should be spin-degenerate. This momentum-dependent pattern, with its unique splitting in one direction and degeneracy in another, is the spectroscopic fingerprint of altermagnetism.

Such a landmark experimental confirmation was achieved by Krempasky \textit{et al.}, who used synchrotron-based ARPES and spin-resolved ARPES on MnTe to demonstrate the altermagnetic lifting of Kramers spin degeneracy~\cite{krempasky2024altermagnetic}. Their central results are presented in Fig.~\ref{fig_arpes_exp}. The measured band splitting and its nodal structure matched the predicted $g$-wave symmetry of the NiAs-type structure. Concurrently, Lee \textit{et al.} observed broken Kramers degeneracy in MnTe thin films~\cite{lee2024broken}, and Osumi \textit{et al.} reported a giant band splitting in single-crystal MnTe~\cite{osumi2024observation}. As we have seen in Section~\ref{sec:material_classes}, CrSb, with its much higher N\'eel temperature of nearly 705 K, has emerged as a particularly attractive system. Reimers \textit{et al.} observed altermagnetic band splitting of ~0.6 eV in epitaxial CrSb thin films using soft X-ray ARPES~\cite{reimers2024direct}. Subsequently, Yang \textit{et al.} mapped the full three-dimensional altermagnetic spin splitting up to nearly 1.0 eV near the Fermi level in CrSb single crystals, unambiguously confirming bulk-type $g$-wave altermagnetism through systematic $k_z$-dependent mapping and spin-resolved measurements~\cite{yang2025three}. The ARPES evidence for altermagnetic splitting in RuO$_2$ has been reported by Fedchenko \textit{et al.}~\cite{fedchenko2024observation}, though the nature of the magnetic ground state in this material remains strongly debated~\cite{liu2024absence,choi2026exploring}.

Several practical challenges are worth mentioning here. First, ARPES is a surface-sensitive technique (probing depth of order 5–10 \AA{}), so the surface must be representative of the bulk magnetic structure. Any surface reconstructions or termination effects can modify or destroy the altermagnetic splitting. Second, altermagnetic materials are typically multidomain. Therefore, regions of the crystal in which the N\'eel vector points in different directions will contribute incoherently to the ARPES signal, potentially washing out the spin splitting. Krempasky \textit{et al.} showed that single-domain regions produce clear twofold anisotropy in constant-energy ARPES maps, while multi-domain regions recover sixfold symmetry~\cite{krempasky2024altermagnetic}. Detwinning the sample, by applying uniaxial strain or by exploiting a magnetoelastic coupling, is often necessary to observe a clean signal. Third, the spin-resolved mode of ARPES has significantly lower count rates than the spin-integrated mode, making high-quality SARPES measurements very demanding in terms of both beam time and sample quality.

Soft X-ray ARPES offers a partial solution to the surface-sensitivity problem by increasing the probing depth to several nanometers, at the cost of reduced momentum resolution. Time-resolved ARPES can additionally probe the dynamics of the altermagnetic state on femtosecond timescales, providing access to the ultrafast spin-lattice coupling that is expected to be relevant for THz applications.

\begin{figure}[t]
\centerline{\includegraphics[scale=0.55]{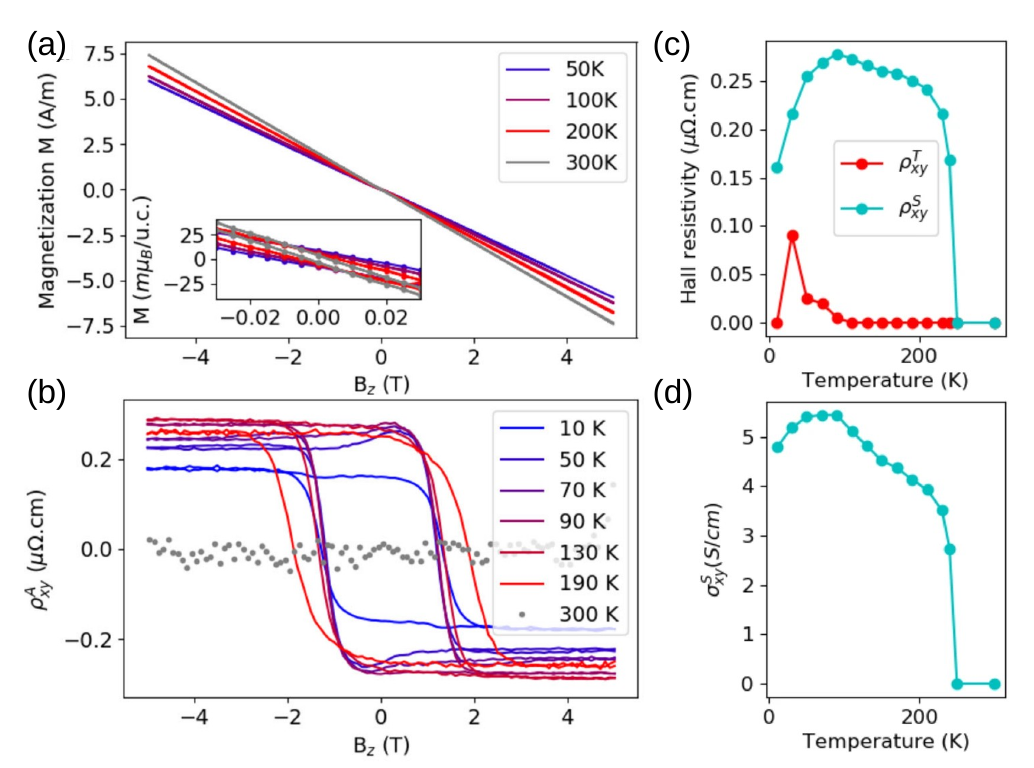}}
\caption{\textbf{Anomalous Hall effect in Mn$_5$Si$_3$.} (a) Magnetization with varying magnetic field along the [0001] direction at different temperatures. (b) Anomalous Hall resistivity as a function of the applied magnetic field at different temperatures. (c) The decomposition of the spontaneous (zero-field) anomalous Hall resistivity as a function of temperature. (d) Anomalous Hall conductivity from $d$-wave altermagnetism as a function of temperature. Reproduced with permission from Reichlova \textit{et al.}~\cite{reichlova2024observation}}
\label{fig_ahe_exp}
\end{figure}

\subsection{Transport measurements}

Transport measurements provide bulk-sensitive probes of the altermagnetic state and are often more accessible experimentally than ARPES, particularly for materials that are difficult to cleave or grow as large single crystals.

The anomalous Hall effect (AHE) is perhaps the most striking transport signature. In a ferromagnet, the AHE arises from the spin-orbit-coupled Berry curvature of the spin-split bands and produces a transverse voltage proportional to the magnetization~\cite{nagaosa2010anomalous}. In a conventional antiferromagnet with ${PT}$ symmetry, the Berry curvature is even under this symmetry and the Hall effect vanishes. In an altermagnet, the ${PT}$ symmetry is broken, and a nonzero anomalous Hall conductivity is permitted even though the net magnetization is zero~\cite{vsmejkal2020crystal}. Reichlova \textit{et al.} provided the first experimental observation of a spontaneous AHE in the $d$-wave altermagnet candidate Mn$_5$Si$_3$, reporting an anomalous Hall conductivity of 5-20 S/cm in epitaxial thin films with vanishingly small net magnetization~\cite{reichlova2024observation}. Their main results are summarized in Fig.~\ref{fig_ahe_exp}. The anisotropy of the AHE signal with respect to the N\'eel vector orientation was subsequently characterized by Leiviska \textit{et al.}~\cite{leiviska2024anisotropy}.

Similar to the AHE, another anomalous transport phenomenon that can arise in altermagnets is the anomalous Nernst effect (ANE). The ANE, often regarded as a thermo-electric counterpart of AHE, is a phenomena in which transverse charge current is generated by a temperature gradient in the absence of an external magnetic field. Traditionally, ANE has been observed in ferromagnets, where it arises from the combined effects of spontaneous magnetization and spin-orbit coupling. In altermagnets, despite the absence of a net magnetization, a finite ANE can occur due to broken time-reversal symmetry and the presence of finite Berry curvature. In contrast to the AHE , which is determined by the integrated Berry curvature of all occupied electronic states, the ANE is governed by the interplay between the Berry curvature and the electronic density of states near the Fermi level~\cite{xiao2006berry,li2025large,xiao2010berry}. Significant ANE has been theoretically predicted in several altermagnets such as RuO$_2$~\cite{zhou2024crystal}, CrSb~\cite{li2025large}, V$_2$Te$_2$O~\cite{liu2025anomalous}, KV$_2$Se$_2$O~\cite{yang2025magnetic} and strained MnTe~\cite{chen2026strain}. Beyond theoretical predictions, experimental observations have also been reported in a handful of altermagnets~\cite{badura2025observation,han2025nonvolatile,li2025large}.

The spin-splitter effect is unique to altermagnets and has no analog in ferromagnets or antiferromagnets. When a charge current flows through an altermagnet, the anisotropic spin splitting generates a transverse spin current. As a result, spin-up electrons are deflected in one direction and spin-down electrons in the opposite direction, even in the absence of spin-orbit coupling. This effect has been predicted theoretically~\cite{gonzalez2021efficient} and its experimental detection would constitute strong evidence for the altermagnetic state. The magnitude of the spin-splitter torque is proportional to the altermagnetic splitting and is expected to be largest in $d$-wave altermagnets with large Fermi-surface anisotropy.

Magnetoresistance measurements provide another route to detecting altermagnetism. Giant magnetoresistance (GMR) and tunnelling magnetoresistance (TMR) in heterostructures containing an altermagnetic layer are predicted to be nonzero because the spin-dependent density of states at the Fermi level differs between the two spin channels~\cite{vsmejkal2022giant}. Anisotropic magnetoresistance (AMR), which relates the dependence of the longitudinal resistivity on the direction of the N\'eel vector, has been measured in MnTe by Gonzalez Betancourt \textit{et al.}, providing a simpler experimental observable than the Hall effect~\cite{gonzalez2024anisotropic}. The AMR in an altermagnet should exhibit the symmetry of the wave type. For example, $d$-wave altermagnets show a $\cos(2\phi)$ angular dependence, where $\phi$ is the angle between the current and the N\'eel vector. Such signatures can provide indirect transport evidence of altermagnetism.

\subsection{Optical and magneto-optical probes}

Optical techniques offer non-contact, spatially resolved probes of the altermagnetic state, with the additional advantage of being potentially applicable to thin films, heterostructures, and devices.

X-ray magnetic linear dichroism (XMLD) measures the difference in X-ray absorption between linearly polarized light aligned parallel and perpendicular to the antiferromagnetic axis. Unlike X-ray magnetic circular dichroism (XMCD), which is proportional to the net magnetization and therefore vanishes in compensated magnets, XMLD is sensitive to the axis of the ordered moment and can detect the N\'eel vector direction in both antiferromagnets and altermagnets~\cite{vaz2025x}. Notably, Hariki \textit{et al.} demonstrated that XMCD is in fact permitted in altermagnetic MnTe due to the broken time-reversal symmetry, providing an additional contrast mechanism beyond XMLD~\cite{hariki2024x}. By combining XMLD and XMCD with photoemission electron microscopy, Amin \textit{et al.} achieved the first nanoscale imaging of altermagnetic domain textures in MnTe, resolving 100-nanometre-scale vortex–antivortex pairs and mapping the local N\'eel vector orientation with six-colour vector maps~\cite{amin2024nanoscale}.

The magneto-optical Kerr effect (MOKE) is traditionally associated with ferromagnets, where it measures the rotation of the polarization plane of reflected light due to the magnetization. In a conventional antiferromagnet, the Kerr rotation vanishes by symmetry. In an altermagnet, however, the broken ${PT}$ symmetry permits a nonzero Kerr effect even at zero net magnetization~\cite{zhou2021crystal}. If confirmed to be broadly applicable, MOKE would provide a convenient laboratory-based probe for screening altermagnetic candidates, requiring only a polished surface and a laser, in contrast to the synchrotron-based XMLD-based techniques.

Optical second-harmonic generation (SHG) is sensitive to the breaking of spatial symmetries and can detect the magnetic point group of a material~\cite{fiebig2005second}. Since altermagnets break different symmetries than conventional antiferromagnets, their SHG response is expected to differ, providing a symmetry-specific optical fingerprint.

\begin{figure*}
\centerline{\includegraphics[scale=0.6]{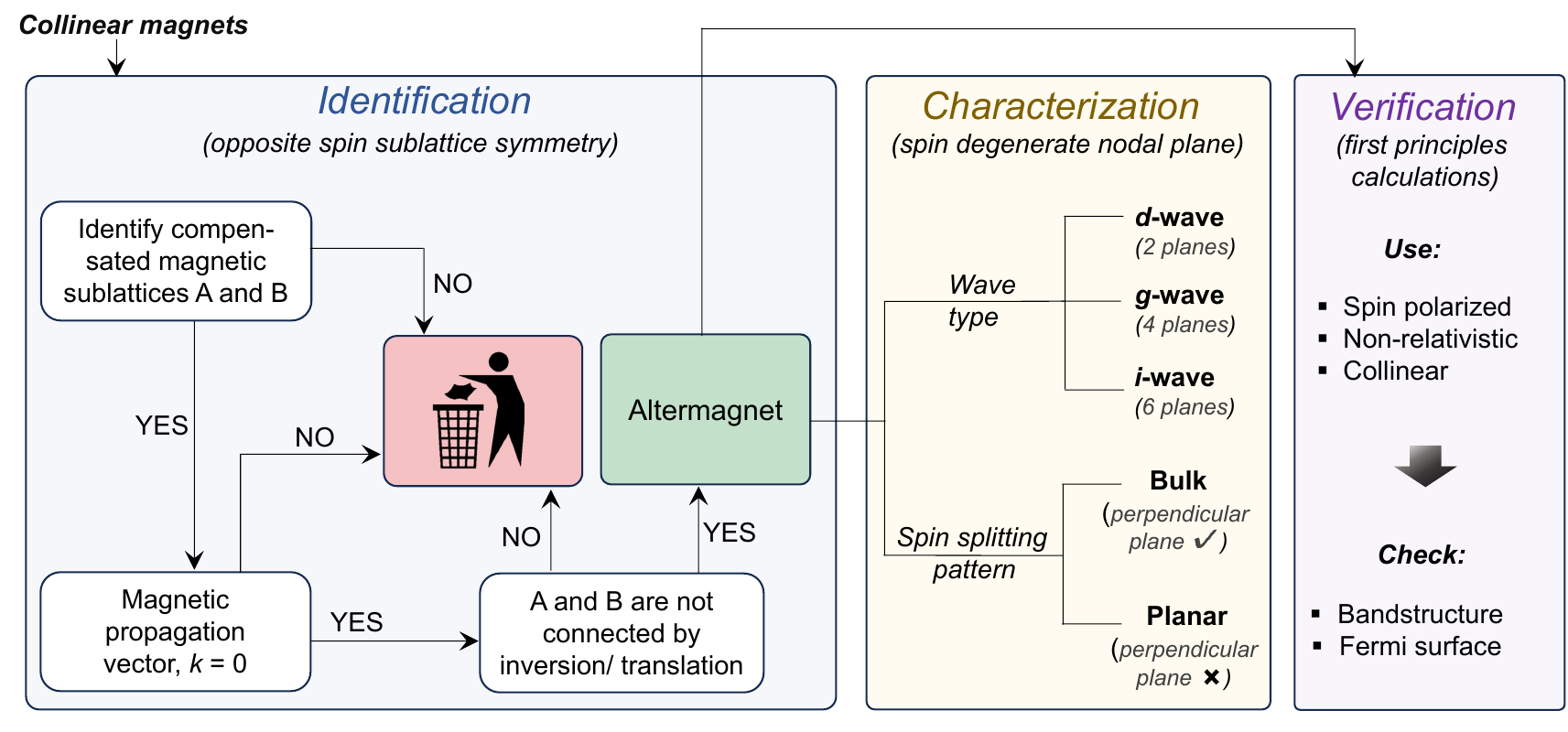}}
\caption{\textbf{Workflow for the identification and characterization of altermagnetic candidates.} The first stage (sky-blue shaded region) employs symmetry-based criteria to identify potential altermagnetic materials. The second stage (light-yellow shaded region) characterizes the resulting altermagnetic states by determining their spin-splitting patterns and corresponding wave symmetries. Finally, first-principles calculations (light-purple shaded region) are performed to validate the symmetry analysis and spin-splitting characteristics through spin-polarized electronic-structure calculations within the non-relativistic and collinear magnetic framework.}
\label{identification}
\end{figure*}

\section{Theoretical material searches}
\label{sec:theoretical_search}

After having looked at the experimental approaches to characterize altermagnets, in this section, we present a systematic framework, combining symmetry analysis and first-principles calculations, for identifying whether a synthesized or computationally-predicted collinear antiferromagnet exhibits altermagnetic order and further deciphering its wave type and spin splitting pattern. The procedure is schematically illustrated in Fig.~\ref{identification} and consists of three stages: identification, characterization, and verification. The first two stages rely on symmetry arguments, while the final stage validates these symmetry-based predictions through first-principles electronic structure calculations.

The identification procedure begins by partitioning the magnetic ions into spin-up ($A$) and spin-down ($B$) sublattices while ensuring complete magnetic compensation and vanishing net magnetization. Next step involves determining the magnetic propagation vector $\bm{k}$ from neutron diffraction experiments or computational ground-state calculations. In most altermagnetic systems, $\bm{k} = 0$, indicating that the magnetic and crystallographic unit cells coincide, whereas nonzero propagation vectors typically correspond to conventional antiferromagnetic order. The key aspect of the identification part is addressed in following step, where the symmetry operations of the paramagnetic space group connecting the $A$ and $B$ sublattices are identified. If the sublattices are related solely by lattice translation or inversion symmetry, the material remains a conventional antiferromagnet with preserved spin degeneracy across the entire Brillouin zone, enforced by translation or combined ${PT}$ symmetry. In contrast, if the sublattices are connected through proper or improper rotational symmetries that cannot be reduced to translation or inversion alone, the material is classified as an altermagnet. In crystal structures, such rotational relations can be inferred directly from the relative orientations of the coordination polyhedra surrounding the magnetic ions. It is important to note that for two-dimensional materials however the symmetry criteria are more stringent as discussed in Section~\ref{sec:2D_altermagnets}.

Once altermagnetism has been established, the next step is to characterize the form of the spin splitting. This can be achieved by identifying the number of spin-degenerate nodal planes passing through the $\Gamma$-point, which is determined by the corresponding spin Laue group ($R_s^{\mathrm{III}}$). The number of nodal planes defines the spin-group integer and provides a simple classification of the altermagnetic state: two, four, and six nodal planes correspond to $d$-, $g$-, and $i$-wave spin-splitting symmetries, respectively. A further distinction can be made between planar and bulk spin-momentum locking based on the dependence of the spin splitting on the out-of-plane momentum component $k_z$, which is governed by the presence or absence of perpendicular spin-degenerate nodal planes. When such perpendicular nodal planes are present, they induce a sign change in the spin-splitting energy upon crossing them, resulting in a $k_z$-dependent, bulk-type spin splitting. In contrast, when these nodal planes are absent, the spin splitting retains the same sign across $k_z$, giving rise to planar spin–momentum locking. Importantly, planar spin–momentum locking can occur in both two-dimensional and three-dimensional systems, whereas bulk spin–momentum locking occurs only in three-dimensional crystals.

In the final part of the material screening, the symmetry-based predictions must be validated through spin-polarized electronic structure calculations performed in the absence of spin–orbit coupling. Conventional antiferromagnets retain globally spin-degenerate electronic bands, whereas altermagnets exhibit characteristic momentum-dependent spin splitting with nodal structures consistent with the predicted wave symmetry. A common limitation of usual band structure analysis is that the conventional high-symmetry $k$-paths may often coincide with nodal planes, thereby concealing the altermagnetic spin splitting. It is therefore essential to probe additional generic $k$-points within the Brillouin zone, which must be examined to clearly resolve the altermagnetic splitting. Alternatively, calculations of the spin-resolved Fermi surface can provide a more direct visualization and confirmation of the predicted altermagnetic splitting patterns.

Recent advances in symmetry-based identification have enabled the development of high-throughput screening for discovering altermagnetic materials. By combining crystallographic databases, magnetic symmetry analysis, and first-principles calculations, several large-scale screening studies have identified numerous candidate altermagnets in both three-dimensional~\cite{guo2023spin,bhattarai2025high,wan2025high, sufyan2026high} and two-dimensional~\cite{sodequist2024two, zeng2024description, wang2025pentagonal,haddadi2026exploring,bose2026symmetry} material families.
{\v{S}}mejkal \textit{et al.} have introduced a symmetry-based classification tool that distinguishes altermagnetic from non-altermagnetic three-dimensional materials using their crystal and magnetic symmetries~\cite{smolyanyuk2024tool}. Building on this framework, Bose \textit{et al.} have developed a fully automated AiiDA-based workflow that combines this symmetry-based classification with first-principles DFT+$U$ calculations to identify ground-state altermagnetism in two-dimensional materials~\cite{bose2026symmetry}. Such automated workflows provide an efficient approach for verifying altermagnetism in predicted or synthesized materials and enable rapid screening of candidate materials in high-throughput searches.

\section{Effects of structural and external perturbations on altermagnetism}
\label{sec:effect_external_perturbation}
Altermagnetism is a symmetry-dependent phenomenon, and its spin-splitting is anisotropic in the Brillouin zone, appearing only along specific momentum-space directions. This raises questions about the robustness and preservation of altermagnetism and its properties under external perturbations and structural modifications. Factors such as chemical doping, defects, strain, surface orientation, and substrate interfacing can all affect crystal symmetry and electronic structure. Addressing the influence of these effects on altermagnets is conceptually important and also essential for assessing the detectability of altermagnetism and its practical viability for various applications. Multiple recent theoretical and experimental studies have investigated the response of altermagnetism to these perturbations.

\subsection{Effect of surface termination and interfaces}
Since altermagnetic spin splitting arises from the interplay of crystal symmetry and magnetic order, truncating a crystal to form a surface or interface can locally break or modify the altermagnetic symmetry conditions. Moreover, surfaces and interfaces are the regions most readily accessible to experimental probes, making them a natural focus for detecting and controlling altermagnetism. Research in this direction on altermagnets is therefore important. Sattigeri \textit{et al.} investigated the surface electronic structure of representative tetragonal, orthorhombic, and hexagonal altermagnets using first-principles calculations~\cite{sattigeri2023altermagnetic}. By projecting the three-dimensional bulk Brillouin zone onto different surface Brillouin zones, they showed that two of the three principal surface orientations lose their altermagnetic character because opposite-spin-split momentum points overlap upon projection, resulting in a conventional spin-degenerate antiferromagnetic surface state. Only one surface orientation preserves the bulk altermagnetic spin splitting, with the retained orientation determined by the magnetic space group, magnetic ordering, and the Wyckoff positions of the magnetic atoms. Furthermore, they demonstrated that applying an electric field perpendicular to a surface that has lost its altermagnetic character can restore spin splitting by breaking the relevant inversion symmetry, providing a practical route to realize altermagnetism on otherwise unfavorable surfaces. Similar conclusions were independently reported by Jha \textit{et al.} in their first-principles study of CrTe, FeSb$_2$, and MnO$_2$. These studies highlight that the manifestation of altermagnetism at surfaces is highly sensitive to surface orientation and symmetry, while also demonstrating that external symmetry-breaking perturbations can recover the characteristic spin splitting.

Lange \textit{et al.} demonstrated an intriguing route to realizing altermagnetism by showing that altermagnetic states can emerge at the surfaces of materials that are conventionally antiferromagnetic in the bulk~\cite{lange2026emergent}. They established a general, symmetry-driven framework for engineering two-dimensional altermagnetism through the controlled termination of bulk collinear antiferromagnets. By introducing the concepts of surface spin groups and surface spin Laue groups, they developed a systematic classification scheme to identify surface terminations capable of hosting altermagnetic symmetries. These findings reveal that symmetry breaking at surfaces can fundamentally alter the magnetic character of a material, enabling altermagnetic spin splitting even when the bulk lacks altermagnetic order. Such theoretical advances provide valuable insights for experimental investigations of altermagnets, particularly for detecting their characteristic spin-split electronic structure via ARPES and STM. With this framework, they could resolve the apparently conflicting experimental observations on KV$_2$Se$_2$O, which had been identified as a $d$-wave altermagnet by the surface-sensitive ARPES measurements reported in Ref.~\cite{jiang2025metallic}, whereas neutron diffraction experimental results indicate a conventional G-type AFM order in the bulk~\cite{sun2025antiferromagnetic}. Hodt \textit{et al.} showed that itinerant electrons can generate a real-space magnetization localized near edges and interfaces in altermagnets~\cite{hodt2024interface}.

These predictions are supported experimentally. Nanoscale quantum sensing of epitaxial MnTe films directly imaged an evanescent, interface-localized magnetization that correlates with the AHE, confirming its surface rather than bulk origin~\cite{zhou2026imaging}. Similarly, preliminary XMCD measurements on MnF$_2$(110) revealed a sign reversal under opposite field cooling at grazing incidence~\cite{curbelo2026probing}. On the transport side, ARPES on MnTe films grown on InP(111) revealed distinct surface states crossing the Fermi level alongside a large bulk spin splitting, with accompanying transport measurements showing that the AHE arises specifically from the Berry curvature of these surface states~\cite{zhou2026surface}. In another work, MnTe integrated on Si(111) similarly exhibited a robust, hysteretic AHE arising from uncompensated Berry curvature generated by symmetry breaking at the thin-film surface~\cite{sarkar2026anomalous}. Complementing these experimental observations, Ghorbani \textit{et al.} showed through combined molecular beam epitaxy and DFT calculations that interfacial coupling with a nonmagnetic substrate provides an extrinsic route to perturb the compensated magnetic order in altermagnetic MnTe~\cite{ghorbani2026transport}. They grew epitaxial MnTe(0001) thin films on InP(111) substrates using molecular beam epitaxy and observed magnetotransport signatures indicative of a finite net magnetization in the thin films. Their findings suggest that the local structural and chemical environment at the interface can modify the magnetic properties of an altermagnet, with such effects expected to be more pronounced in ultrathin films. Collectively, these studies demonstrate that surface termination and interfaces play a major role in determining altermagnetic spin splitting, magnetization, and anomalous Hall response, even though these effects remain fundamentally tied to the bulk magnetic symmetry.

\subsection{Effect of defects and doping}
Defects inevitably occur in materials and doping has been widely employed as a method to engineer material properties. The effects of doping have been analyzed in detail using density functional theory calculations, symmetry analysis, and model studies in Ref.~\cite{devaraj2026unlocking}, focusing on the prototypical altermagnet MnTe. The authors identified the general conditions under which altermagnetism persists or deviates under non-magnetic doping. They show that even when ideal symmetry conditions are broken, altermagnets can exhibit momentum-dependent spin splitting while deviating from ideal altermagnetic characteristics, yet still retaining essential altermagnetic features. The authors term these ``quasi-altermagnets.'' They further demonstrate that doping can be used as an effective route not only to tune the value of the intrinsic AHC, but also to induce new AHC components absent in the pristine compound.

Maiani \textit{et al.} investigated how nonmagnetic impurities modify altermagnetic superconductors using a microscopic theoretical model~\cite{maiani2025impurity}. They showed that such impurities create spin-polarized subgap states whose structure reflects the underlying altermagnetic order, making them a useful local probe of altermagnetism and a potential handle for STM-based detection and device control. In another theoretical study, Burkard \textit{et al.} showed that at the single-defect level, vacancies act as local, symmetry-sensitive probes rather than merely disruptive perturbations~\cite{burkard2026anisotropic}. An isolated vacancy induces anisotropic, real-space magnetization textures whose structure directly encodes the altermagnetic order parameter, offering a route to detect altermagnetism using local probe techniques such as STM. Wakabayashi \textit{et al.}~\cite{wakabayashi2026vacancy} demonstrated that structural reconstruction driven by vacancies offers a general microscopic pathway to $d$-wave altermagnetism in two-dimensional systems. Starting from trigonal VX$_2$, vacancy-driven reconstruction transforms the lattice into V$_2$X$_2$ (X = S, Se) monolayers with an inverse Lieb geometry, where two inequivalent edge vanadium sites carry opposite exchange fields. Despite individually breaking time-reversal and combined inversion-time-reversal (PT) symmetries, these sites are connected by C$_4$ rotational symmetry, resulting in a net magnetization of zero.

These studies illustrate that defects and doping can act as constructive design principles for realizing or tuning altermagnetism.

\subsection{Effects of strain and pressure}
Similar to defects and doping, strain is another external stimulus that may be used to tune and engineer materials. Strain may arise due to substrate mismatch, thin-film growth, or deliberate loading. While retaining the same chemical composition, strain induces changes in bond lengths, bond angles, and octahedral distortions, and its coupling with crystal symmetry, which is an important determining factor for altermagnetism, makes the effect of strain on altermagnetism an interesting topic of research.

In both two- and three-dimensional materials that are already altermagnetic, strain has been shown to tune the magnitude, sign, and even the momentum-space texture of the nonrelativistic spin splitting~\cite{khodas2025tuning,huang2024emerging,leon2025hybrid,lin2026tailoring}. Strain also plays a role in determining the relative ground-state energies of competing magnetic configurations, and can make the altermagnetic phase the energetically favored one, where the antiferromagnetic phase is otherwise more stable. In ReO$_2$~\cite{chakraborty2024strain} and Ca$_3$Ru$_2$O$_7$~\cite{leon2025strain}, the altermagnetic phase becomes more stable than the antiferromagnetic phase once sufficient strain is applied. Using symmetry analysis, a minimal model, and DFT calculations, Karetta \textit{et al.} showed that shear strain lowers the spin symmetry of CrSb, driving a transition from $g$-wave to $d$-wave altermagnetism or an uncompensated phase, with an accompanying spin-splitter effect~\cite{karetta2025strain}. In certain cases strain induces qualitatively new effects beyond magnitude tuning, such as valley polarization, piezomagnetism, and topological phase transitions~\cite{xun2025stacking,zhang2025multiple,subhan2026large,tirth2026multi,tirth2026multi}.

Hydrostatic pressure produces qualitatively similar effects, since both strain and pressure modulate lattice parameters and bond lengths. As a result, pressure effectively tunes altermagnetic properties in several classes of altermagnetic materials~\cite{devaraj2024interplay,bhandari2026effect}. It can also induce phase transition from an antiferromagnetic phase to an altermagnetic phase in certain cases~\cite{fan2025high,li2025layered,zhao2026pressure,zhao2026structural}.

Overall, surfaces, interfaces, defects, dopants, strain, and pressure all act as useful tools for controlling and enhancing altermagnetic properties. Even though altermagnetism comes from the symmetry of the bulk crystal, these external tuning factors can weaken or strengthen it or even create it in materials where it did not exist before. This means such effects must be considered carefully when interpreting experiments on altermagnets. Furthermore, they also open up practical ways to design and tune altermagnetic materials for potential future devices. \\
\begin{figure*}[t]
\centering
\includegraphics[scale=0.55]{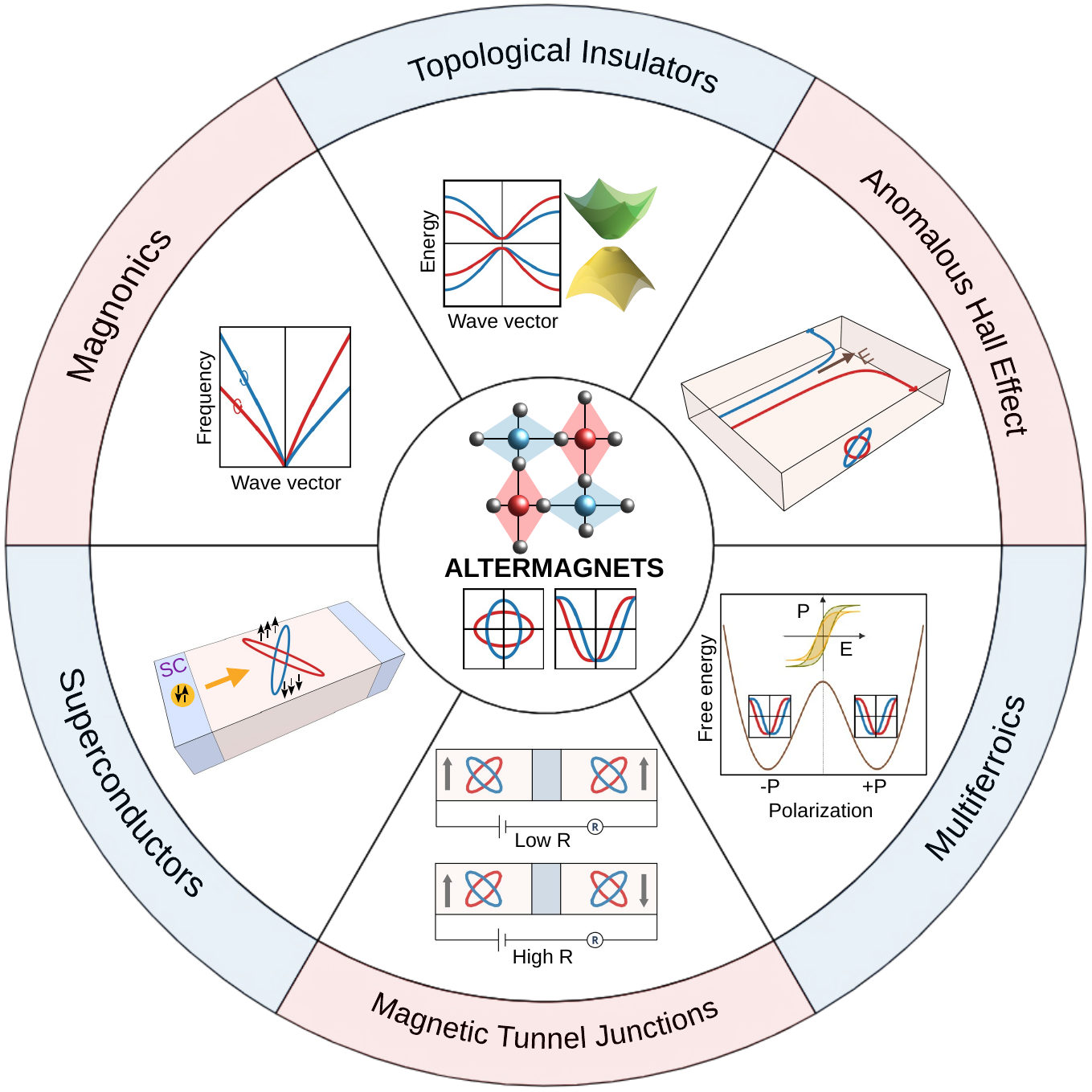}
\caption{\textbf{Applications and emerging cross coupling of altermagnetism with other fields.} Altermagnets exhibits rich interplay with a broad range of physical phenomena (clockwise), including topological insulators, the anomalous Hall effect, ferroelectrics, magnetic tunnel junctions, superconductivity, and magnonics, highlighting its potential for both fundamental studies and spintronic applications.}
\label{fig_application}
\end{figure*}
\section{Potential applications}
\label{sec:potential_applications}

The co-existence of spin splitting and compensated magnetization, along with the emergence of phenomena such as the anomalous Hall effect, spin splitter effect, spin-transfer torque, and other spin-dependent transport effects, make altermagnets promising candidates for a wide range of applications beyond fundamental studies. In this section, we discuss selected potential applications of altermagnets.

\subsection{Spintronics}

Altermagnets have emerged as promising class of magnets for spintronics by uniquely combining the advantages of ferromagnets and antiferromagnets~\cite{jungwirth2025altermagnetic,bai2024altermagnetism,song2025altermagnets,tamang2025altermagnetism}. Unlike ferromagnets, which generate strong non-relativistic spin-polarized currents and large magnetoresistive effects but suffer from stray fields, altermagnets exhibit vanishing net magnetization and zero stray fields. Unlike conventional antiferromagnets, which are free from stray fields but rely on weak relativistic spin-orbit torque effects due to their time-reversal-invariant, spin-degenerate band structure, altermagnets break time-reversal symmetry in a non-relativistic manner despite compensated magnetization~\cite{vsmejkal2022emerging}.

This non-relativistic spin splitting in altermagnetic bands produces spin-dependent anisotropic conductivities, enabling spin-polarized electrical currents analogous to ferromagnets. In $d$-wave altermagnets, this leads to a giant magnetoresistance (GMR) effect in altermagnetic multilayers, with \textit{ab initio} calculations predicting GMR comparable to those achieved in ferromagnetic spin valve devices~\cite{vsmejkal2022giant,jiang2023prediction}. Furthermore, altermagnetic tunnel junctions are also predicted to exhibit very large tunneling magnetoresistance (TMR) as demonstrated in RuO$_2$, Mn$_5$Si$_3$, and CrSb~\cite{sun2025tunneling,samanta2024tunneling,chi2024crystal, liu2024giant,shi2025spin}. Furthermore, Noh \textit{et al.} experimentally observed spin dependent TMR in RuO$_2$ based magnetic tunnel junctions~\cite{noh2025tunneling}. Interestingly, TMR is expected in all altermagnetic spin groups, while GMR is allowed specifically in $d$-wave altermagnets~\cite{vsmejkal2022giant,shao2021spin}.\\

Spin-charge interconversion plays a crucial role in spintronics. In altermagnets, the non-relativistic spin splitting, crystal orientation, and the direction of the Néel vector give rise to efficient and anisotropic spin-charge conversion\cite{lai2025d,wang2024inverse,zhang2026ultrahigh,sheoran2026tuning}. Among these materials, RuO$_2$ has emerged as a promising platform, offering both high spin-charge conversion efficiency and long spin transport length~\cite{zhang2024simultaneous,bai2022observation}. Besides these, spin-charge conversion has also been theoretically investigated in the chiral altermagnets Mn$_3$IrSi~\cite{hu2025spin} and TM$_3$X$_6$~\cite{tenzin2025persistent}.\\

Altermagnetic semiconductors, such as MnTe, offer opportunities for gate-tunable spintronics, overcoming the fundamental challenge of dilute ferromagnetic semiconductors, that of Curie temperatures well below room temperature, by exhibiting intrinsic magnetic ordering above room temperature with semiconducting properties. This opens possibilities for spin field-effect transistors and voltage-controlled spintronic devices~\cite{zhu2025altermagnetoelectric,craco2025theory,song2025altermagnets}.

Magnetoresistive Random-Access Memories (MRAMs) are non-volatile memory devices that store information in the form of magnetic states rather than electric charges.
Scalability of ferromagnetic MRAMs is limited by stray magnetic fields, which can lead to crosstalk and stability issues in densely packed devices. Antiferromagnetic MRAMs eliminate stray fields due to their zero magnetization. However, their spin-degenerate band structures limit the generation of spin-polarized currents and large GMR or TMR signals. Altermagnets provide a promising alternative by combining momentum-space spin splitting with compensated magnetization~\cite{dong2025field, sun2025tunneling}. Consequently, altermagnetic tunnel junctions have been proposed as potential building blocks for next-generation MRAMs~\cite{guo2025magnetic}. Several studies have demonstrated the potential of altermagnets as efficient electrode materials for magnetic tunnel junctions~\cite{vsmejkal2022giant,shao2021spin,xu2023spin}. Beyond the applications discussed above, altermagnets have also been proposed as candidates for advanced technological platforms, including neuromorphic computing architectures~\cite{kurenkov2020neuromorphic}.

\subsection{Cross coupling with other properties}
Altermagnets provide unexpected and exciting prospects for cross coupling with a number of other ordered phases. A very recent direction has been the exploration of coupling with multiferroic materials, where electric and magnetic degrees of freedom coexist and couple. Recent theoretical and computational efforts have established several distinct pathways to achieve ferroelectric altermagnets, where altermagnetic spin splitting is either driven by or coupled to ferroelectric polarization~\cite{sun2026unified,gu2025prl}. \v{S}mejkal demonstrated altermagnetic spin polarization in ferroelectric materials and proposed the altermagnetoelectric effect, which is a non-relativistic cross-coupling between altermagnetic spin splitting and ferroelectric polarization~\cite{vsmejkal2024altermagnetic}. In BaCuF$_4$ ($T_N \sim 275$ K) and Ca$_3$Mn$_2$O$_7$ ($T_N \sim 110$ K), octahedral rotations generating geometric ferroelectricity simultaneously produce $d$-wave altermagnetism. Reversing the octahedral rotation reverses both the polarization and the spin splitting. Symmetry analysis identified 11 polar altermagnetic spin point groups, including BiFeO$_3$ as an $i$-wave altermagnet, which could offer a pathway towards high-temperature magnetoelectric multiferroics. Hexagonal wurtzite MnSe is predicted to be a robust multiferroic altermagnet with a direct band gap of 1.98 eV and a spin splitting of up to 245 meV~\cite{bezzerga2025high}. Zhu and colleagues established a universal symmetry-based design principle for two-dimensional ferroelectric altermagnets, where lattice distortions break translational symmetry while preserving rotation-related symmetry to enable altermagnetism~\cite{zhu2025two}. Reversing ferroelectric polarization flips the spin splitting sign, and the magneto-optical Kerr effect has been proposed to allow experimental detection. Janus Cr$_2$SeO ($T_N \sim 280$ K)~\cite{khan2025altermagnetism} and V$_2$SeTeO ($T_N \sim 510$ K)~\cite{bezzerga2024giant} monolayers both exhibit altermagnetic spin splitting and significant out-of-plane piezoelectric responses, with piezoelectric coefficients of $d_{31}=1.18~\mathrm{pm\,V^{-1}}$ and $d_{31}=0.245~\mathrm{pm\,V^{-1}}$, respectively. The coexistence of altermagnetism, ferroelectricity, and piezoelectricity makes Janus altermagnets a class of multifunctional materials for next‑generation spintronics, ferroelectric memories, and piezoelectric devices.\\

The interplay between altermagnetism and superconductivity provides a unique platform for realizing unconventional superconducting states and advancing superconducting spintronics~\cite{mazin2025notes}. The coexistence of altermagnetic and superconducting phases, as proposed in La$_2$CuO$_4$ ~\cite{song1995electron,vsmejkal2022beyond},  Sr$_2$RuO$_4$\cite{autieri2025conditions} and SrRbCuO$_2$Cl$_2$\cite{li2025exploring} provides an opportunity to study the interplay between magnetism and superconductivity. The interplay between altermagnetism and superconductivity has been comprehensively reviewed in Ref.~\cite{liu2025altermagnetism}, with particular emphasis on the non-relativistic spin-momentum-locked Fermi surface and the multipole expansion framework. This has led to intriguing proposals for a field-free Josephson diode effect~\cite{cheng2024field,banerjee2024altermagnetic} and perfect superconducting diode effect~\cite{chakraborty2025perfect}. The orientation-dependent anisotropic Andreev reflection in altermagnets has led to proposals for direct-current Josephson junctions exhibiting controllable $0-\pi$ oscillations, which may be promising for superconducting qubits~\cite{sun2023andreev,beenakker2023phase,papaj2023andreev,lu2024varphi}. By manipulating the N\'eel vector direction, the momentum-split bands of an altermagnet can be aligned with those of a superconductor, effectively switching Andreev reflection on and off. Additionally, the interfacial proximity between superconductors and altermagnets is predicted to generate Majorana bound states in the absence of net magnetization, potentially allowing a path towards realizing topological quantum computing platforms\cite{ghorashi2024altermagnetic}. Furthermore, altermagnets can intrinsically support finite-momentum Cooper pairing~\cite{zhang2024finite} and spin-triplet Cooper pairs, enabling spin transmission without net magnetization~\cite{mazin2025notes}.

The cross-coupling of altermagnetism and topological band theory has emerged as exciting new frontier in this field. Owing to their characteristic momentum-dependent spin splitting in the absence of spin-orbit coupling, altermagnets naturally host Berry-curvature hot spots and nontrivial band topology. Altermagnetic heterostructures provide a  promising platform for engineering and controlling Chern insulating phases. By combining altermagnetic layers with suitable substrates, gate dielectrics, or other two-dimensional materials, the interfacial symmetry can be modified. The resulting symmetry breaking opens a topological gap, giving rise to Chern insulator phases~\cite{jiang2026altermagnetism,li2025stacking,chen2026altermagnets}. Tagani \textit{et al.} showed that an external magnetic field can induce a topological phase transition from a trivial altermagnetic semiconductor to a Chern insulator~\cite{bagheri2026quantum}. Beyond the quantum anomalous Hall regime, the quantum spin Hall effect has been predicted in altermagnets~\cite{gonzalez2025model,antonenko2025mirror}. Several quantized transport and optical phenomena, such as the photogalvanic effect~\cite{yoshida2026quantization} and the topological piezomagnetic effect~\cite{radhakrishnan2026topological}, also emerge from the nontrivial topology of altermagnetic band structures. Layered altermagnets exhibit a wide variety of Hall responses. Qin and Chen predicted a layer Hall effect, in which an in-plane electric field generates Hall voltages of opposite sign in adjacent layers~\cite{qin2026layer}. In addition, the planar Hall effect in altermagnets can approach half-quantized values. Ezawa demonstrated that three-dimensional altermagnets exhibit an almost half-quantized planar Hall conductivity plateau arising from the Berry curvature of spin-split Fermi surfaces~\cite{ezawa2025almost}. Complementing this work, Korrapati \textit{et al.} derived an analytical expression for the half-quantized planar Hall effect in two-dimensional altermagnets~\cite{korrapati2025approximate}. Other intriguing topological phases, such as axion insulators~\cite{pournaghavi2021realization} and higher-order topological phases~\cite{huo2026altermagnetism}, have also been proposed in altermagnets. The application of light has emerged as a powerful approach for controlling the topological properties of altermagnets. Zou \textit{et al.} demonstrated a Floquet-driven quantum anomalous Hall effect in Janus V$_2$XTeO (X = Se, S) monolayers with in-plane magnetization~\cite{zou2025floquet}. Floquet engineering has also been shown to realize quantum anomalous Hall insulating phases with tunable Chern numbers, dynamically generated higher-order spin-orbit couplings, and controllable spin textures~\cite{ghorashi2025dynamical,cheraghchi2026floquet,ganguli2026tunable}. These studies establish a connection between altermagnetism and Floquet engineering, highlighting altermagnets as promising platforms for realizing and manipulating nontrivial topological phases. This emerging interplay between altermagnetism and topology not only enriches the fundamental understanding of symmetry‑protected topological phases but also paves the way for potentially realizing spintronic and quantum devices that exploit dissipationless edge, surface, and corner states.

Altermagnets offer an exceptionally promising avenue for magnonics. Magnons are the quanta of spin waves, and analogous to electronic band dispersions, magnon bands also exhibit a similar kind of splitting in altermagnets~\cite{vsmejkal2023chiral,hoyer2025altermagnetic,zhang2025chiral,morano2025absence}. For instance, using inelastic neutron scattering, Liu \textit{et al.} have experimentally confirmed the existence of chiral magnon splitting in MnTe~\cite{liu2024chiral}. In altermagnets, the degeneracy of chiral magnon bands is lifted without an external magnetic field, which distinguishes them from conventional antiferromagnets. The non-relativistic spin splitting arises from direction-dependent exchange interactions that are influenced by the local environment~\cite{bhowal2025non}. Altermagnets can simultaneously support ultrafast terahertz spin waves and efficient spin-current generation through spin pumping \cite{sun2023spin,hodt2024spin} or thermal effects\cite{cui2023efficient,liao2024separation}. These properties reveal altermagnets as promising materials for low-power, ultrafast magnonic devices~\cite{vsmejkal2022emerging,song2025altermagnets}. Finally, we note that heterostructures of altermagnets with conventional ferromagnets have been predicted to host non-Hermitian phenomena, including exceptional points~\cite{reja2024emergence,reja2025neel}. Explorations along this direction will uncover new possibilities for engineering dissipative phenomena in unconventional magnetic junctions.

\section{Future directions and prospects}
\label{sec:future_directions}

Altermagnetism has rapidly grown from a theoretical classification into a vibrant experimental field, yet many fundamental and applied questions remain open. In this final section, we highlight the most pressing challenges and outline how chemical design principles may help address them.

\subsection{The gap in experimental confirmation}

The most immediate challenge facing the field is a quantitative imbalance between prediction and confirmation. High-throughput computational surveys have identified hundreds of candidate altermagnets from magnetic structure databases, but only a handful have been confirmed experimentally through spectroscopic or transport measurements. For the vast majority of predicted candidates, neither the altermagnetic spin splitting nor the associated transport signatures have been measured. This gap reflects practical bottlenecks in materials synthesis. Many predicted altermagnets are difficult to grow as high-quality single crystals, and the thin-film growth conditions required for ARPES or transport studies have been optimized for only a few systems.

Chemistry can help close this wide experiment-theory gap. The techniques of solid-state synthesis are well matched to the challenge of producing phase-pure, single-domain samples of altermagnetic candidates. A systematic effort to grow crystals of the most promising computationally predicted altermagnets, guided by the chemical design principles that we outlined in Section~\ref{sec:chemical_design}, would substantially accelerate experimental progress.

\subsection{The debate on RuO$_2$ and its lessons}

The debate surrounding the magnetic ground state of RuO$_2$ has been an instructive episode in the short history of altermagnetism. Early neutron diffraction measurements were interpreted as evidence for itinerant antiferromagnetic order in RuO$_2$~\cite{berlijn2017itinerant}. This conclusion was further reinforced by resonant x-ray scattering experiments, which reported signatures consistent with antiferromagnetism~\cite{zhu2019anomalous}. In 2020 \v{S}mejkal \textit{et al.} proposed rutile RuO$_2$ as a prototypical altermagnet on the basis of DFT calculations~\cite{vsmejkal2020crystal}. Subsequent experiments appeared to support this interpretation. Between 2022 and 2024, transport studies reported anomalous Hall responses~\cite{feng2022anomalous} and efficient spin-current generation~\cite{bai2022observation,karube2022observation,bai2023efficient,zhang2024simultaneous,guo2024direct,liao2024separation}. In 2024, ARPES measurements revealed momentum-dependent spin splitting and a characteristic $d$-wave spin texture~\cite{lin2024observation}, while magnetic circular dichroism experiments reported signatures of time-reversal symmetry breaking~\cite{fedchenko2024observation}. However, a series of recent studies have challenged this consensus. Muon spin rotation measurements on high-purity crystals detected negligible local magnetic moments~\cite{hiraishi2024nonmagnetic}, and subsequent neutron diffraction experiments on stoichiometric single crystals found no evidence of static magnetic order~\cite{kessler2024absence}. Independent ARPES measurements further reported electronic structures consistent with a nonmagnetic state~\cite{liu2024absence}. Furthermore, optical conductivity~\cite{wenzel2025fermi} and quantum oscillation measurements~\cite{wu2025fermi} indicated a conventional Fermi-liquid ground state. Moreover, spin-charge conversion effects previously attributed to inverse spin splitting were reinterpreted in terms of an anisotropic inverse spin Hall effect~\cite{wang2024inverse}.

The emerging picture remains fairly complex. Stoichiometric bulk RuO$_2$ is likely non-magnetic, but the energy difference between the non-magnetic and antiferromagnetic states is quite small, only approximately 23 meV per formula unit according to high-accuracy diffusion quantum Monte Carlo calculations~\cite{ahn2026nonmagnetic}. This places RuO$_2$ near a strain-tunable magnetic instability. A modest compressive strain can stabilize the antiferromagnetic (and hence altermagnetic) state. Epitaxial thin films, which are subject to substrate-induced strain, may therefore be magnetic even when the bulk is not. Defects, vacancies, and non-stoichiometry further complicate the picture~\cite{smolyanyuk2024fragility}.

The RuO$_2$ case carries several lessons for this rapidly growing field. DFT predictions of altermagnetic order should be treated with caution when the energy difference between magnetic and non-magnetic states is small and sensitive to the choice of exchange-correlation functional or Hubbard-$U$ parameter. Meta-GGA calculations on RuO$_2$ confirm that the non-magnetic ground state is correctly identified when a sufficiently accurate functional is used~\cite{meinert2025meta}. Moreover, experimental claims of altermagnetism must be supported by multiple independent techniques. A transport anomaly alone is insufficient, as conventional mechanisms can produce similar signals as altermagnetism. Finally, strain engineering may provide a potential route to stabilizing altermagnetic order in materials that are borderline magnetic.

\subsection{Altermagnets from molecular chemistry}

The vast majority of known and predicted altermagnets are extended inorganic solids. These include oxides, fluorides, chalcogenides, pnictides, and intermetallics. Whether molecular chemistry can produce altermagnetic order is an open and tantalizing question.

As we discussed in Section~\ref{sec:material_classes}, MOFs and COFs are promising platforms because their crystal structures are built from discrete building blocks whose geometry can be rationally designed. If the framework topology enforces a rotational relationship between the coordination environments of opposite-spin metal centers, the symmetry conditions for altermagnetism would be satisfied. The tunability of linker length, functionality, and flexibility in MOFs and COFs offers a degree of chemical control over the sublattice-connecting symmetry that is difficult to achieve in dense inorganic structures.

Single-molecule magnets and molecular crystals present a different possibility. If a molecular crystal packs antiferromagnetically ordered paramagnetic molecules with a rotation relating opposite-spin sites, the molecular crystal as a whole could in principle be altermagnetic. The challenge is that molecular crystals tend to have low symmetry and weak intermolecular exchange, making it difficult both to achieve the required symmetry and to sustain magnetic order at accessible temperatures.

Hybrid organic-inorganic perovskites, in which organic cations template the inorganic magnetic sublattice, offer an intermediate approach. The organic cation can control the octahedral tilt pattern, and hence the sublattice-connecting symmetry, without directly participating in the exchange. This provides a potentially useful chemical handle for switching between antiferromagnetic and altermagnetic order.

\subsection{Beyond conventional altermagnets}

Although even-parity altermagnets have attracted most of the attention so far, recent studies have expanded the concept to odd-parity altermagnets. Furthermore, the framework of altermagnetism has been generalized beyond collinear magnetic order to encompass noncollinear spin configurations. These emerging developments have opened new research directions and will be briefly discussed next.

The altermagnetic classification applies to collinear magnetic structures. A natural question is whether the same symmetry-based framework can be extended to non-collinear magnets. Hellenes \textit{et al.} have shown that non-collinear, coplanar magnets in noncentrosymmetric crystals can host $p$-wave spin polarization~\cite{hellenes2023p}. This is the magnetic analogue of unconventional $p$-wave superfluidity. In these $p$-wave magnets, the spin splitting has odd parity, unlike the even parity $d/g/i$-wave splitting of altermagnets. Notably, the spin texture in the odd parity case is qualitatively different from the even parity setting~\cite{brekke2024minimal}.  
Sim and Rachel established an interacting microscopic model for $p$-wave magnetism based on the Hubbard model, demonstrating its relevance to the honeycomb magnet Ni$_2$Mo$_3$O$_8$~\cite{sim2026quantum}. The interplay between non-relativistic spin--orbit coupling and inversion symmetry in $p$-wave magnets has been explored in Ref.~\cite{fakhredine2026interplay,liu2026nonrelativistic}. These studies provide symmetry-based design principles for engineering tunable spin textures and realizing $p$-wave magnetism. First-principles calculations on the $p$-wave candidate material CeNiAsO reveal a non-relativistic Edelstein effect that is approximately 25 times larger than the highest reported relativistic Edelstein effect, as demonstrated by Chakraborty \textit{et al.}~\cite{chakraborty2025highly}. The work establishes $p$‑wave magnets as a highly efficient platform for charge‑to‑spin conversion. Notably, the electrical switching of a $p$-wave magnets has been experimentally demonstrated very recently~\cite{song2025electrical,yamada2025metallic}. 

The spin-group classification that underpins altermagnetism is emerging as a unified framework for all collinear and coplanar magnetic phases. For chemists, this opens additional design space where materials with non-collinear but coplanar spin arrangements, stabilized for example by geometric frustration or Dzyaloshinskii–Moriya interactions, can host spin-polarized electronic states without net magnetization and with $p$-wave nodal structure.

Another very recent extension has been to $s$-wave altermagnets. In contrast to conventional altermagnets, whose spin splittings alternate in sign across momentum space according to even-parity ($d$-, $g$-, or $i$-wave) or odd-parity ($p$- or $f$-wave) symmetries, $s$-wave altermagnets exhibit a nodeless spin splitting that may retain the same sign throughout the entire Brillouin zone. As a consequence, no spin-degenerate nodal planes are present in momentum space. This behavior originates from the fact that the two oppositely polarized magnetic sublattices are not related by any crystal symmetry. Despite this uniform momentum-space spin polarization, the total magnetization remains strictly zero, thereby distinguishing $s$-wave altermagnets from ferromagnets, which also lack nodal planes but possess a finite magnetic moment. Several mechanisms have been proposed to realize this state. One route exploits chemical inequivalence between the magnetic sublattices~\cite{spaldin2026there}. More generally, antiferromagnets can exhibit exact magnetic compensation when the moments on oppositely aligned sublattices are equal in magnitude. A representative example is provided by symmetry-inequivalent ions such as Fe$^{3+}$ and Mn$^{2+}$, each carrying a local moment of $5 \mu_{\mathrm{B}}$, whose antiferromagnetic alignment leads to a fully compensated state. When at least one spin channel is insulating, the vanishing net magnetization becomes robust~\cite{pickett1998spin}. An alternative mechanism, proposed by Durrnagel \textit{et al.}, relies on valley-exchange symmetries acting as momentum-space translations, thereby extending the symmetry framework for compensated spin-split states beyond conventional crystallographic spin-group classifications~\cite{durrnagel2025extended}. Beyond the $p$-wave and $s$-wave classes, the concept of altermagnetism can also be extended to systems with noncollinear spin configurations~\cite{cheong2024altermagnetism,hu2025spin,chen2024mn}.

For all these classes, known material candidates remain scarce and chemistry-guided ideas are needed to discover these unconventional altermagnetic features in real materials.

\subsection{Challenges of N\'eel temperature}

For any technological application, the altermagnetic state must be stable at or above room temperature. The N\'eel temperatures of confirmed altermagnets span a wide range. MnTe (307 K) is marginally above room temperature, while CrSb (705 K) is comfortably above it, but many predicted candidates have their N\'eel temperatures well below 300 K. The altermagnetic spin splitting itself decreases with temperature roughly as the sublattice magnetization, so even in materials with N\'eel temperature higher than 300 K the splitting at room temperature may be significantly reduced from its zero-temperature value.

Chemical strategies for raising the N\'eel temperature are well established. These include strengthening the superexchange, increasing the magnetic moment, and enhancing the dimensionality of the exchange network. What is less clear is whether these strategies can be applied without simultaneously destroying the crystal-field anisotropy that drives the altermagnetic splitting. The ideal material has both a strong exchange and a large anisotropic orbital overlap, and these two requirements are not obviously correlated. As a unique example, CrSb appears to achieve both but whether this combination is rare or common among candidate materials remains to be established by systematic experimental studies.

Chemical pressure and isovalent substitution offer routes to tune the N\'eel temperature while preserving crystal symmetry. The magnetovolume coupling in MnTe, for example, is among the largest known for any antiferromagnet, and neutron diffraction under applied pressure has shown that the N\'eel temperature increases significantly while the ordered moment decreases~\cite{carlisle2025tuning}. This suggests that the magnetic properties of altermagnets may be tuned mechanically as well.

\section{Outlook}
\label{sec:outlook}

We close with a forward-looking perspective on what chemistry and chemical design principles can contribute to this emerging field of altermagnetism. The central insight of this review is that altermagnetism is not an exotic electronic state requiring unusual ingredients. Rather, it emerges from the well-established chemistry of coordination environments and exchange pathways, whenever the crystal structure places opposite-spin sublattices in a rotational relationship. This means that the vast knowledge base of crystal chemistry, coordination chemistry, and solid-state synthesis accumulated over the past century is directly applicable to the design and discovery of new altermagnets. Many concrete opportunities arise in this young and rapidly developing field. These include the rational design of altermagnetic MOFs and COFs through framework topology engineering, the use of high-pressure and epitaxial strain to tune materials across the antiferromagnet–altermagnet boundary, the exploitation of Jahn-Teller and orbital ordering effects to maximize the crystal-field anisotropy and the development of molecular altermagnets through crystal engineering of coordination compounds. We expect the next decade to bring a rapid expansion of altermagnetic materials, driven increasingly by chemical intuition and design.

\section*{Author contributions}

All authors conceptualized the review and contributed to its writing.

\section*{Conflicts of interest}

There are no conflicts to declare.

\section*{Data availability}

No primary research results, software or code have been included and no new data were generated or analyzed as part of this review.

\section*{Acknowledgments}

AN acknowledges support from DST CRG grant (CRG/2023/000114) and DST Nanomission (DST/NM/QM-10/2019). MAR acknowledges the graduate fellowship support received from the Indian Institute of Science.

\bibliography{rsc}
\end{document}